\documentclass[varenna]{cimento}
\usepackage{graphicx,amsmath}

\makeatletter
\def\alt{\mathrel{\mathpalette\vereq<}}
\def\vereq#1#2{\lower3pt\vbox{\baselineskip1.5pt \lineskip1.5pt
\ialign{$\m@th#1\hfill##\hfil$\crcr#2\crcr\sim\crcr}}}
\def\agt{\mathrel{\mathpalette\vereq>}}
\makeatother

\def\I{{\rm i}}

\shorttitle{Neutrinos and the stars}
\title{\hbox to 0pt{\rm\normalsize
\vbox to 0pt{\vskip-4cm \noindent Proceedings ISAPP School ``Neutrino Physics and Astrophysics,''
26 July--5 August 2011, Villa Monastero, Varenna, Lake Como, Italy\vfil}\hfilneg}Neutrinos and the Stars}
\author{Georg G.~Raffelt}
\institute{Max-Planck-Institute f\"ur Physik
(Werner-Heisenberg-Institut)\\
F\"ohringer Ring 6, 80805 M\"unchen, Germany}

\begin{document}

\maketitle

\begin{abstract}
The role of neutrinos in stars is introduced for students with little
prior astrophysical exposure. We begin with neutrinos as an
energy-loss channel in ordinary stars and conversely, how stars
provide information on neutrinos and possible other low-mass
particles. Next we turn to the Sun as a measurable source of
neutrinos and other particles. Finally we discuss supernova (SN)
neutrinos, the SN~1987A measurements, and the quest for a
high-statistics neutrino measurement from the next nearby SN. We also
touch on the subject of neutrino oscillations in the high-density SN
context.
\end{abstract}

\section{Introduction}                        \label{sec:introduction}

Neutrinos were first proposed in 1930 by Wolfgang Pauli to explain,
among other problems, the missing energy in nuclear beta decay.
Towards the end of that decade, the role of nuclear reactions as an
energy source for stars was recognized and the hydrogen fusion
chains were discovered by Bethe~\cite{Bethe1939} and von
Weizs\"acker~\cite{Weizsacker1938}. It is intriguing, however, that
these authors did not mention neutrinos---for example, Bethe writes
the fundamental pp reaction in the form ${\rm H}+{\rm H}\to{\rm
D}+\epsilon^+$. It was Gamow and Schoenberg in 1940 who first
stressed that stars must be powerful neutrino sources because beta
processes play a key role in the hydrogen fusion reactions and
because of the feeble neutrino interactions that allow them to
escape unscathed~\cite{Gamow1940}. Moreover, the idea that supernova
explosions had something to do with stellar collapse and
neutron-star formation had been proposed by Baade and Zwicky in 1934
\cite{Baade:1934}, and Gamow and Schoenberg (1941) developed a first
neutrino theory of stellar collapse~\cite{Gamow1941}. Solar
neutrinos were first measured by Ray Davis with his Homestake
radiochemical detector that produced data over a quarter century
1970--1994 \cite{Cleveland:1998nv} and since that time solar
neutrino measurements have become routine in many experiments. The
neutrino burst from stellar collapse was observed only once when the
star Sanduleak $-69\,202$ in the Large Magellanic Cloud, about
160,000 light years away, exploded on February 23, 1987 (Supernova
1987A). The Sun and SN~1987A remain the only measured astrophysical
neutrino sources.

Stars for sure are prime examples for neutrinos being of practical
relevance in nature. The smallness of neutrino masses compared with
stellar temperatures ensures their role as radiation. The weak
interaction strength ensures that neutrinos freely escape once
produced, except for the case of stellar core collapse where even
neutrinos are trapped, but still emerge from regions where nothing
else can directly carry away information except gravitational waves.
The properties of stars themselves can sometimes provide key
information about neutrinos or the properties of other low-mass
particles that may be emitted in analogous ways. The Sun is used as
a source of experimental neutrino or particle measurements. The
SN~1987A neutrino burst has provided a large range of
particle-physics limits. Measuring a high-statistics neutrino light
curve from the next nearby supernova will provide a bonanza of
astrophysical and particle-physics information. The quest for such
an observation and measuring the diffuse neutrino flux from all past
supernovae are key targets for low-energy neutrino astronomy.

The purpose of these lectures is to introduce an audience of young
neutrino researchers, with not much prior exposure to astrophysical
concepts, to the role of neutrinos in stars and conversely, how
stars can be used to gain information about neutrinos and other
low-mass particles that can be emitted in similar ways. We will
describe the role of neutrinos in ordinary stars and concomitant
constraints on neutrino and particle properties
(Section~\ref{sec:ordinarystars}). Next we turn to the Sun as a
measurable neutrino and particle source (Section~\ref{sec:sun}). The
third topic are collapsing stars and the key role of supernova
neutrinos in low-energy neutrino astronomy (Section~\ref{sec:SN}).

\section{Neutrinos from ordinary stars}      \label{sec:ordinarystars}

\subsection{Some basics of stellar evolution}

An ordinary star like our Sun is a self-gravitating ball of hot gas.
It can liberate gravitational energy by contraction, but of course
its main energy source is nuclear binding energy. During the initial
phase of hydrogen burning, the effective reaction is
\begin{equation}
4p+2e^-\to{}^4{\rm He}+2\nu_e+26.7~{\rm MeV}\,.
\end{equation}
In detail, the reactions can proceed through the pp chains
(Table~\ref{tab:ppchains}) or the CNO cycle (Table~\ref{tab:CNO}).
The latter contributes only a few percent in the Sun, but dominates
in slightly more massive stars due to its steep temperature
dependence. Neutrinos carry away a few percent of the energy, in
detail depending on the reaction channels. Based on the solar photon
luminosity of \footnote{Following astrophysical convention, we will
use cgs units, often mixed with natural units, where $\hbar=c=k_{\rm
B}=1$.} $L_\odot=3.839\times10^{33}~{\rm erg}~{\rm s}^{-1}$ one can
easily estimate the solar neutrino flux at Earth to be about
$6.6\times10^{10}~{\rm cm}^{-2}~{\rm s}^{-1}$.

\begin{table}
  \caption{Hydrogen burning by pp chains.}
  \label{tab:ppchains}
  \begin{tabular}{lllll}
    \hline
    Termination&Reaction&Branching&Neutrino&Name\\
    &&(Sun)&Energy [MeV]&\\
    \hline
    &$p+p\to d+e^++\nu_e$   &99.6\%&$<0.423$&pp\\
    &$p+e^-+p\to d+\nu_e$    &0.44\%&1.445&pep\\
    &$d+p\to{}^3{\rm He}+\gamma$\\
    PP I&${}^3{\rm He}+{}^3{\rm He}\to{\bf{}^4{\bf He}}+2p$&85\%\\
    \hline
    &${}^3{\rm He}+{}^4{\rm He}\to{}^7{\rm Be}$&15\%\\
    &${}^7{\rm Be}+e^-\to{}^7{\rm Li}+\nu_e$&90\%&0.863&Beryllium\\
    &${}^7{\rm Be}+e^-\to{}^7{\rm Li}^*+\nu_e$&10\%&0.385&Beryllium\\
    PP II&${}^7{\rm Li}+p\to{\bf{}^4{\bf He}+{}^4{\bf He}}$\\
    \hline
    &${}^7{\rm Be}+p\to{}^8{\rm B}+\gamma$&0.02\%\\
    &${}^8{\rm B}+p\to{}^8{\rm Be}^*+e^++\nu_e$&&$<15$&Boron\\
    PP III&${}^8{\rm Be}^*\to{\bf{}^4{\bf He}+{}^4{\bf He}}$\\
    \hline
    hep&${}^3{\rm He}+p\to{\bf{}^4{\bf He}}+e^++\nu_e$&
    $3\times10^{-7}$&$<18.8$&hep\\
    \hline
  \end{tabular}
\end{table}

In the simplest case we model a star as a spherically symmetric
static structure, excluding phenomena such as rotation, convection,
magnetic fields, dynamical evolution such as supernova explosion,
and so forth. Stellar structure is then governed by three
conditions. The first is hydrostatic equilibrium, i.e.\ at each
radius $r$ the pressure $P$ must balance the gravitational weight of
the material above, or in differential form
\begin{equation}\label{eq:hydroequilibrium}
\frac{dP}{dr}=-\frac{G_{\rm N}M_r\rho}{r^2}\,,
\end{equation}
where $G_{\rm N}$ is Newton's constant, $\rho$ the local mass
density, and $M_r=\int_0^r dr'\,\rho(r')\,4\pi r'^2$ the integrated
stellar mass up to radius $r$.

Energy conservation implies that the energy flux $L_r$ flowing
through a spherical surface at radius $r$ can only change if there
are local sources or sinks of energy,
\begin{equation}
\frac{dL_r}{dr}=4\pi r^2\,\epsilon\,\rho\,.
\end{equation}
The local rate of energy generation $\epsilon$, measured in ${\rm erg}~{\rm
  g}^{-1}~{\rm s}^{-1}$, is the sum of nuclear and gravitational energy
release, reduced by neutrino losses,
$ 
\epsilon=\epsilon_{\rm nuc}+\epsilon_{\rm grav}-\epsilon_{\nu}\,.
$ 

Finally the flow of energy is driven by a temperature gradient. If
most of the energy is carried by electromagnetic
radiation---certainly true at the stellar surface---we may express
the thermal energy density by that of the radiation field in the form
$\rho_\gamma= a T^4$ where the radiation-density constant is
$a=7.57\times10^{-15}~{\rm erg}~{\rm cm}^{-3}~{\rm K}^{-4}$ or in
natural units $a=\pi^2/15$. The flow of energy is then
\begin{equation}\label{eq:energytransfer}
L_r=\frac{4\pi r^2}{3\kappa\rho}\,\frac{d(aT^4)}{dr}\,,
\end{equation}
where $\kappa$ (units ${\rm cm}^2~{\rm g}^{-1}$) is the opacity. The
photon contribution (radiative opacity) is
$ 
\kappa_\gamma\rho=\langle\lambda_\gamma\rangle^{-1}_{\rm Rosseland}\,.
$ 
In other words, $(\kappa_\gamma\rho)^{-1}$ is a  spectral average
(``Rosseland mean'') of the photon mean free path $\lambda_\gamma$.
Radiative transfer corresponds to photons carrying energy in a
diffusive way with typical step size $\lambda_\gamma$. Energy is
also carried by electrons (``conduction''), the total opacity being
$\kappa^{-1}=\kappa_{\gamma}^{-1}+\kappa_{\rm c}^{-1}$.

In virtually all stars there are regions that are convectively
unstable and energy transport is dominated by convection, a
phenomenon that breaks spherical symmetry. In practice, convection
is treated with approximation schemes. In our Sun, the outer layers
beyond about $0.7\,R_\odot$ (solar radius) are convective.

\begin{table}
  \caption{Hydrogen burning by the CNO cycle.}
  \label{tab:CNO}
  \begin{tabular}{ll}
    \hline
    Reaction&Neutrino Energy [MeV]\\
    \hline
    ${}^{12}{\rm C}+p\to{}^{13}{\rm N}+\gamma$\\
    ${}^{13}{\rm N}\to{}^{13}{\rm C}+e^++\nu_e$&$< 1.199$\\
    ${}^{13}{\rm C}+p\to{}^{14}{\rm N}+\gamma$\\
    ${}^{14}{\rm N}+p\to{}^{15}{\rm O}+\gamma$\\
    ${}^{15}{\rm O}\to{}^{15}{\rm N}+e^++\nu_e$&$< 1.732$\\
    ${}^{15}{\rm N}+p\to{}^{12}{\rm C}+{}^{\bf 4}{\bf He}$\\
    \hline
  \end{tabular}
\end{table}

The stellar structure equations must be solved with suitable
boundary condition at the center and stellar surface. From nuclear,
neutrino and atomic physics calculations one needs the
energy-generation rate $\epsilon$ and the opacity $\kappa$, both
depending on density, temperature and chemical composition. In
addition one needs the equation of state, relating the thermodynamic
quantities $P$, $\rho$ and $T$, again depending on chemical
composition. For detailed discussions we refer to the textbook
literature~\cite{Clayton:1968,Kippenhahn:1990}.

However, simple reasoning can reveal deep insights without solving
the full problem. For a self-gravitating system, the virial theorem
is one of those fundamental propositions that explain many puzzling
features. One way of deriving it in our context is to begin with the
equation of hydrostatic equilibrium in
eq.~(\ref{eq:hydroequilibrium}) and integrate both sides over the
entire star, $\int_0^R dr\,4\pi r^3\,P'=-\int_0^R dr\,4\pi
r^3\,G_{\rm N} M_r\rho/r^2$ where $P'=dP/dr$. The rhs is the
gravitational binding energy $E_{\rm grav}$ of the star. After
partial integration of the lhs with the boundary condition $P=0$ at
the surface, one finds $-3\int_0^R dr\,4\pi r^2 P=E_{\rm grav}$. If
we model the stellar medium as a monatomic gas we have the
relationship $P=\frac{2}{3}\,U$ between pressure and density of
internal energy, so the lhs is simply twice the total internal energy
which is the sum over the kinetic energies of the gas particles. Then
the average energy of a single ``atom'' of the gas and its average
gravitational energy are related by
\begin{equation}
\langle E_{\rm kin}\rangle=-\frac{1}{2}\,\langle E_{\rm grav}\rangle\,.
\end{equation}
This is the virial theorem for a simple self-gravitating system and
can be applied to everything from stars to clusters of galaxies.

In the latter case, Fritz Zwicky (1933) was the first to study the
motion of galaxies that form gravitationally bound systems. We may
write $E_{\rm kin} =\frac{1}{2}\,m\,v^2$ and $E_{\rm grav}=G_{\rm
N}M_r\,m\,r^{-1}$ so that the virial theorem reads $\langle
v^2\rangle=G_{\rm N}M \langle r^{-1}\rangle$. The lhs is the velocity
dispersion revealed by Doppler shifts of spectral lines whereas the
geometric size of the cluster is directly observed. This allowed
Zwicky to estimate the total gravitating mass $M$ of the Coma
cluster. It turned out to be far larger than luminous matter, leading
to the proposition of large amounts of dark matter in the
universe~\cite{Zwicky:1933}.

We next apply the virial theorem to the Sun and estimate its
interior temperature. We approximate the Sun as a homogeneous sphere
of mass $M_\odot=1.99\times10^{33}~{\rm g}$ and radius
$R_\odot=6.96\times10^{10}~{\rm cm}$. The gravitational potential of
a proton near the center is $E_{\rm grav}=-\frac{3}{2}\,G_{\rm N}
M_\odot m_p/R_\odot=-3.2~{\rm keV}$. In thermal equilibrium we have
$\langle E_{\rm kin}\rangle=\frac{3}{2}\,k_{\rm B} T$, so the virial
theorem implies $\frac{3}{2}\,k_{\rm B} T=-\frac{1}{2}\,E_{\rm
grav}=-3.2~{\rm keV}$ or $T\sim 1.1~{\rm keV}$. This is to be
compared with $T_{\rm c}=1.56\times10^7~{\rm K}=1.34~{\rm keV}$ for
the central temperature in standard solar models. Without any
detailed modeling we have correctly estimated the thermal energy
scale relevant for the solar interior and thus for hydrogen burning.

A crucial feature of a self-gravitating system is its ``negative
heat capacity.'' The total energy $\langle E_{\rm kin}+E_{\rm
grav}\rangle=\frac{1}{2}\langle E_{\rm grav}\rangle$ is negative.
Extracting energy from such a system and letting it relax to virial
equilibrium leads to contraction and an {\it increase} of the
average kinetic energy, i.e.\ to heating. Conversely, pumping energy
into the system leads to expansion and cooling. In this way a star
self-regulates its nuclear burning processes. If the ``fusion
reactor'' overheats, it builds up pressure, expands and thereby
cools, or conversely, if it underperforms it loses pressure,
contracts, heats, and thereby increases the fusion rates and thus
the pressure.

\begin{figure}
\centering
\includegraphics[width=0.8\textwidth]{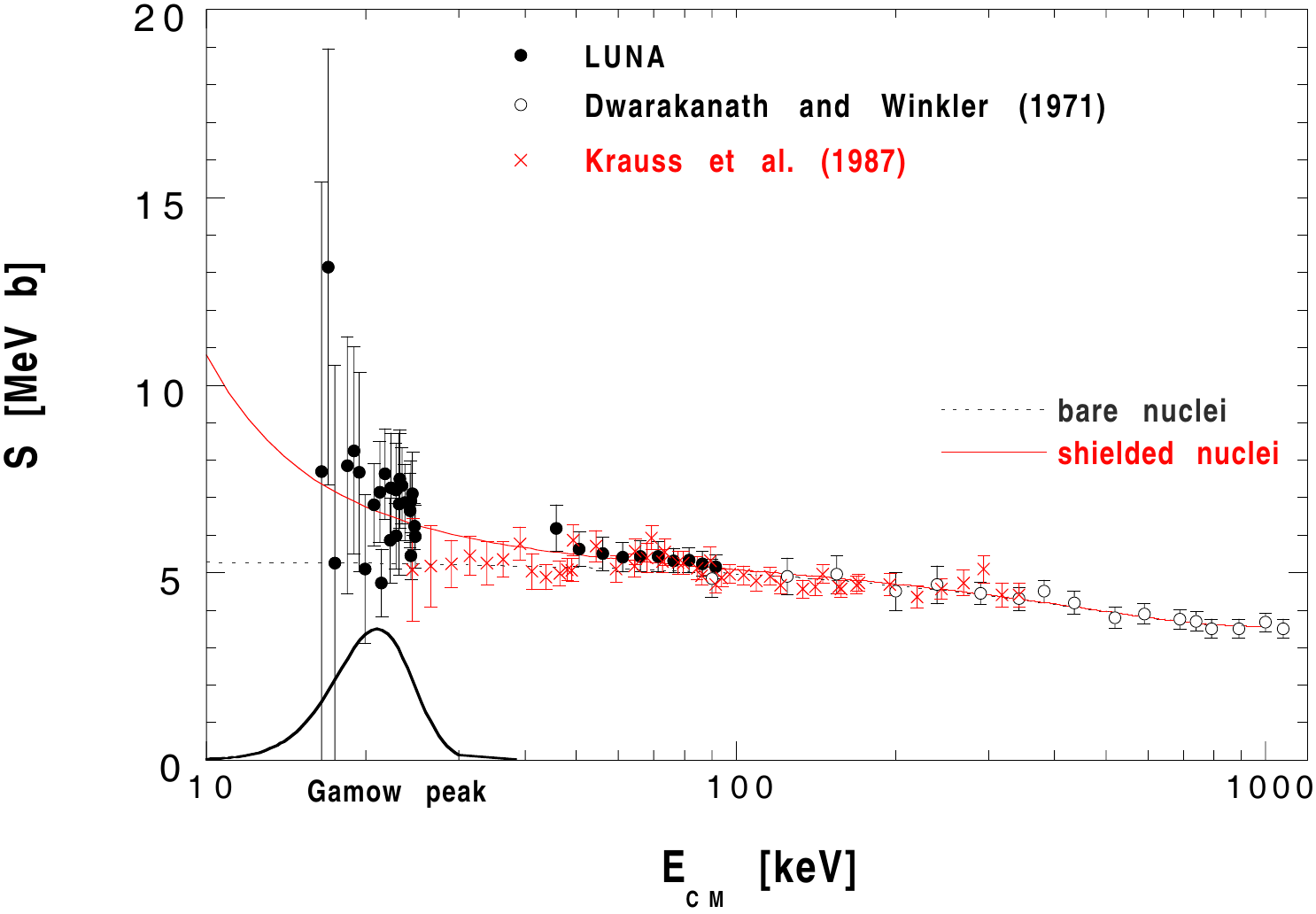}
\caption{First measurements of the ${}^3{\rm He}+{}^3{\rm He}\to{}^4{\rm He}+2p$
cross section by the LUNA collaboration~\cite{Bonetti:1999yt},
together with some previous measurements.
The solar Gamow peak is shown in arbitrary units.\label{fig:luna}}
\end{figure}

Nuclear reactions can only occur if the participants approach each
other enough for nuclear forces to come into play. To this end
nuclei must penetrate the Coulomb barrier. The quantum-mechanical
tunneling probability is proportional to $E^{-1/2}\,e^{-2\pi\eta}$
where $\eta=(m/2E)^{1/2}Z_1Z_2e^2$ is the Sommerfeld parameter with
$m$ the reduced mass of the two-body system with nuclear charges
$Z_1e$ and $Z_2 e$. Usually one expresses the relevant nuclear cross
sections in terms of the astrophysical S-factor
$S(E)=\sigma(E)\,E\,e^{2\pi\eta(E)}$ which is then a slowly varying
function of CM energy $E$. Thermonuclear reactions take place in a
narrow range of energies (``Gamow peak'') that arises from the
convolution of the tunneling probability with the thermal velocity
distribution. For more than a decade, the relevant low-energy cross
sections have been measured in the laboratory, notably the LUNA
experiment in the Gran Sasso underground laboratory. Their first
results for the ${}^3{\rm He}+{}^3{\rm He}$ fusion cross
section~\cite{Bonetti:1999yt} are shown in fig.~\ref{fig:luna}
together with the solar Gamow peak. The temperature is about 1~keV,
whereas the reaction probability peaks for CM energies of some
20~keV. Thermonuclear reactions depend steeply on temperature: If it
is too low, nothing happens, if it were too high, energy generation
would be explosive.

One consequence is that hydrogen burning always occurs at roughly
the same $T\sim 1$~keV. As discussed earlier, $T$ in the star
essentially corresponds to a typical gravitational potential by the
virial theorem. Since $E_{\rm grav}\propto M/R$ where $M$ is the
stellar mass and $R$ its radius, this ratio should be roughly the
same for all hydrogen burning stars and thus the stellar radius
scales roughly linearly with mass.

Once a star has burnt its hydrogen, helium burning sets in which
proceeds by the triple alpha reaction ${}^4{\rm He}+{}^4{\rm
He}+{}^4{\rm He}\to{}^{8}{\rm Be}+{}^4{\rm He}\to{}^{12}{\rm C}$.
There is no stable isotope of mass number 8 and $^{8}{\rm Be}$
builds up with a very small concentration of about $10^{-9}$.
Additional reactions are ${}^{12}{\rm C}+{}^4{\rm He}\to{}^{16}{\rm
O}$ and ${}^{16}{\rm O}+{}^4{\rm He}\to{}^{20}{\rm Ne}$. Helium
burning is extremely temperature sensitive and occurs approximately
at $T\sim 10^8$~K, corresponding roughly to 10~keV. The next phase
is carbon burning which proceeds by many reactions, for example
${}^{12}{\rm C}+{}^{12}{\rm C}\to{}^{23}{\rm Na}+p$ or ${}^{12}{\rm
C}+{}^{12}{\rm C}\to{}^{20}{\rm Ne}+{}^4{\rm He}$. It burns at
$T\sim 10^9$~K, corresponding roughly to 100~keV.

\begin{figure}
\centering
\includegraphics[width=0.8\textwidth]{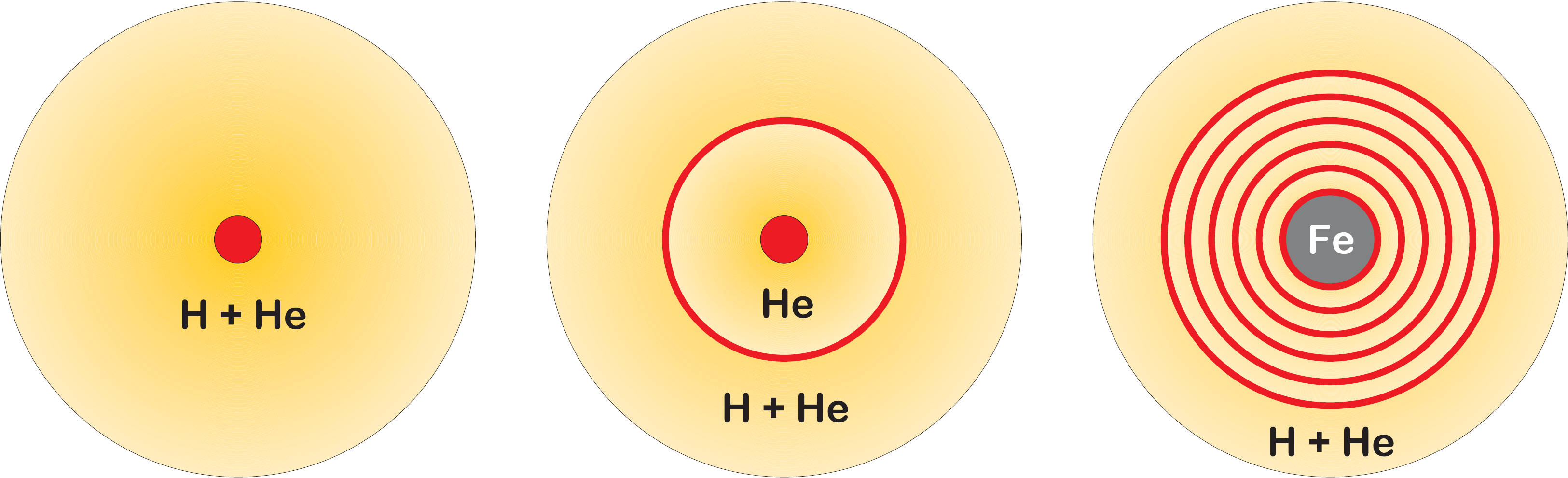}
\caption{Schematic structure of hydrogen and helium burning stars and
final ``onion skin structure'' before core collapse.
\label{fig:onionshell}}
\end{figure}

Stable thermonuclear burning, for the different burning phases,
occurs in a characteristic narrow range of temperatures, but broad
range of densities. Every star initially contains about 25\% helium,
originating from the big bang, and builds up more by hydrogen
burning, but helium burning will not occur at the hydrogen-burning
temperatures, and conversely, at the helium-burning $T$, hydrogen
burning would be explosive. Different burning phases must occur in
separate regions with different $T$. When a star exhausts hydrogen
in its center, it will make a transition to helium burning which
then occurs in its center, but hydrogen burning continues in a shell
inside of which we have only helium, outside a mixture of hydrogen
and helium (fig.~\ref{fig:onionshell}). When helium is exhausted in
the center, carbon burning is ignited, and so forth. A star more
massive than about 6--$8\,M_\odot$ goes through all possible burning
stages until an iron core is produced. As iron is the most tightly
bound nucleus, no further burning phase can be ignited.

A normal star is supported by thermal pressure, allowing for
self-regulated nuclear burning as explained earlier. A stable
configuration without nuclear burning is also possible when the star
supports itself by electron degeneracy pressure (white dwarfs). The
number density of a cold electron gas is related to the maximum
momentum, the Fermi momentum $p_{\rm F}$, by $n_e=p_{\rm
F}^3/(3\pi^3)$. A typical electron velocity is then $v=p_{\rm
F}/m_e$, assuming electrons are non-relativistic. The pressure $P$
is proportional to the number density times a typical momentum times
a typical velocity and thus $P\propto p_{\rm F}^5\propto
\rho^{5/3}\propto M^{5/3} R^{-5}$ where we have used that
$\rho\propto M\,R^{-3}$. If we approximate the pressure gradient as
$dP/dR\sim P/R$, together with the equation of hydrostatic
equilibrium, leads to $P\propto G_{\rm N} M\rho R^{-1}\propto M^2
R^{-4}$. We have already found $P\propto M^{5/3} R^{-5}$ and the two
conditions are consistent for $R\propto M^{-1/3}$. In contrast to
normal stars, white dwarfs are smaller for larger mass. From
polytropic stellar models one finds numerically
\begin{equation}
R=10,500~{\rm km}\,\left(\frac{0.6 M_\odot}{M}\right)^{1/3}\,(2Y_e)^{5/3}\,,
\end{equation}
where $Y_e$ is the number of electrons per nucleon. In other words, a
white dwarf is roughly the size of the Earth for roughly the mass of
the Sun.

The inverse mass-radius relation fundamentally derives from
electrons producing more pressure if they are squeezed into smaller
space, a manifestation of Heisenberg's uncertainty relation together
with Pauli's exclusion principle. However, if the white-dwarf mass
becomes too large and therefore its size very small, eventually
electrons become relativistic. In this case their typical velocity
is the speed of light and no longer $v=p_{\rm F}/m_e$. We lose one
power of $p_{\rm F}$ in the expression for the pressure that becomes
$P\propto p_{\rm F}^4\propto\rho^{4/3}\propto M^{4/3} R^{-4}$. We no
longer obtain a relation between $M$ and $R$, meaning that there is
no stable configuration. In polytropic models one finds explicitly
for the limiting white-dwarf mass, the Chandrasekhar limit,
\begin{equation}
M_{\rm Ch}=1.457\,M_\odot (2Y_e)^2\,.
\end{equation}
This result, combining quantum mechanics with relativistic effects,
was derived by the young Subrahmanyan Chandrasekhar on his way from
India to England in 1930 and was published the following
year~\cite{Chandrasekhar:1931}. This fundamental finding was
initially ridiculed by the experts, but later helped Chandrasekhar
win the 1983 physics Nobel prize.

We finally mention ``giant stars'' as another important phenomenon of
stellar structure. A normal star like our Sun has a monotonically
decreasing density from the center to the surface, but on the crudest
level of approximation could be described as a homogeneous sphere. On
the other hand a star with a core, especially with a small degenerate
core, tends to have a hugely inflated envelope and is then a giant
star. This behavior follows from the stellar structure equations, but
cannot be explained in a few sentences with a simple physical reason.
When a low-mass hydrogen-burning star like our Sun has exhausted
hydrogen in its center, it will develop a degenerate helium core and
at the same time expand its envelope and become a red giant. (For a
given luminosity and an expanding surface area, the surface
temperature must decline because thermal radiation, by the
Stefan-Boltzmann-law, is proportional to the surface area and $T^4$.)

\begin{table}[b]
  \caption{Evolution of stars, depending on their initial mass.}
  \label{tab:StellarEvolution}
  \begin{tabular}{llll}
    \hline
    Mass Range&Evolution&End State\\
    \hline
    $M\alt 0.08\,M_\odot$&Hydrogen burning never ignites&Brown Dwarf\\[1ex]
    $0.08\,M_\odot\alt M\alt0.8\,M_\odot$&Hydrogen burning not&Low-mass\\
    &completed in Hubble time&main-sequence star\\[1ex]
    $0.8\,M_\odot\alt M\alt 2\,M_\odot$&Degenerate helium core&Carbon-oxygen\\
    &after hydrogen exhaustion&white dwarf surrounded\\
    $2\,M_\odot\alt M\alt\hbox{6--8}\,M_\odot$&Helium ignition non-degenerate&by planetary nebula\\[1ex]
    $\hbox{6--8}\,M_\odot\alt M$&All burning phases&Neutron star (often pulsar)\\
    &$\to$ Onion skin structure&Sometimes black hole\\
    &$\to$ Core-collapse supernova&Supernova remnant (SNR)\\
    &&e.g.\ crab nebula\\
    \hline
  \end{tabular}
\end{table}

\begin{figure}
\centering
\includegraphics[width=1.0\textwidth]{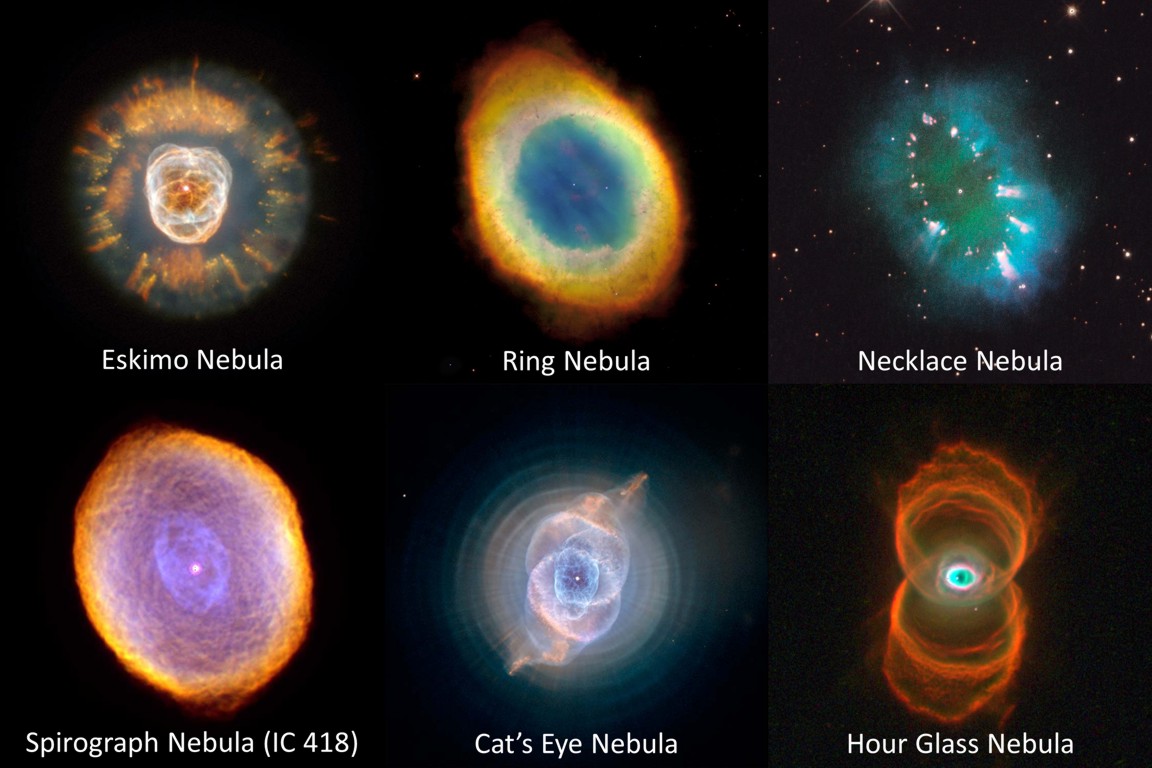}
\caption{Several planetary nebulae, the remnants of stars
with initial masses of a few $M_\odot$. Image credits: Necklace and
Cat's Eye Nebula: NASA, ESA, HEIC, and The Hubble Heritage Team
(STScI/AURA). Ring Nebula and IC418: NASA and The Hubble Heritage
Team (STScI/AURA). Hour Glass Nebula: NASA, R.~Sahai, J.~Trauger
(JPL), and The WFPC2 Science Team. Eskimo Nebula: NASA, A.~Fruchter
and the ERO Team (STScI).\label{fig:nebulae}}
\end{figure}

We can now roughly understand how stars live and die. If the mass is
too small, roughly below 8\% of the solar mass, hydrogen burning
never ignites, the star contracts and ``browns out'', eventually
forming a degenerate hydrogen star
(table~\ref{tab:StellarEvolution}). For masses up to about
$0.8\,M_\odot$, hydrogen burning will not finish within the age of
the universe and even the oldest such stars are still around today.
For masses up to a few $M_\odot$, stars ignite helium burning. After
its completion they develop a degenerate carbon-oxygen core and
inflate so much that they shed their envelope, forming what is
called a ``planetary nebula'' with a carbon-oxygen white dwarf as a
central star. Planetary nebulae are among the most beautiful
astronomical objects (fig.~\ref{fig:nebulae}). White dwarfs then
cool and become ever darker with increasing age. For initial masses
above 6--$8\,M_\odot$, stars will go through all burning phases and
eventually develop a degenerate iron core which will grow in mass
(and shrink in size) until it reaches the Chandrasekhar limit and
collapses, leading to a core-collapse supernova to be discussed
later.

\subsection{Neutrino emission processes}

During hydrogen burning, for every produced helium nucleus one needs
to convert two protons into two neutrons, so inevitably two neutrinos
with MeV-range energies emerge. Advanced burning stages consist
essentially of combining $\alpha$ particles to larger nuclei and do
not produce neutrinos in nuclear reactions. However, neutrinos are
still produced by several ``thermal processes'' that actually
dominate the stellar energy losses for carbon burning and more
advanced phases.

\begin{figure}
\centering
\includegraphics[width=0.6\textwidth]{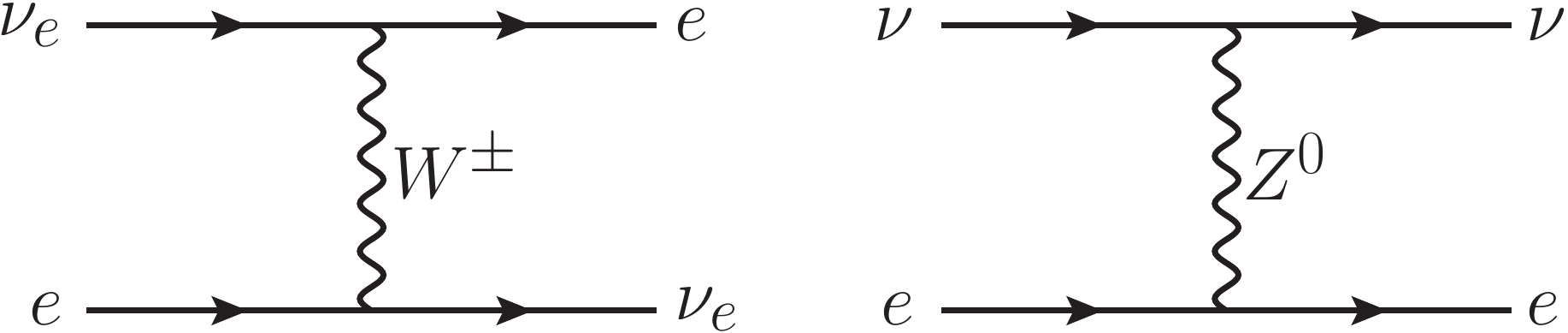}
\caption{Interaction of neutrinos with electrons by $W$ exchange (charged current)
and $Z$ exchange (neutral current).
\label{fig:neutrino-electron-coupling}}
\end{figure}

Thermal neutrino emission arises from processes involving electrons,
nuclei and photons of the medium and are based on the neutrino
interaction with electrons. Fundamentally this corresponds to either
$W$ or $Z$ exchange (fig.~\ref{fig:neutrino-electron-coupling}). For
the low energies characteristic of stellar interiors and even in the
collapsed core of a supernova, one can integrate out $W$ and $Z$ and
describe neutrino interactions with electrons and nucleons by an
effective four-fermion neutral-current interaction of the form
\begin{equation}\label{eq:NChamiltonian}
{\cal H}_{\rm int}=\frac{G_{\rm F}}{\sqrt{2}}\,
\overline\psi_f\gamma_\mu(C_V-C_A\gamma_5)\psi_f\,
\overline\psi_\nu\gamma^\mu(1-\gamma_5)\psi_\nu\,,
\end{equation}
where $G_{\rm F}=1.16637\times10^{-5}~{\rm GeV}^{-2}$ is the Fermi
constant. When $f$ is a charged lepton and $\nu$ the corresponding
neutrino, this effective neutral-current interaction includes a
Fierz-transformed contribution from $W$ exchange. The compound
effective $C_{V,A}$ values are given in table~\ref{tab:couplings}.
(Note that the $C_{V,A}$ for neutral currents are typically $\pm
1/2$, a factor that is sometimes pulled out so that the overall
coefficient becomes $G_{\rm F}/2\sqrt{2}$ and $C_{V,A}$ are twice the
values shown in table~\ref{tab:couplings}.) For neutrinos interacting
with the same flavor, a factor 2 for an exchange amplitude for
identical fermions was included. The $C_A$ values for nucleons are
often taken to be $\pm 1.26/2$, derived by isospin invariance from
the charged-current values. However, the strange-quark contribution
to the nucleon spin implies an isoscalar piece as
well~\cite{Raffelt:1993ix}, giving rise to the values shown in
table~\ref{tab:couplings}. For the effective weak mixing angle a
value $\sin^2\Theta_{\rm W}=0.23146$ was
used~\cite{Nakamura:2010zzi}.

\begin{table}
  \caption{Effective neutral-current couplings for the interaction
  Hamiltonian of eq.~(\ref{eq:NChamiltonian}).\label{tab:couplings}}
  \begin{tabular}{llllll}
    \hline
    Fermion $f$&Neutrino&$C_V$&$C_A$&$C_V^2$&$C_A^2$\\
    \hline
    Electron          &$\nu_e$           &$+1/2+2\sin^2\Theta_{\rm W}$&$+1/2$   &0.9376&0.25\\
                      &$\nu_{\mu,\tau}$  &$-1/2+2\sin^2\Theta_{\rm W}$&$-1/2$   &0.0010&0.25\\
    Proton            &$\nu_{e,\mu,\tau}$&$+1/2-2\sin^2\Theta_{\rm W}$&$+1.37/2$&0.0010&0.47\\
    Neutron           &$\nu_{e,\mu,\tau}$&$-1/2$                      &$-1.15/2$&0.25  &0.33\\
    Neutrino ($\nu_a$)&$\nu_{a}$         &$+1$                        &$+1$     &1.00  &1.00\\
                      &$\nu_{b\not=a}$   &$+1/2$                      &$+1/2$   &0.25  &0.25\\
    \hline
  \end{tabular}
\end{table}

\begin{figure}
\centering
\includegraphics[width=1.0\textwidth]{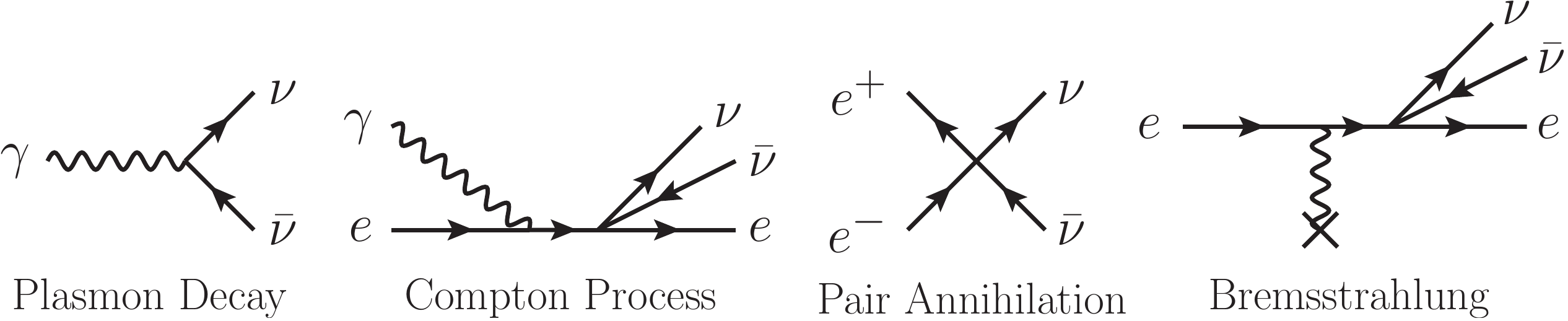}
\caption{Thermal neutrino emission processes in stars.\label{fig:processes}}
\end{figure}

\begin{table}
  \caption{Major burning stages of a $15\,M_\odot$ star and thermal neutrino
  losses~\cite{Weaver:1978zz}.}
  \label{tab:phases}
  \begin{tabular}{llrlllll}
    \hline
    Burning&Dominant&$T_{\rm c}$ [keV]&$\rho_{\rm c}$ [g/cm$^3$]&
    $L_\gamma$ [$10^4\,L_\odot$]&$L_\nu/L_\gamma$&Duration\\
    stage&process&&&&&[years]\\
    \hline
    Hydrogen&H  $\to$ He    &~~3&$5.9$          &2.1&---               &$1.2\times10^7$\\
    Helium  &He $\to$ C, O  &~14&$1.3\times10^3$&6.0&$1.7\times10^{-5}$&$1.3\times10^6$\\
    Carbon  &C  $\to$ Ne, Mg&~53&$1.7\times10^5$&8.6&$1.0$             &$6.3\times10^3$\\
    Neon    &Ne $\to$ O, Mg &110&$1.6\times10^7$&9.6&$1.8\times10^3$   &7.0\\
    Oxygen  &O  $\to$ Si    &160&$9.7\times10^7$&9.6&$2.1\times10^4$   &1.7\\
    Silicon &Si $\to$ Fe, Ni&270&$2.3\times10^8$&9.6&$9.2\times10^5$   &6~days\\
    \hline
  \end{tabular}
\end{table}

In the early history of neutrino physics it was thought that
neutrinos would be produced only in nuclear $\beta$-decay. After
Fermi formulated the $V{-}A$ theory in the late 1950s, however, it
became clear that neutrinos could have a direct coupling to
electrons, which today we understand as an effective low-energy
interaction. Around 1961--63 these ideas led to the proposition of
thermal neutrino processes in stars shown in
fig.~\ref{fig:processes}, i.e.\ plasmon decay, the photo or Compton
production process, pair annihilation, and bremsstrahlung by
electrons interacting with nuclei or other electrons. While thermal
neutrino emission is negligible in the Sun, the steep temperature
dependence of the emission rate implies large neutrino losses in
more advanced burning stages where neutrino losses are much more
important than surface photon emission (table~\ref{tab:phases}).
This means that without neutrino losses such giant stars  should
live much longer and hence one should see more of them in the sky
relative to ordinary stars than are actually observed. Richard
Stothers (1970) used this argument to show that indeed the direct
neutrino-electron interaction should be roughly governed by the same
constant $G_{\rm F}$ as nuclear $\beta$
decay~\cite{Stothers:1970ap}. Neutral-current interactions were
first experimentally observed in 1973 in the Gargamelle bubble
chamber at CERN \cite{Hasert:1973cr}.

Once neutrinos have a direct coupling to electrons (in the sense of
our low-energy effective theory), the existence of these processes
is obvious, except for the plasmon decay which seems impossible
because the decay of massless particles (photons) is kinematically
forbidden and neutrinos do not interact with photons. However, a
photon propagating in a medium has a nontrivial dispersion relation
that can be ``time like'', $\omega^2-k^2>0$, or ``space like'',
$\omega^2-k^2<0$. In the former case, typical for a stellar plasma,
one may say that the photon has an effective mass in the medium and
a decay $\gamma\to\nu\bar\nu$ is kinematically allowed. In the
latter case, typical for visible light in air or water, the process
$e\to e+\gamma$ is kinematically allowed and is identical with the
well-known Cherenkov effect: a high-energy charged particle moving
in water or air emits detectable light.

In a non-relativistic plasma, typical for ordinary stars, the photon
dispersion relation is that of a particle with a mass corresponding
to the plasma frequency,
\begin{equation}\label{eq:plasmafrequency}
\omega^2-k^2=\omega_{\rm pl}^2\quad\hbox{where}\quad
\omega_{\rm pl}^2=\frac{4\pi\alpha\,n_e}{m_e}\,.
\end{equation}
Here $m_e$ and $n_e$ are the electron mass and number density. The
general dispersion relation in a relativistic and/or degenerate
medium is more complicated~\cite{Braaten:1993jw}, but for large
photon energies always that of a massive particle. A photon in a
medium is sometimes called ``transverse plasmon.'' In addition there
exists a propagating mode with longitudinal polarization called
``longitudinal plasmon'' or simply ``plasmon.'' It has no counterpart
in vacuum and corresponds to the negative and positive electric
charges of the plasma oscillating coherently against each other.

An effective neutrino-photon coupling is mediated by the electrons
of the medium. Photon decay can be viewed as the Compton process
(fig.~\ref{fig:processes}) when the incoming and outgoing electron
have identical momenta, i.e.\ the electron scatters forward. The
electron can then be integrated out to produce an effective
neutrino-photon interaction. The main contribution arises from the
neutrino-electron vector coupling, so that the truncated matrix
element producing the photon mass and the neutrino-photon coupling
are actually the same.

\begin{figure}[b]
\centering
\includegraphics[width=0.4\textwidth]{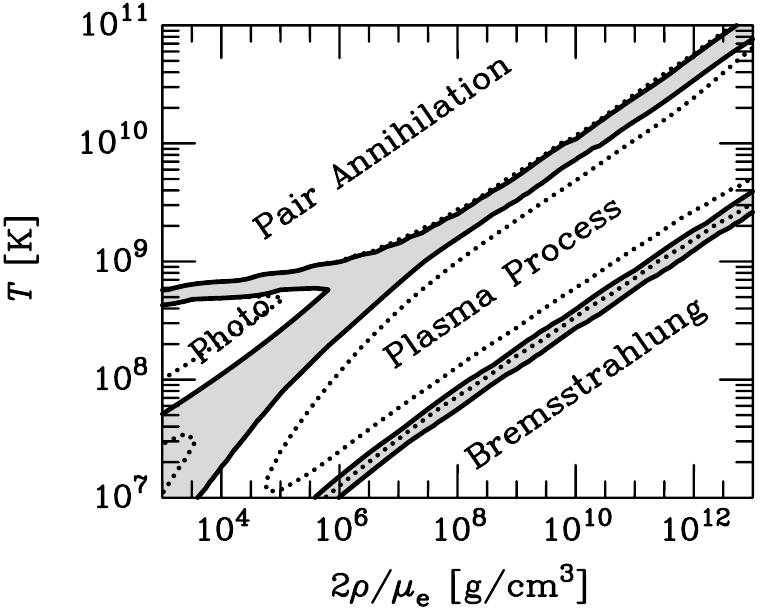}
\hskip0.7cm
\includegraphics[width=0.455\textwidth]{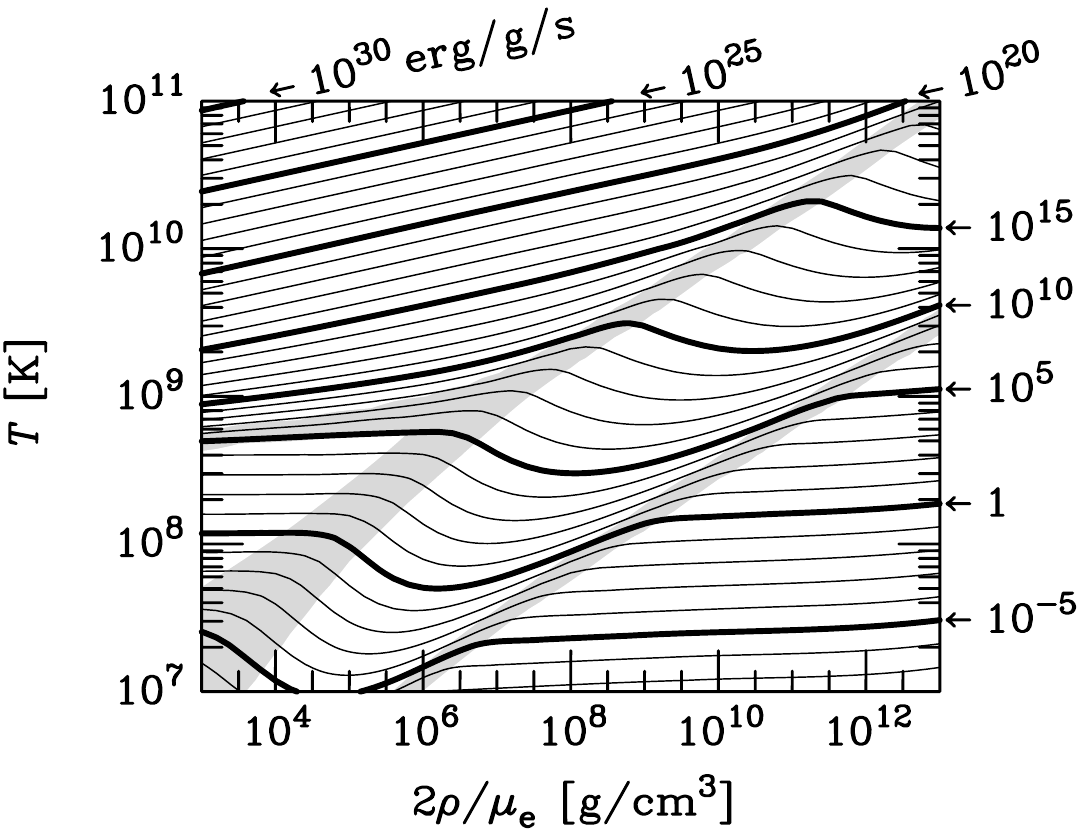}
\caption{Relative dominance of different neutrino emission processes (left)
and contours for total energy-loss rate (right).
$\mu_e$ is the electron ``mean molecular weight,'' i.e.\ roughly the
number of baryons per electron. Bremsstrahlung depends on the
chemical composition (solid lines for helium, dotted lines for
iron, right panel for helium). \label{fig:relativeimportance}}
\end{figure}

Neutrino emission rates have been calculated by different authors
over the years and numerical approximation formulas have been
derived. In a heroic effort over a decade, neutrino emission rates
were calculated and put into numerically useful form for all
relevant conditions and processes by N.~Itoh and
collaborators~\cite{Itoh:1996a}, for the plasma process see
Refs.~\cite{Haft:1993jt,Itoh:1996b}. Different processes dominate in
different regions of temperature and density
(fig.~\ref{fig:relativeimportance}). In cold and dense matter as
exists in old white dwarfs, bremsstrahlung dominates where
correlation effects among nuclei become very important.

\subsection{Neutrino electromagnetic properties}

The plasmon decay process is an important neutrino emission process
in a broad range of temperature and density even though neutrinos do
not couple directly to photons. One may speculate, however, that
neutrinos could have nontrivial electromagnetic properties, notably
magnetic dipole moments, allowing the plasma process to be more
efficient. Bernstein, Ruderman and Feinberg (1963) showed that one
can then use the observed properties of stars to constrain the
possible amount of additional energy loss and thus neutrino
electromagnetic properties~\cite{Bernstein:1963qh}.

Considering all possible interaction structures of a fermion field
$\psi$ with the electromagnetic field, one can think of four
different terms,
\begin{eqnarray}
{\cal L}_{\rm eff}=-F_1 \overline\psi\gamma_\mu \psi\, A^\mu
&-&G_1 \overline\psi\gamma_\mu\gamma_5\psi\,\partial_\nu F^{\mu\nu}
\\
&-&\frac{1}{2}\,F_2\,\overline\psi\sigma_{\mu\nu}\psi\,F^{\mu\nu}
-\frac{1}{2}\,G_2\,\overline\psi\sigma_{\mu\nu}\gamma_5\psi\,F^{\mu\nu}\,,
\nonumber
\end{eqnarray}
where $A^\mu$ is the electromagnetic field and $F^{\mu\nu}$ the
field-strength tensor. In a matrix element, the coefficients
$F_{1,2}$ and $G_{1,2}$ are functions of the energy-momentum
transfer $Q^2$ and play the role of form factors. In the limit
$Q^2\to0$, the meaning of the form factors is that of an electric
charge $e_\nu=F_1(0)$, an anapole moment $G_1(0)$, a magnetic dipole
moment $\mu=F_2(0)$ and an electric dipole moment $\epsilon=G_2(0)$.
In the standard model, neutrinos are of course electrically neutral
and $F_1(0)=0$. The anapole moment also vanishes and for
non-vanishing $Q^2$ the form factors $F_1$ and $G_1$ represent
radiative corrections to the tree-level couplings.

The $F_2$ and $G_2$ form factors couple left- with right-handed
fields and vanish if all neutrino interactions are purely left-handed
as would be the case for massless neutrinos in the standard model.
Today we know that neutrinos have small masses, and hence small
dipole moments are inevitable that are proportional to the neutrino
mass. These dipole moments can connect neutrinos of the same flavor
or of different flavors (transition moments). If neutrinos are
Majorana particles, their (diagonal) dipole moments must vanish,
whereas they still have transition moments. A Dirac neutrino mass
eigenstate has a magnetic dipole moment
\begin{equation}
\frac{\mu}{\mu_{\rm B}}=\frac{6\sqrt{2}\,G_{\rm F} m_e}{(4\pi)^2}\,m_\nu
=3.20\times10^{-19}\,\frac{m_\nu}{\rm eV}\,,
\end{equation}
where $\mu_{\rm B}=e/2m_e$ is the Bohr magneton, the usual unit to
express neutrino dipole moments. Standard transition moments are
even smaller because of a ``GIM cancelation'' in the relevant loop
diagram. Diagonal electric dipole moments violate the CP symmetry,
whereas electric dipole transition moments exist for massive mixed
neutrinos even in the standard model. Large neutrino dipole moments
would signify physics beyond the standard model and are thus
important to measure or constrain.

Neutrino dipole moments would have a number of phenomenological
consequences. In a magnetic field, these particles spin precess,
turning left-handed states into right-handed ones and vice versa.
Since neutrino flavor mixing is now established, it is clear that
such processes would also couple neutrinos of different flavor,
leading to spin-flavor oscillations~\cite{Schechter:1981hw,
Akhmedov:1988uk, Lim:1987tk}. Stars usually have magnetic fields
that can be very large and would induce spin and spin-flavor
oscillations. It is now clear that the solar neutrino observations
are explained by ordinary flavor oscillations, not by spin-flavor
oscillations. Still, if one were to observe a small $\bar\nu_e$ flux
from the Sun, which produces only $\nu_e$ in its nuclear reactions,
this could be explained by spin-flavor oscillations of Majorana
neutrinos~\cite{Miranda:2003yh, Miranda:2004nz, Raffelt:2009mm}.
Much larger magnetic fields exist in supernovae, leading to
complicated spin and spin-flavor oscillation
phenomena~\cite{Ando:2003pj}. It would appear almost hopeless to
disentangle spin-flavor oscillations in a supernova neutrino signal,
except if one were to observe a strong burst of antineutrinos in the
prompt de-leptonization burst~\cite{Akhmedov:2003fu}.

A dipole moment contributes to the scattering cross section
$\nu_e+e\to e+\nu$ where the final-state $\nu$ has opposite spin and
may have different flavor. The photon mediating this process renders
the cross section forward peaked, allowing one to disentangle it
from the ordinary weak-interaction process. The difference is most
pronounced for the lowest-energy neutrinos and the most restrictive
limit, $\mu_\nu<3.2\times10^{-11}\,\mu_{\rm B}$ at 90\%~CL, arises
from a reactor neutrino experiment~\cite{Beda:2010hk}. Dipole and
transition moments that do not involve $\nu_e$ are experimentally
less well constrained.

\begin{figure}[b]
\centering
\includegraphics[width=0.5\textwidth]{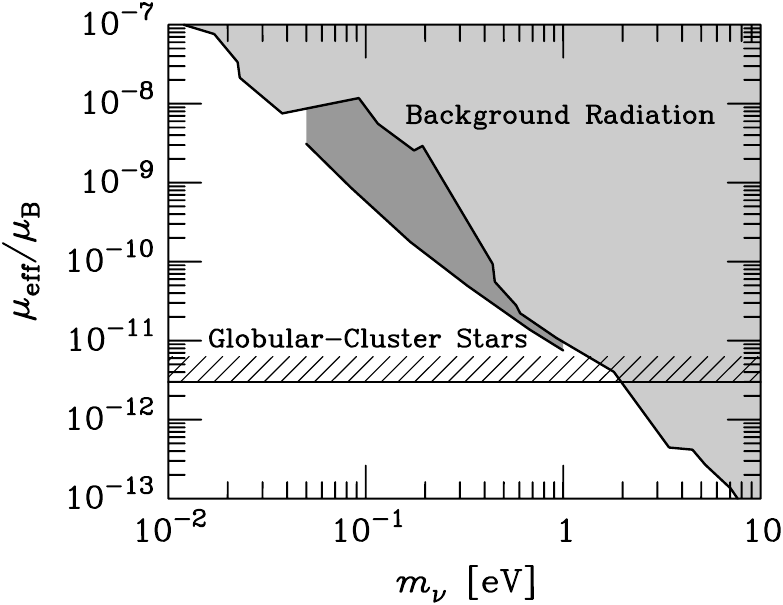}
\caption{Exclusion range for neutrino transition moments~\cite{Raffelt:1998xu}.
The light-shaded region is ruled out by the
contribution of radiative neutrino decays to the cosmic
photon backgrounds~\cite{Ressell:1989rz},
the dark-shaded region is excluded by TeV-gamma ray limits for
the infrared background~\cite{Biller:1998nc}.
Values above the hatched bar are excluded by plasmon decay
in globular-cluster stars.
\label{fig:dipolelimits}}
\end{figure}

Transition moments inevitably allow for the radiative decay
$\nu_2\to\nu_1+\gamma$ between two mass eigenstates $m_2>m_1$. In
terms of the transition moment $\mu_{\nu}$ the decay rate is
\begin{equation}
\Gamma_{\nu_2\to\nu_1\gamma}=\frac{\mu_{\nu}^2}{8\pi}\,
\left(\frac{m_2^2-m_1^2}{m_2}\right)^3
=5.308~{\rm s}^{-1}~\left(\frac{\mu_{\nu}}{\mu_{\rm B}}\right)^2
\left(\frac{m_\nu}{\rm eV}\right)^3\,,
\end{equation}
where the numerical expression assumes $m_\nu=m_2\gg m_1$. Mass
dependent $\mu_\nu$ constraints from the absence of cosmic excess
photons are shown in fig.~\ref{fig:dipolelimits}. They become very
weak for small $\mu_\nu$ due to the $m_\nu^3$ phase-space factor in
the expression for $\Gamma_{\nu_2\to\nu_1\gamma}$.

The most restrictive limit arises from the plasmon decay in low-mass
stars. If $\mu_\nu$ is too large, neutrino emission by
$\gamma\to\nu\bar\nu$ would affect stars more than is allowed by the
observations discussed below. The volume energy loss rates caused by
a putative neutrino ``milli charge'' $e_\nu$, a dipole moment
$\mu_\nu$, and the effective standard coupling caused by the
electrons of the medium are~\cite{Raffelt:1996wa}
\begin{equation}
Q=\frac{8\zeta_3}{3\pi}\,T^3\times
\begin{cases}
\displaystyle\kern1.2em\alpha_\nu\kern1.5em
\frac{\omega_{\rm pl}^2}{4\pi}\kern1.5em Q_1&\text{Millicharge}\\[2ex]
\displaystyle\kern1em\frac{\mu_\nu^2}{2}\kern0.8em
\left(\frac{\omega_{\rm pl}^2}{4\pi}\right)^2Q_2&
\text{Dipole Moment}\\[2ex]
\displaystyle\frac{C_V^2G_{\rm F}^2}{\alpha}\,\left(\frac{\omega_{\rm pl}^2}{4\pi}\right)^3Q_3&
\text{Standard Model}
\end{cases}
\end{equation}
where $Q_{1,2,3}$ are numerical factors that are 1 in the limit of a
very small plasma frequency and if we neglect the contribution of
longitudinal plasmons. Relative to the standard-model (SM) case, the
``exotic'' emission rates are
\begin{eqnarray}
\frac{Q_{\rm charge}}{Q_{\rm SM}}&=&
\frac{\alpha_\nu\alpha\,(4\pi)^2}{C_V^2 G_{\rm F}^2\omega_{\rm pl}^4}
\,\frac{Q_1}{Q_3}
=0.664\,e_{14}^2\,\left(\frac{10~{\rm keV}}{\omega_{\rm pl}}\right)^4
\frac{Q_1}{Q_3}\,,\\
\frac{Q_{\rm dipole}}{Q_{\rm SM}}&=&
\frac{\mu_\nu^2\,\alpha\, 2\pi}{C_V^2G_{\rm F}^2\omega_{\rm pl}^2}
\,\frac{Q_2}{Q_3}
=0.318\,\mu_{12}^2\,\left(\frac{10~{\rm keV}}{\omega_{\rm pl}}\right)^2
\frac{Q_2}{Q_3}\,.
\end{eqnarray}
From these ratios we directly see when the exotic contribution would
roughly dominate. The observations described below finally provide
the limits
\begin{equation}\label{eq:neutrinoemlimits}
e_\nu\alt 2\times10^{-14}\,e\qquad\hbox{and}\qquad
\mu_\nu\alt3\times10^{-12}\,\mu_{\rm B}\,.
\end{equation}
This is the most restrictive limit on diagonal dipole moments. From
fig.~\ref{fig:dipolelimits} we conclude that for $m_\nu\alt 2$~eV
this is also the most restrictive limit on transition moments.

\subsection{Globular clusters testing stellar evolution and particle physics}

The theory of stellar evolution can be quantitatively tested by
using the stars in globular clusters. Our own Milky Way galaxy has
at least 157 of these gravitationally bound ``balls'' of stars that
surround the galaxy in a spherical halo~\cite{Harris:2010}. Each
cluster consists of up to a million stars. Once a globular cluster
has formed, new star formation is quenched because the first
supernovae sweep out the gas from which new stars might otherwise
form. Therefore, as a first approximation we may assume that all
stars in a globular cluster have the same age and chemical
composition and differ only in their mass. Since stellar evolution
proceeds faster for higher-mass stars, in a globular cluster today
we see a snapshot of stars in different evolutionary stages.
Moreover, since the advanced stages after hydrogen burning are fast,
for those stages we essentially see a star of a certain initial mass
simultaneously in all advanced stages of evolution.

\begin{figure}
\centering
\includegraphics[width=0.47\textwidth]{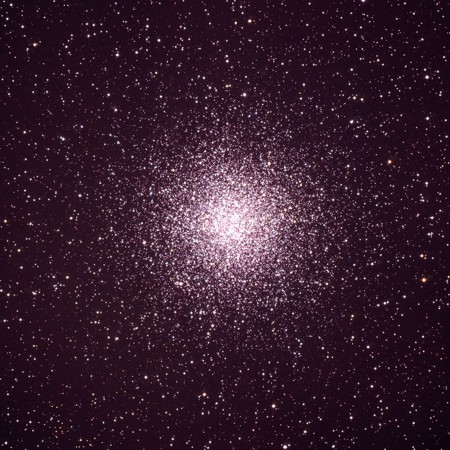}\hfill
\includegraphics[width=0.47\textwidth]{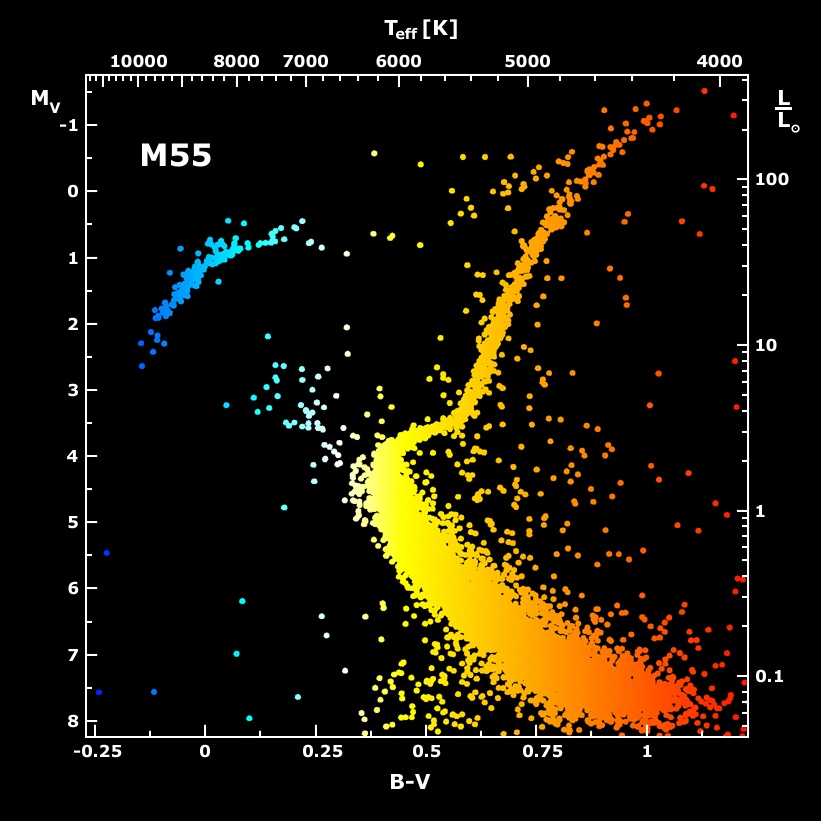}
\caption{Globular cluster M55 (NGC 6809) in the constellation
Sagittarius, as imaged by the ESO 3.6~m telescope on La Silla (Credit: ESO).
Right panel: Color magnitude diagram of M55 (Credit: B.~J.~Mochejska and J.~Kaluzny,
CAMK, see also
Astronomy Picture of the Day,
http://apod.nasa.gov/apod/ap010223.html).\label{fig:M55}}
\end{figure}

\begin{figure}
\centering
\includegraphics[width=\textwidth]{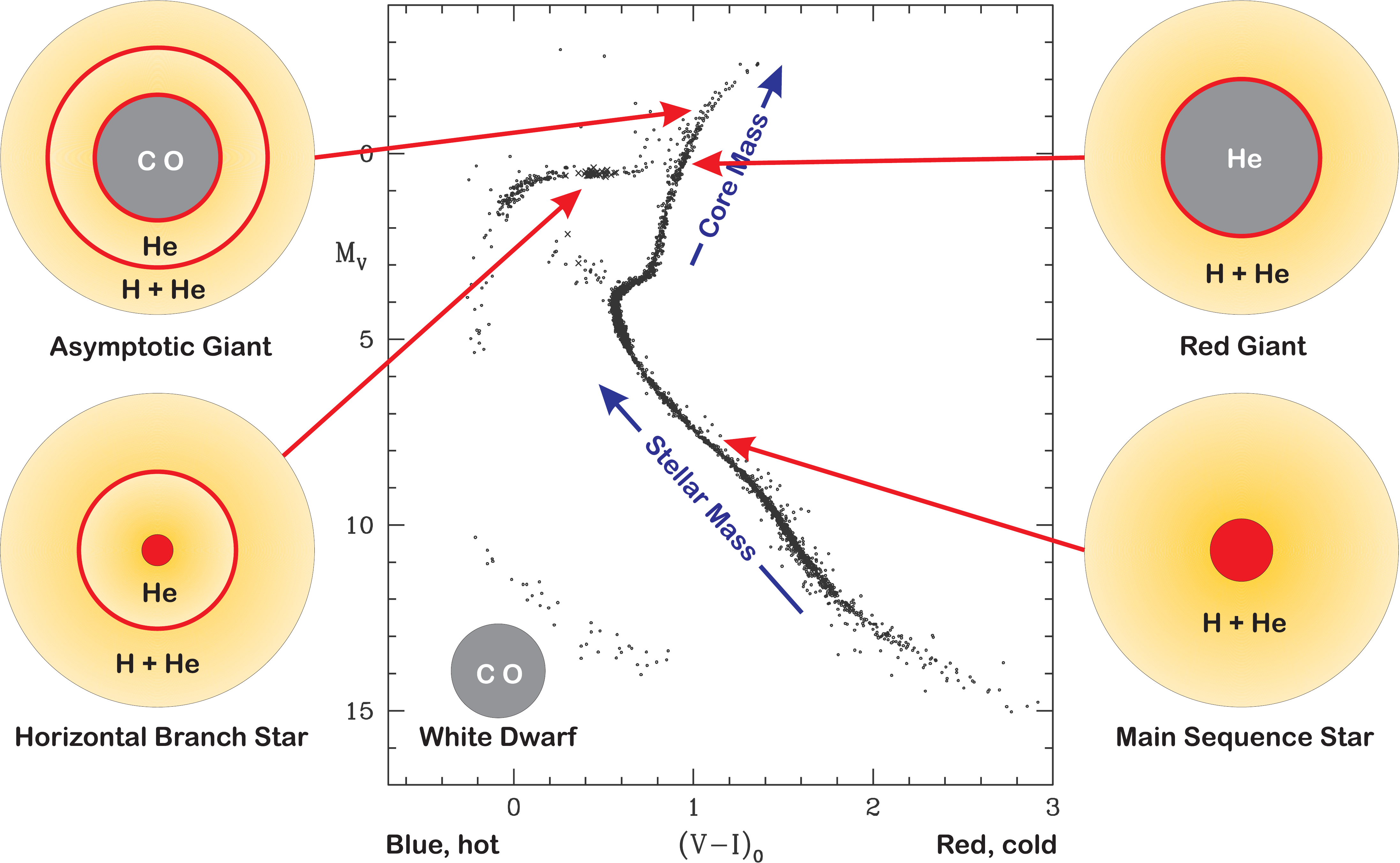}
\caption{Schematic color-magnitude diagram for a globular cluster produced
from selected stars of several galactic globular clusters~\cite{Harris:2000}.
The structure of stars corresponding to the different
branches of the diagram are indicated.\label{fig:hr}}
\end{figure}

As an example we show the large globular cluster M55 in
fig.~\ref{fig:M55}. The theoretically relevant information is
revealed when the stars are arranged in a color-magnitude diagram
where the stellar brightness is plotted on the vertical axis, the
color (essentially surface temperature) on the horizontal axis. (The
brightness is a logarithmic measure of luminosity.) The different
loci in the color-magnitude diagram correspond to different
evolutionary phases as indicated in fig.~\ref{fig:hr}.
\begin{itemize}
\item {\bf Main Sequence (MS).} Hydrogen burning stars like our
    Sun, the lower-mass ones being dimmer and redder. The MS
    turnoff corresponds to a mass of around $0.8\,M_\odot$,
    whereas more massive stars have completed hydrogen burning
    and are no longer on the MS.

\item{\bf Red Giant Branch (RGB).} After hydrogen is exhausted
    in the center, the star develops a degenerate helium core
    with hydrogen burning in a shell. Along the RGB, brighter
    stars correspond to a larger core mass,  smaller core
    radius, and larger gravitational potential, which in turn
    causes hydrogen to burn at a larger $T$ so that these stars
    become brighter as the core becomes more massive. The RGB
    terminates at its tip, corresponding to helium ignition in
    the core.

\item{\bf Horizontal Branch (HB).} Helium ignition expands the
    core which develops a self-regulating non-degenerate
    structure. The gravitational potential decreases, hydrogen
    burns less strongly, and the star dims, even though helium
    has been ignited. The structure of the envelope depends
    strongly on mass and other properties, so these stars spread
    out in $T_{\rm surface}$ at an almost fixed brightness. The
    blue HB downturn is an artifact of the visual filter---if
    measured in total (bolometric) brightness, the HB is truly
    horizontal. For a certain $T_{\rm surface}$, the envelope of
    these stars is not stable and they pulsate: the class of RR
    Lyrae stars.

\item{\bf Asymptotic Giant Branch (AGB).} After helium is
    exhausted, a degenerate carbon-oxygen core develops and the
    star now has two shell sources. As the core becomes more
    massive, it shrinks in size, increases its gravitational
    potential, and thus brightens quickly: the star ascends the
    red giant branch once more. Mass loss is now strong and
    eventually the star sheds all of its envelope to become a
    planetary nebula with a hot white dwarf in its center.

\item{\bf White Dwarfs.} The compact remnants are very small and
    thus very dim, but at first rather hot. White dwarfs then
    cool and become dimmer and redder. They will cross the
    instability strip once more, forming the class of ZZ Ceti
    stars.
\end{itemize}

In any of these phases, a new energy-loss channel modifies the
picture. Increased neutrino losses on the RGB imply an increased
core mass to ignite helium and the tip of the RGB brightens. A
larger core mass at helium ignition also implies a brighter HB.
Excessive particle emission on the HB implies that helium is
consumed faster, the HB phase finishes more quickly for each star,
implying that we see fewer HB stars. Therefore, the number of HB
stars in a globular cluster relative to other phases is a direct
measure for the helium-burning lifetime. Comparing theoretical
predictions with these and other observables for several globular
clusters reveals excellent agreement \cite{Raffelt:1996wa,
Raffelt:1989xu, Catelan:1995ba}. The core mass at helium ignition
agrees with predictions approximately to within 5--10\%. This
implies that the true energy loss can be at most a few times larger
than the standard neutrino losses. The helium burning lifetime
agrees to within 10--20\%.

The helium core before ignition, essentially a helium white dwarf,
has a central density of around $10^6~{\rm g}~{\rm cm}^{-3}$, an
average density of around $2\times10^5~{\rm g}~{\rm cm}^{-3}$, and
an almost constant temperature of $10^8$~K. The average standard
neutrino losses, mainly from the plasma process, are about $4~{\rm
erg}~{\rm g}^{-1}~{\rm s}^{-1}$. To avoid the helium core growing
too massive, the core-averaged emission rate of any novel process
should fulfill
\begin{equation}
\epsilon_x\alt 10~{\rm erg}~{\rm g}^{-1}~{\rm s}^{-1}\,.
\end{equation}
Coincidentally the same constraint applies to the energy losses from
the helium burning core during the HB phase, but now to be
calculated at a typical average density of about $0.6\times10^4~{\rm
g}~{\rm cm}^{-3}$ and $T\sim 10^8$~K, detailed average values given
in Ref.~\cite{Raffelt:1996wa}.

This argument has been applied to many cases of novel particle
emission, ranging from neutrino magnetic dipole moments and milli
charges to new scalar or pseudoscalar
particles~\cite{Raffelt:1996wa, Raffelt:1990yz, Raffelt:1999tx}. The
limits on neutrino electromagnetic properties were already stated in
eq.~(\ref{eq:neutrinoemlimits}). In addition we mention explicitly
the case of axions~\cite{Peccei:2006as, Sikivie:2006ni,
Raffelt:2006cw, Kim:2008hd}, new very low-mass pseudoscalars that
are closely related to neutral pions and could be the dark matter of
the universe. Axions have a two-photon interaction of the form
\begin{equation}\label{eq:axiontwophotoncoupling}
{\cal L}_{a\gamma}=-\frac{g_{a\gamma}}{4}\,F_{\mu\nu}\tilde F^{\mu\nu} a
=g_{a\gamma}{\bf E}\cdot{\bf B}\,a,
\quad\hbox{where}\quad
g_{a\gamma}=\frac{\alpha}{2\pi f_a}\,\left(\frac{E}{N}-\frac{2(4+z)}{3(1+z)}\right)\,.
\end{equation}
Here, $F$ is the electromagnetic field-strength tensor, $\tilde F$
its dual, $a$ the axion field, $z=m_u/m_d\sim0.5$ the up/down quark
mass ratio, and $E/N$ a model-dependent ratio of small integers
reflecting the ratio of electromagnetic to color anomaly of the
axion current. The energy scale $f_a$ is the axion decay constant,
related to the Peccei-Quinn scale of spontaneous breaking of a new
U(1)$_{\rm PQ}$ symmetry of which the axion is the Nambu-Goldstone
boson. By mixing with the $\pi^0$-$\eta$-$\eta'$ mesons, axions
acquire a small mass
\begin{equation}
m_a=\frac{\sqrt{z}}{1+z}\,\frac{m_\pi f_\pi}{f_a}
=6~{\rm meV}\,\frac{10^9~{\rm GeV}}{f_a}\,.
\end{equation}
Finally, they would interact with fermions $f$, notably nucleons and
possibly electrons, with a derivative axial-vector structure
\begin{equation}
{\cal L}_{af}=\frac{C_f}{2f_a}\,\overline\psi_f\gamma^\mu\gamma_5\psi_f\partial_\mu a
\quad\hbox{and}\quad
g_{af}=\frac{C_f m_f}{f_a}\,,
\end{equation}
where $C_f$ is a model-dependent numerical coefficient of order
unity and $g_{af}$ a dimensionless Yukawa coupling of the axion
field to the fermion $f$.

\begin{figure}
\centering
\includegraphics[width=1.0\textwidth]{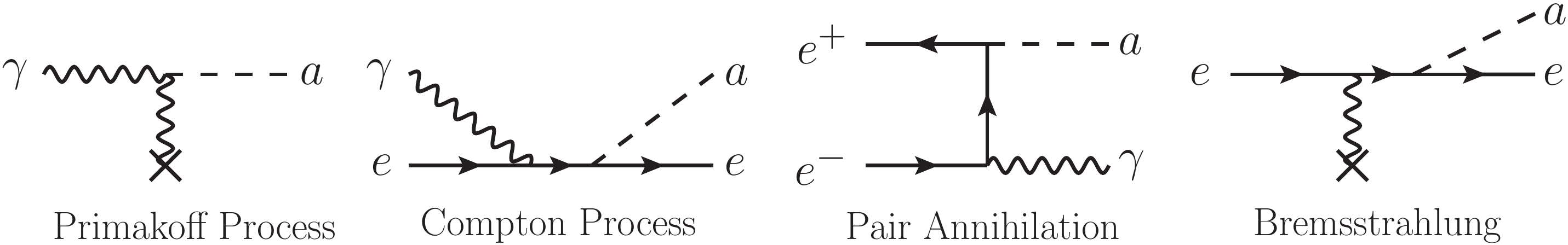}
\caption{Thermal axion emission processes in normal stars.\label{fig:axionprocesses}}
\end{figure}

In normal stars, these interactions allow for the axion emission
processes shown in fig.~\ref{fig:axionprocesses}. The Compton,
pair-annihilation and bremsstrahlung processes are analogous to the
corresponding neutrino processes based on the axial-current
interaction. The main difference is the axion phase space compared
with the two-neutrino phase space, implying a less steep temperature
dependence of axion emission, so the relative importance of axion
losses is greater in cooler stars. The plasmon decay does not exist
for axions, but instead we have the Primakoff conversion of photons
to axions in the electric fields of charged particles in the medium
that is enabled by the two-photon vertex.

In globular-cluster stars, the Primakoff process is much more
effective during the HB phase in the non-degenerate helium core than
during the RGB phase when the helium core is degenerate. Therefore,
the helium-burning lifetime will be shortened by excessive axion
emission without affecting the RGB evolution. As discussed earlier,
the number of HB stars in globular clusters relative to RGB stars
can then be used to constrain the axion-photon interaction strength
and leads to a limit~\cite{Raffelt:2006cw}
\begin{equation}
g_{a\gamma}\alt 1\times10^{-10}~{\rm GeV}^{-1}\,.
\end{equation}
Similar constraints have been established by the CAST experiment
searching for solar axions to be discussed later. For axion models
with $E/N=0$ this corresponds to $f_a\agt2\times10^7$~GeV or
$m_a\alt0.3$~eV.

Axions are a QCD phenomenon, but in a broad class of models they
also interact with electrons, the DFSZ model~\cite{Dine:1981rt,
Zhitnitsky:1980tq} being the usual benchmark example for which
$E/N=8/3$. The limit on $g_{a\gamma}$ then translates into the
weaker constraint $m_a\alt0.8$~eV. The axion-electron coupling is
determined by $C_e=\frac{1}{3}\,\cos^2\beta$ with $\cos\beta$ a
model-dependent parameter. The dominant effect on globular cluster
stars is axion emission by bremsstrahlung and the Compton process
from degenerate red giant cores, delaying helium ignition. The
established core mass at helium ignition then leads to the
bound~\cite{Raffelt:1994ry} $ g_{ae}\alt3\times10^{-13}$,
translating to $m_a\alt 9~{\rm meV}/\cos^2\beta$ and
$g_{a\gamma}\alt 1.2\times10^{12}~{\rm GeV}/\cos^2\beta$.

\subsection{White dwarf cooling}

\label{sec:whitedwarfcooling}

\begin{figure}
\centering
\includegraphics[width=0.6\textwidth]{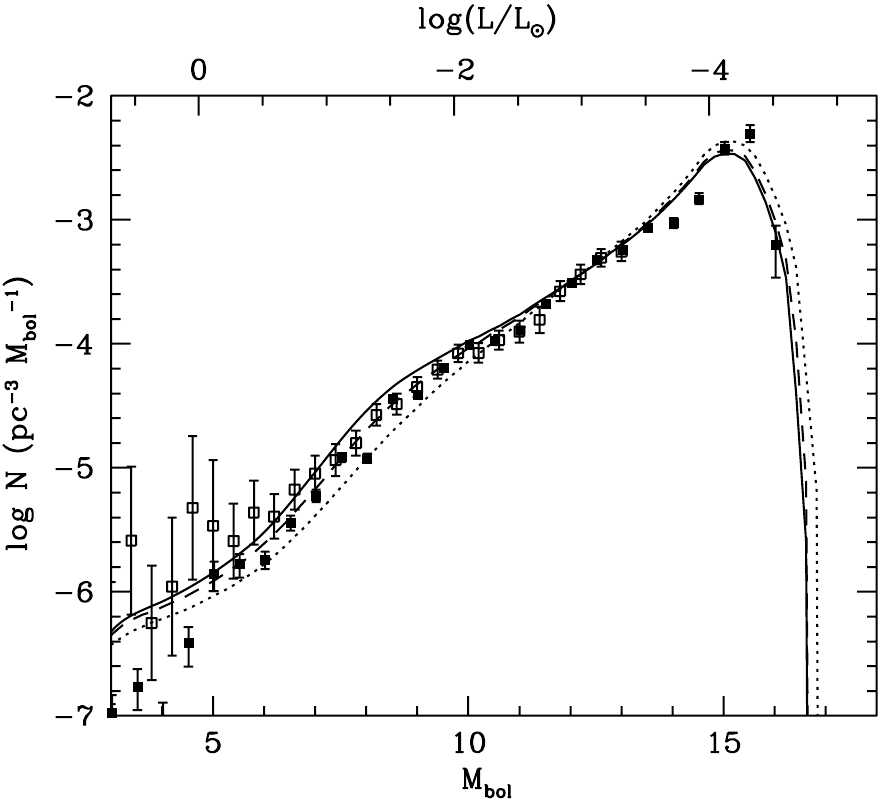}
\caption{White dwarf luminosity function~\cite{Isern:2008nt}.
Open and filled squares correspond to different methods for identifying
white dwarfs. Solid line: Theoretical luminosity function for a constant
formation rate and 11~Gyr for the age of the galactic disk.
Dashed and dotted lines: Including axion cooling corresponding
to $m_a\cos^2\beta=5$~meV and 10~meV.
\label{fig:wd}}
\end{figure}

More restrictive limits on the axion-electron interaction arise from
white-dwarf (WD) cooling. When a WD has formed after an asymptotic
red giant has shed its envelope, forming a planetary nebula, the
compact remnant is a carbon-oxygen WD. It is supported by degeneracy
pressure and simply cools and dims without igniting carbon burning.
Assuming WDs are born at a constant rate in the galactic disk, the
number of observed WDs per brightness interval, the ``luminosity
function'' (fig.~\ref{fig:wd}), then represents the cooling speed of
an average WD. Any new energy-loss channel accelerates the cooling
speed and, more importantly, deforms the luminosity function. A new
energy-loss channel mostly affects hot WDs, whereas late-time
cooling is dominated by surface photon emission.

An early application of this argument provided a limit on the
axion-electron coupling of $g_{ae}\alt4\times10^{-13}$
\cite{Raffelt:1985nj}, comparable to the globular cluster limit.
Revisiting WD cooling with modern data and cooling simulations
\cite{Isern:2008nt, Isern:2008fs} reveals that the standard theory
does not provide a perfect fit (solid line in fig.~\ref{fig:wd}). On
the other hand, including a small amount of axion cooling
considerably improves the agreement between observations and cooling
theory (dashed line in fig.~\ref{fig:wd}). If interpreted in terms
of axion cooling, a value $g_{ae}=0.6$--$1.7\times10^{-13}$ is
implied, not in conflict with any other limit.

In the early 1990s it became possible to test the cooling speed of
pulsating WDs, the class of ZZ Ceti stars, by their measured period
decrease $\dot P/P$. In particular, the star G117-B15A was cooling
too fast, an effect that could be attributed to axion losses if
$g_{ae}\sim2\times10^{-13}$ \cite{Isern:1992}. Over the past twenty
years, observations and theory have improved and the G117-B15A
cooling speed still favors a new energy-loss
channel~\cite{Isern:2010wz, Corsico:2011vy}.

It is perhaps premature to be certain that these observations truly
require a new WD energy-loss channel. Moreover, the interpretation
in terms of axion emission is, of course, speculative. Still, these
findings suggest that one should investigate other consequences of
the ``meV frontier'' of axion physics, for example for
supernovae~\cite{Raffelt:2011ft}.

\section{Neutrinos from the Sun}                       \label{sec:sun}

\subsection{Solar neutrino measurements and flavor oscillations}

\begin{figure}[b]
\centering
\includegraphics[width=0.6\textwidth]{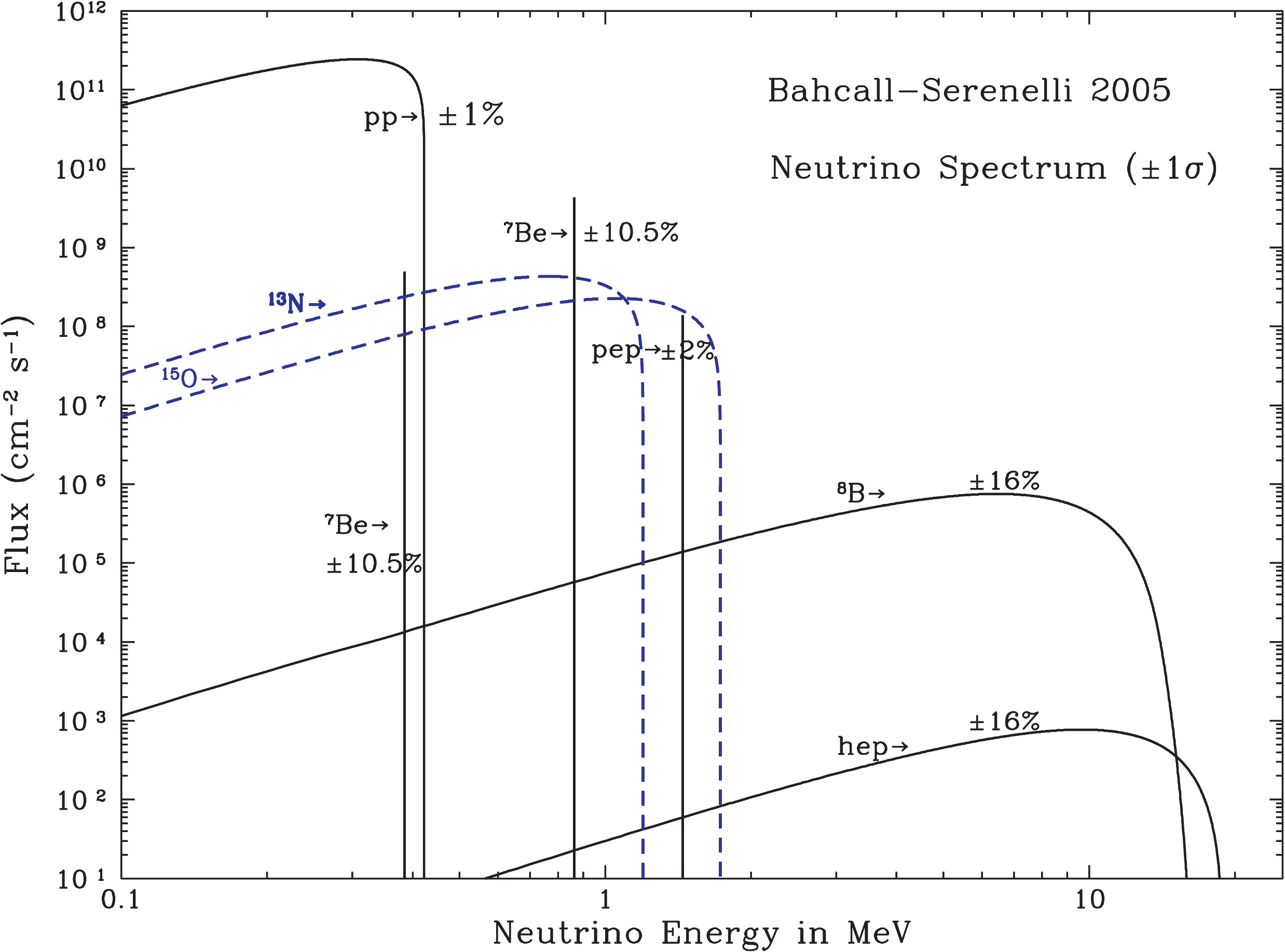}
\caption{Predicted solar neutrino spectrum~\cite{Bahcall:2004pz}
according to the solar model of Bahcall and Serenelli (2005) \cite{Bahcall:2004mq},
based on traditional opacities.\label{fig:solnuspec}}
\end{figure}

The Sun produces energy by fusing hydrogen to helium, primarily by
the pp chains (table~\ref{tab:ppchains}) and a few percent through
the CNO cycle (table~\ref{tab:CNO}), emitting $\nu_e$ fluxes by the
tabulated processes. In addition, a low-energy flux of keV-range
thermal neutrinos emerges~\cite{Haxton:2000xb} which is negligible
for energy loss. The predicted flux spectrum is shown in
fig.~\ref{fig:solnuspec}. The largest flux consists of the
low-energy pp neutrinos, whereas the $^8$B flux with the largest
energies is much smaller. The predicted fluxes
(table~\ref{tab:solarfluxes}) depend somewhat on the assumed solar
abundance of CNO elements which is not entirely settled
(section~\ref{sec:opacity}), but this uncertainty is not crucial for
our present discussion.

\begin{table}
  \caption{Solar neutrino fluxes predicted with the GS98 and AGSS09
  opacities compared with experimentally inferred fluxes, assuming neutrino
  flavor oscillations~\cite{Serenelli:2011py}.
  \label{tab:solarfluxes}}
  \begin{tabular}{lllllll}
    \hline
    Source&\multicolumn{2}{l}{Old opacities (GS98)}&\multicolumn{2}{l}{New opacities (AGSS09)}&\multicolumn{2}{l}{Best measurements}\\
    \cline{2-7}
    &Flux&Error&Flux&Error&Flux&Error\\
    &cm$^{-2}$~s$^{-1}$&\%  &cm$^{-2}$~s$^{-1}$&\%&cm$^{-2}$~s$^{-1}$&\%\\
    \hline
    pp      &$5.98\times10^{10}$&$\pm0.6$&$6.03\times10^{10}$&$\pm0.6$&$6.05\times10^{10}$&$+0.3/{-}1.1$\\
    pep     &$1.44\times10^{8}$ &$\pm1.1$&$1.47\times10^{8}$ &$\pm1.2$&$1.46\times10^{8}$ &$+1/{-}1.4$\\
    hep     &$8.04\times10^{3}$ &$\pm30$ &$8.31\times10^{3}$ &$\pm30$ &$18\times10^{3}$   &$+40/{-}50$ \\
    $^7$Be  &$5.00\times10^{9}$ &$\pm7$  &$4.56\times10^{9}$ &$\pm7$  &$4.82\times10^{9}$ &$+5/{-}4$\\
    $^8$B   &$5.58\times10^{6}$ &$\pm14$ &$4.59\times10^{6}$ &$\pm14$ &$5.00\times10^{6}$ &$\pm3$  \\
    $^{13}$N&$2.96\times10^{8}$ &$\pm14$ &$2.17\times10^{8}$ &$\pm14$ &$<6.7\times10^{8}$ &        \\
    $^{15}$O&$2.23\times10^{8}$ &$\pm15$ &$1.56\times10^{8}$ &$\pm15$ &$<3.2\times10^{8}$ &        \\
    \hline
  \end{tabular}
\end{table}

The first solar neutrino experiment was proposed by Ray Davis in
1964 \cite{Davis:1964hf}, accompanied by the first solar flux
predictions by John Bahcall~\cite{Bahcall:1964gx}. The detection
principle, going back to an idea of Bruno Pontecorvo in 1946, is
based on the radiochemical technique where a tank is filled with
carbon tetrachloride, allowing for the reaction $\nu_e+{}^{37}{\rm
Cl}\to{}^{37}{\rm Ar}+e^-$. The argon noble gas atoms can be washed
out, concentrated, collected in a counter, and finally one can count
them by observing their electron capture decay, emitting several
Auger electrons. Davis used such a detector to establish in 1955 an
upper limit on the $\nu_e$ flux from a reactor~\cite{Davis:1955bi},
which of course emits primarily $\bar\nu_e$. Around the same time,
Reines and Cowan observed the first $\bar\nu_e$ events in their
detector and in this way were the first to observe neutrinos. Davis
then turned to measuring solar neutrinos with a much bigger tank,
holding 615 tons of tetrachlorethylene, ${\rm C}_2{\rm Cl}_4$, that
was located deep underground in the Homestake gold mine in South
Dakota. First solar neutrino results were published in 1968
\cite{Davis:1968cp}. After some improvements, the finally used data
were taken during a quarter century 1970--1994
\cite{Cleveland:1998nv}, producing in 108 extractions a total of
around 800 registered argon atoms. This heroic effort was awarded
with the physics nobel prize of 2002, shared between Ray Davis and
Masatoshi~Koshiba who built the first water Cherenkov detector
(Kamiokande) to see solar neutrinos.

\begin{figure}
\centering
\includegraphics[width=0.6\textwidth]{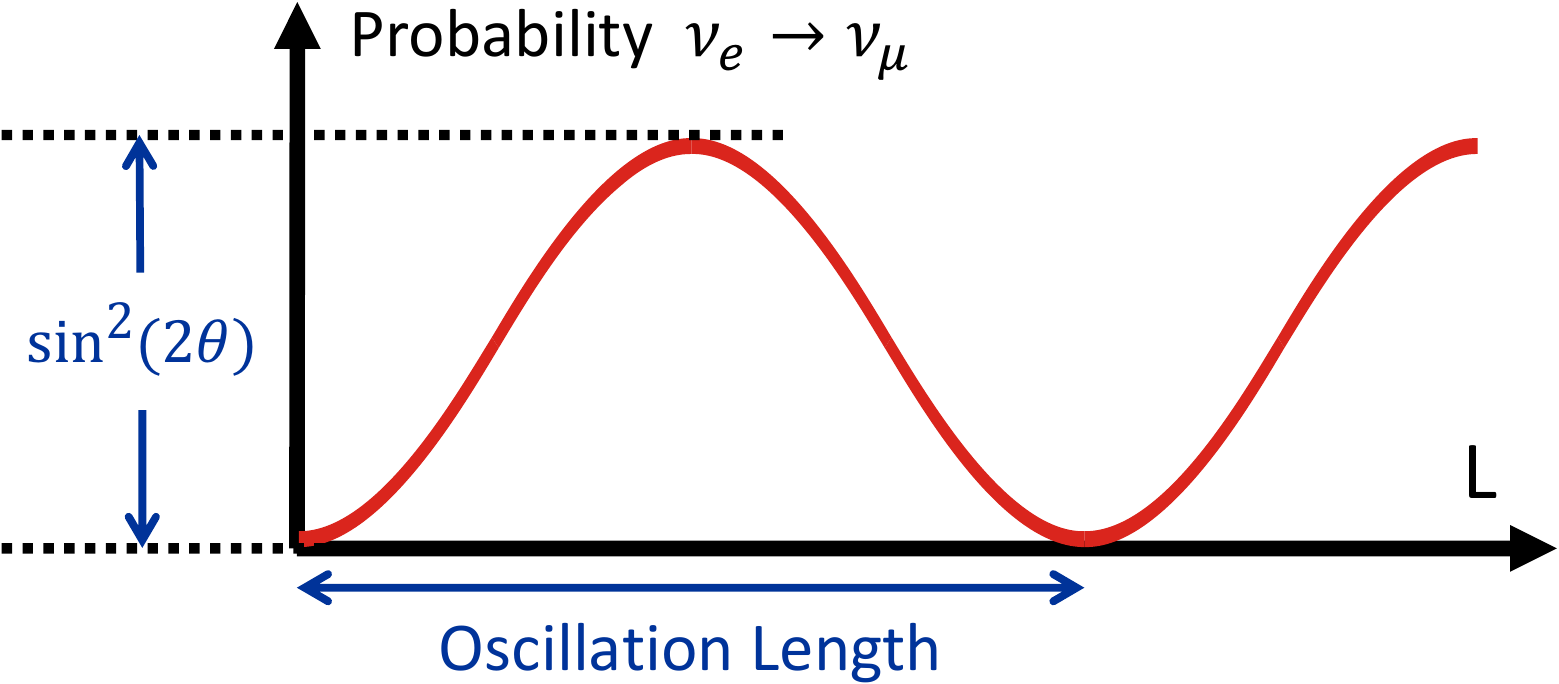}
\caption{Flavor oscillations.\label{fig:appearance}}
\end{figure}

For a given exposure, only a handful of argon atoms is produced so
that the measurements show huge statistical fluctuations. Still, it
quickly became clear that there was a deficit of measured $\nu_e$
relative to predictions. The detection threshold of 0.814~MeV means
that one picks up primarily the rather uncertain $^8$B flux, so for
a long time the ``solar neutrino problem'' was widely attributed to
solar model, nuclear cross section, and experimental uncertainties.
However, already in 1969 Gribov and Pontecorvo proposed neutrino
flavor oscillations $\nu_e\to\nu_\mu$ as a possible
interpretation~\cite{Gribov:1968kq}. It is assumed that the flavor
and mass eigenstates are related by a rotation with mixing angle
$\theta$
\begin{equation}
\begin{pmatrix}
\nu_e\\ \nu_\mu
\end{pmatrix}
=
\begin{pmatrix}
\cos\theta&\sin\theta\\ -\sin\theta&\cos\theta
\end{pmatrix}
\begin{pmatrix}
\nu_1\\ \nu_2
\end{pmatrix}\,.
\end{equation}
At $\nu_e$ production, actually a coherent superposition of the mass
eigenstates $\nu_1$ and $\nu_2$ emerges which propagate with
different momenta $p_{1,2}=(E^2-m_{1,2}^2)^{1/2}\approx
E-m_{1,2}^2/2E$, so that after some distance $L$ their interference
provides for a nonvanishing $\nu_\mu$ amplitude. It is easy to work
out that the $\nu_\mu$ appearance probability is
(fig.~\ref{fig:appearance})
\begin{equation}\label{eq:oscillationprobability}
P_{\nu_e\to\nu_\mu}=\sin^2(2\theta)\,\sin^2\left(\frac{\Delta m^2}{4E}L\right)
\quad\hbox{and}\quad
L_{\rm osc}=\frac{4\pi E}{\Delta m^2}=2.5~{\rm m}\,\frac{E}{\rm MeV}\,\frac{{\rm eV}^2}{\Delta m^2}\,,
\end{equation}
where $\Delta m^2=m_2^2-m_1^2$ and $L_{\rm osc}$ is the oscillation
length.

One reason for being skeptical about the flavor oscillation
hypothesis was the required large mixing angle to achieve a large
$\nu_e$ deficit, in contrast to the known small mixing angles among
quarks. This perception changed when the impact of matter on flavor
oscillations was recognized. Wolfenstein (1978) showed that neutrino
refraction in matter strongly influences flavor oscillations if
neutrino mass differences are indeed
small~\cite{Wolfenstein:1977ue}. Neutrinos in normal unpolarized
matter feel an effective weak potential
\begin{equation}\label{eq:weakpotential}
V_{\rm weak}=\pm\sqrt{2}\,G_{\rm F}\times
\begin{cases}
n_e-\frac{1}{2}\,n_n&\hbox{for $\nu_e$,}\\
-\frac{1}{2}\,n_n&\hbox{for $\nu_{\mu,\tau}$,}\\
\end{cases}
\end{equation}
where $n_e$ and $n_n$ are the electron and neutron densities. The
potential depends on flavor because $\nu_e$ has an additional
contribution to its effective neutral-current interaction with $e$
from $W$ exchange (fig.~\ref{fig:neutrino-electron-coupling}). The
positive sign applies to neutrinos, the negative sign to
antineutrinos. In the Earth, taking a typical density of $5~{\rm
g}~{\rm cm}^{-3}$, the $\nu_e$-$\nu_\mu$ weak potential difference
is $\Delta V_{\rm weak}=\sqrt{2}\,G_{\rm F}n_e\sim 2\times
10^{-13}~{\rm eV}=0.2~{\rm peV}$. The flavor variation along the
propagation direction $z$ is now governed by the Schr\"odinger-like
equation
\begin{equation}
{\rm i}\,\frac{\partial}{\partial z}
\begin{pmatrix}\nu_e\\ \nu_\mu\end{pmatrix}=
{\sf H}\begin{pmatrix} \nu_e\\ \nu_\mu\end{pmatrix}
\end{equation}
where the Hamiltonian $2{\times}2$ matrix is
\begin{equation}\label{eq:ham}
{\sf H}=
\frac{\Delta m^2}{4E}
\begin{pmatrix}
-\cos2\theta&\sin2\theta\\
\sin2\theta&\cos2\theta
\end{pmatrix}
\pm\sqrt{2}G_{\rm F}
\begin{pmatrix}
n_e-n_n/2&0\\0&-n_n/2\,.
\end{pmatrix}
\end{equation}
The first term is the neutrino mass-squared matrix in the
weak-interaction basis. In the matter term, the neutron contribution
is the same for both flavors. It only provides an overall common
phase and thus is usually removed.

\begin{figure}
\centering
\includegraphics[width=0.7\textwidth]{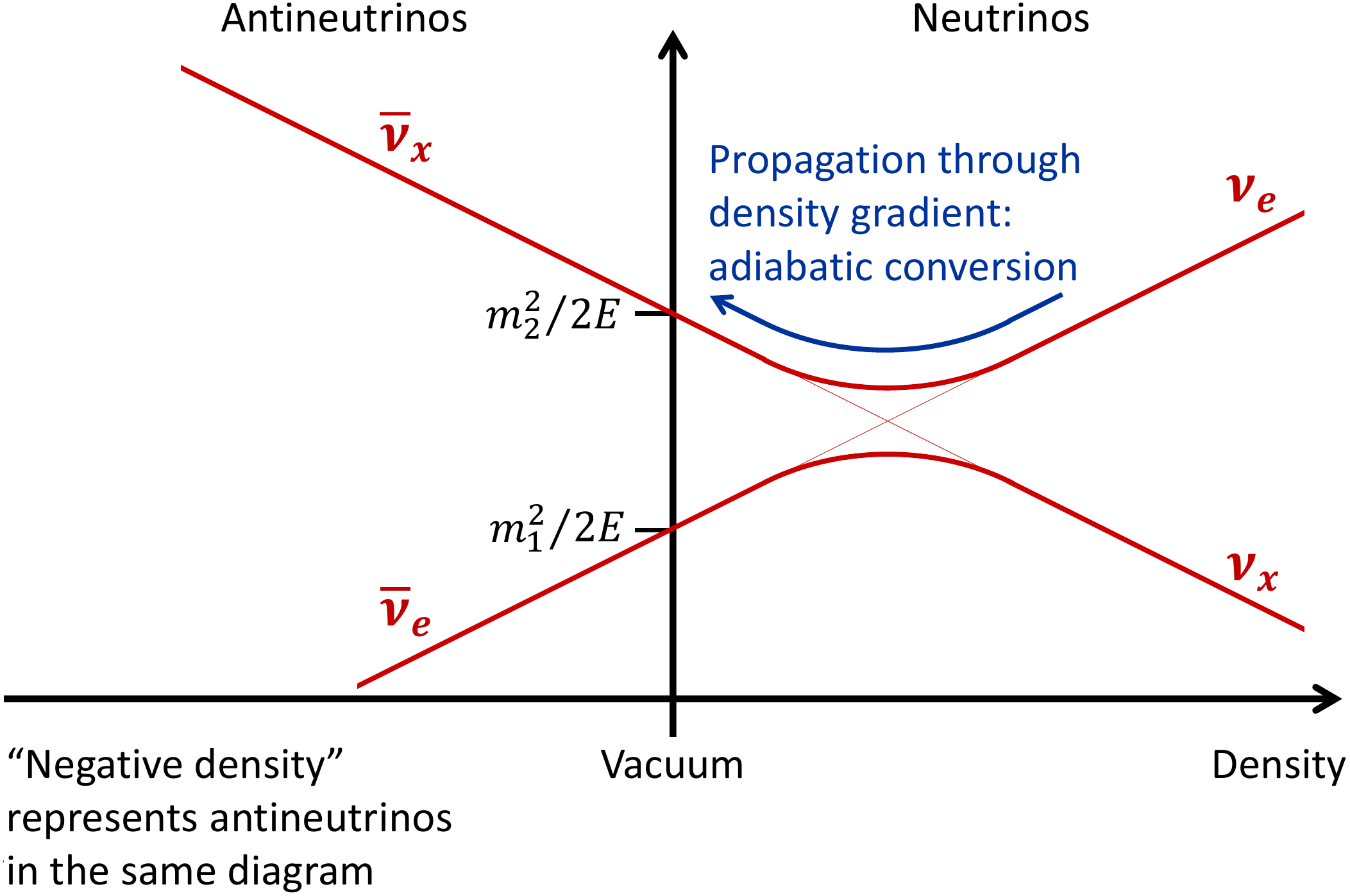}
\caption{Eigenvalue diagram of the $2{\times}2$ Hamiltonian matrix
for 2-flavor oscillations in matter.\label{fig:eigenvalue}}
\end{figure}

The matter contribution has the effect that the eigenstates of ${\sf
H}$, the propagation eigenstates, are not identical with the vacuum
mass eigenstates. In particular, when the density is large,
propagation and flavor eigenstates become more and more similar and
neutrinos are essentially ``un-mixed.'' A completely new effect
arises when neutrinos propagate through a density gradient as in the
Sun. What happens is best explained if one plots the energy
eigenvalues of ${\sf H}$ in eq.~(\ref{eq:ham}) as a function of
density (fig.~\ref{fig:eigenvalue}). The sign of the matter term
changes for antineutrinos, so we can extend the plot to ``negative
densities'' to include neutrinos and antineutrinos in the same plot.
Neutrinos propagating through a density gradient amount to solving
the Schr\"odinger equation with a slowly changing Hamiltonian. If a
system is prepared in an eigenstate of the Hamiltonian and if the
latter changes adiabatically, then the system will always stay in an
eigenstate that slowly changes. So if the neutrino is born as
$\nu_e$ at high density, it is essentially in a propagation
eigenstate. As the density slowly decreases on the neutrino's way
out of the Sun, it always stays in a propagation eigenstate and thus
emerges at the surface (vacuum) as the mass eigenstate $\nu_2$
connected to $\nu_e$ in the level diagram
(fig.~\ref{fig:eigenvalue}). If it were prepared as a $\bar\nu_e$ at
high density (far to the left on the plot), it would emerge as a
$\nu_1$ eigenstate. The crucial point is that the eigenvalues are
unique and do not cross as a function of density---they ``repel''
and ``avoid each other.'' If the mixing angle is small and $\nu_e$
is essentially the lower mass eigenstate $\nu_1$, it still emerges
as $\nu_2$ and thus essentially as $\nu_\mu$, i.e.\ we obtain a
large flavor conversion effect even though the mixing angle is
small. This is the celebrated Mikheev-Smirnov-Wolfstein (MSW) effect
that was discovered in 1985 by Stanislav Mikhheev and Alexei
Smirnov~\cite{Mikheev:1986gs}. The interpretation in terms of an
``avoided level crossing'' as in fig.~\ref{fig:eigenvalue} was given
in the same year by Hans Bethe \cite{Bethe:1986ej}. These results
completely changed the particle physicists' attitude toward the
solar neutrino problem in that a beautiful mechanism had been found
where a small mixing angle could cause large flavor conversion.

\begin{figure}
\centering
\includegraphics[width=1.0\textwidth]{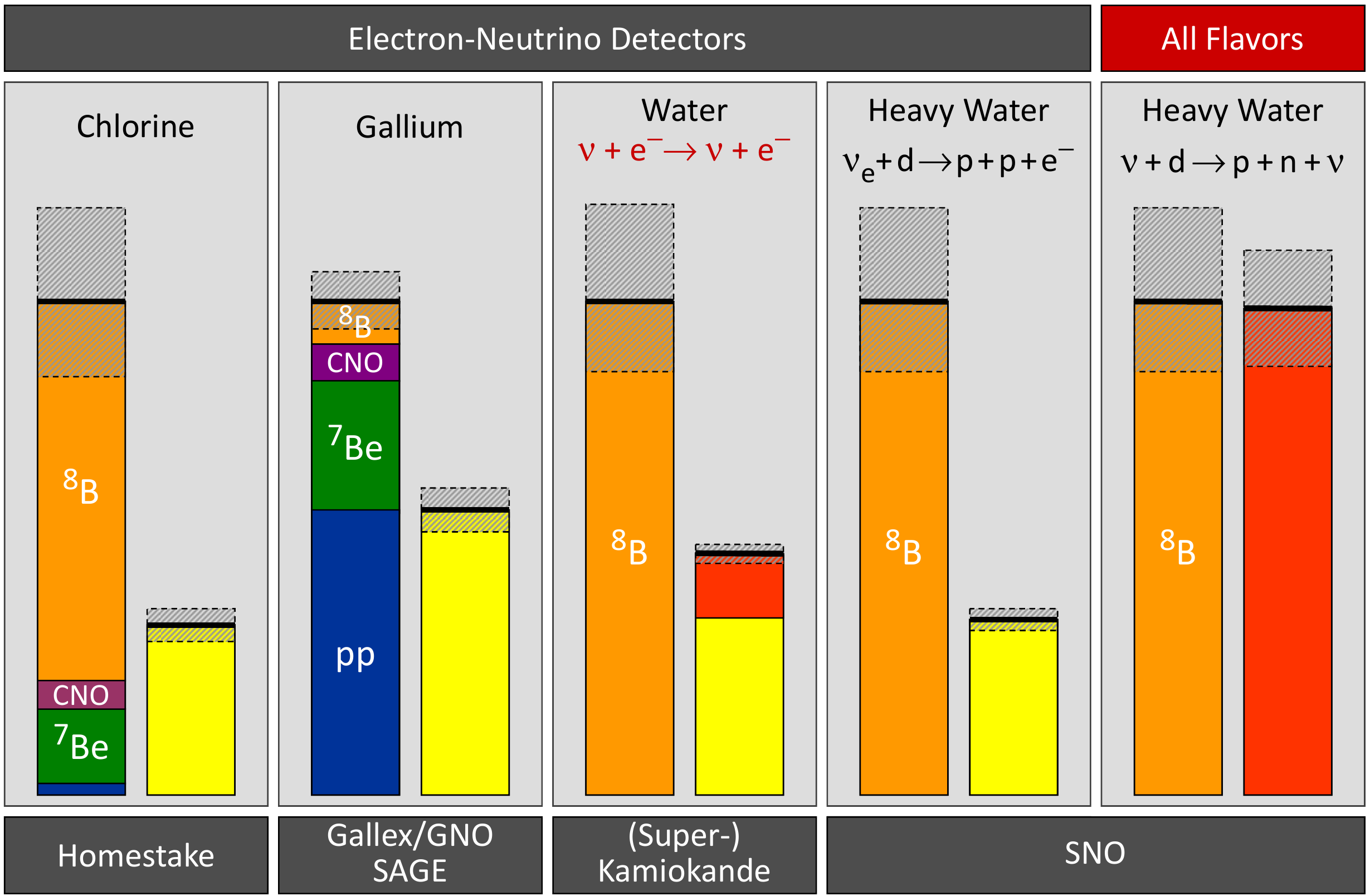}
\caption{Solar neutrino predictions and measurements in different
experiments circa 2002. For each experiment, the total prediction
(in arbitrary units normalized to one) and its error bar are shown
as well as the fractional contribution of different source reactions.
Juxtaposed is the experimental measurement with its uncertainties.
Yellow experimental bars are for $\nu_e$, red bars for
all flavors. (Adapted after a similar plot frequently shown by John Bahcall.)
\label{fig:solarrates}}
\end{figure}

After more than 20 years of data taking with the Homestake Cl
detector, new experiments were coming online. The radiochemical
technique was used with gallium as a target, $\nu_e+{}^{71}{\rm
Ga}\to{}^{71}{\rm Ge}+e^-$. The low energy threshold of 233~keV
allows one to pick up neutrinos from all source reactions, including
the dominant pp flux. The GALLEX experiment, later Gallium Neutrino
Observatory (GNO), used dissolved gallium and was located in the
Gran Sasso laboratory. GALLEX/GNO took data 1991--2003 and confirmed
the solar neutrino problem \cite{Altmann:2005ix}. The Soviet
American Gallium Experiment (SAGE) uses metallic gallium. It took
its first extraction in 1990 and is still running today, with
1990--2007 data published~\cite{Abdurashitov:2009tn}. The expected
contribution of the different source reactions juxtaposed with the
measured rate is shown in fig.~\ref{fig:solarrates}.

The next step forward was the advent of water Cherenkov detectors,
measuring electron scattering $\nu+e\to e+\nu$ where all flavors
contribute, although the $\nu_ee$ cross section is much larger. The
challenge was to lower the energy threshold enough to pick up up
solar $^8$B neutrinos. This feat was first achieved with the
Japanese Kamiokande detector, originally built in 1982--1983 to
search for proton decay. It was ready for solar neutrino detection
in January 1987, consisting of 2140 tons of pure water viewed by 948
photomultipliers, providing 20\% photosensitive area. Almost
immediately, on 23 February 1987, it saw the neutrino burst from
Supernova 1987A. Solar neutrino data were taken January
1987--February 1995 and yielded an $^8$B neutrino flux of
$2.80\pm0.19({\rm stat})\pm0.33({\rm syst})\times10^6~{\rm
cm}^{-2}~{\rm s}^{-1}$, about 49--64\% of standard solar model
predictions, if a pure $\nu_e$ flux is assumed.

\begin{figure}
\centering
\includegraphics[width=1.0\textwidth]{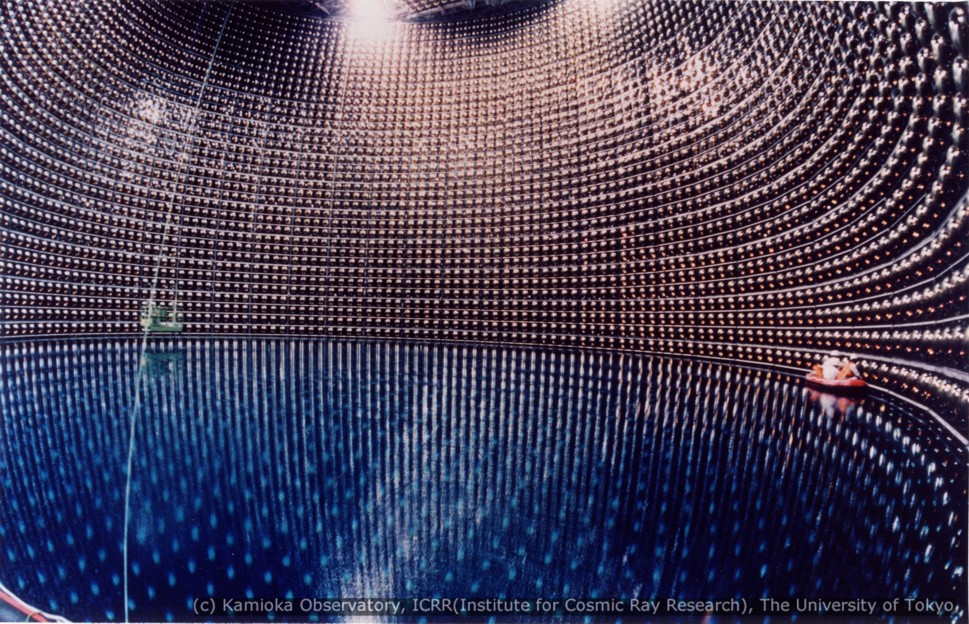}
\caption{Super-Kamiokande water Cherenkov detector being filled in
January 1996 (Copyright: Kamioka Observatory, ICRR, The University of Tokyo).
\label{fig:sk}}
\end{figure}

\begin{figure}
\centering
\includegraphics[width=0.47\textwidth]{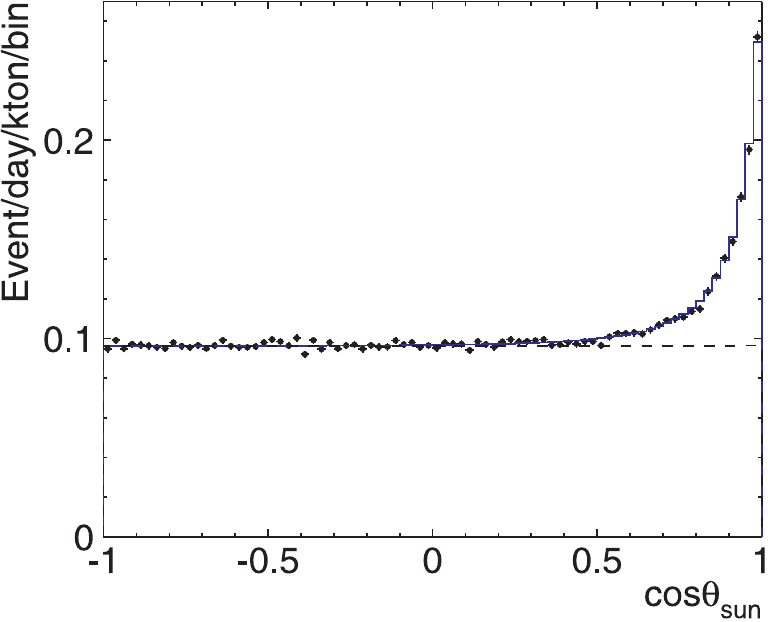}\hfill
\includegraphics[width=0.49\textwidth]{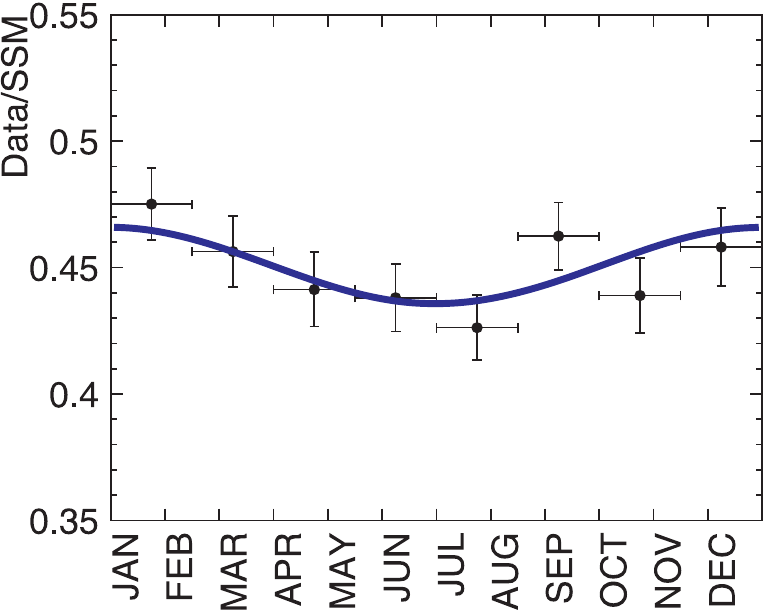}
\caption{Solar neutrino measurements with 1258 days of Super-Kamiokande
\cite{Fukuda:2001nj}. Left: Positron direction relative to Sun, including
a uniform background on the level of 0.1.
Right: Seasonal variation of the total flux.\label{fig:sksolar}}
\end{figure}

The era of high-statistics solar neutrino measurements began when
the 50~kton water Cherenkov detector Super-Kamiokande
(fig.~\ref{fig:sk}) took up operation on 1~April 1996 and has taken
data since with some interruptions for repairs and upgrades. Super-K
registers about 15 solar neutrinos per day, i.e.\ about as many in
two months as Homestake did in a quarter century. The latest
published results are those of Super-K phase III that ended in
August 2008 \cite{arXiv:1010.0118}, when the electronics was
replaced, giving way to Super-K IV as the currently operating
detector. The $^8$B flux, under the assumption of pure $\nu_e$, was
measured by Super-K III to be $2.32\pm0.04({\rm stat})\pm0.05({\rm
syst})\times10^6~{\rm cm}^{-2}~{\rm s}^{-1}$.

With such high statistics one can perform true neutrino astronomy.
The electron recoil events crudely maintain the neutrino direction
and therefore statistically point back to the Sun
(fig.~\ref{fig:sksolar}, left panel). Likewise, the annual neutrino
flux variation reveals the ellipticity of the Earth orbit around the
Sun (fig.~\ref{fig:sksolar}, right panel).

\begin{figure}[b]
\centering
\includegraphics[width=0.45\textwidth]{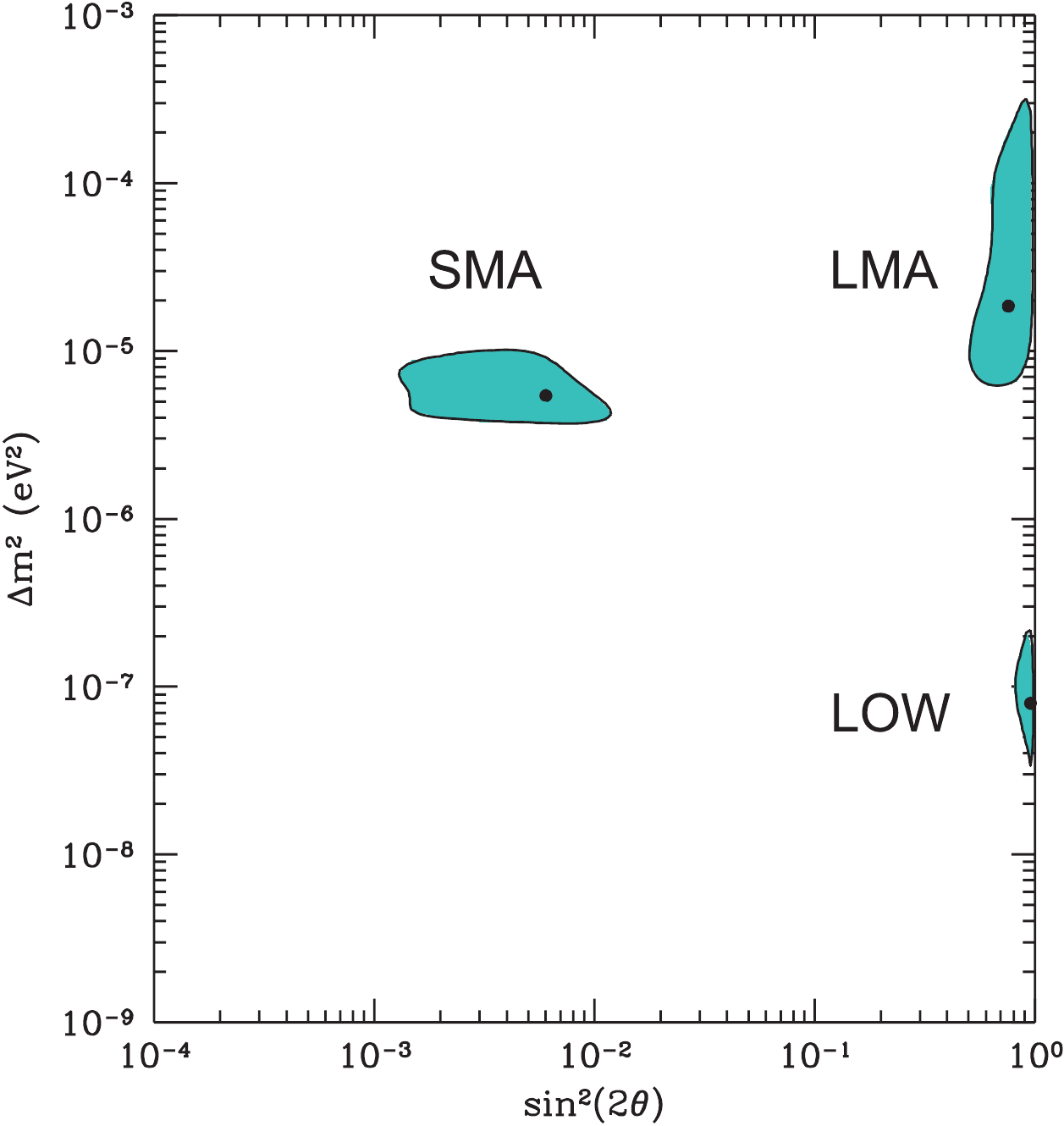}\hfil
\includegraphics[width=0.45\textwidth]{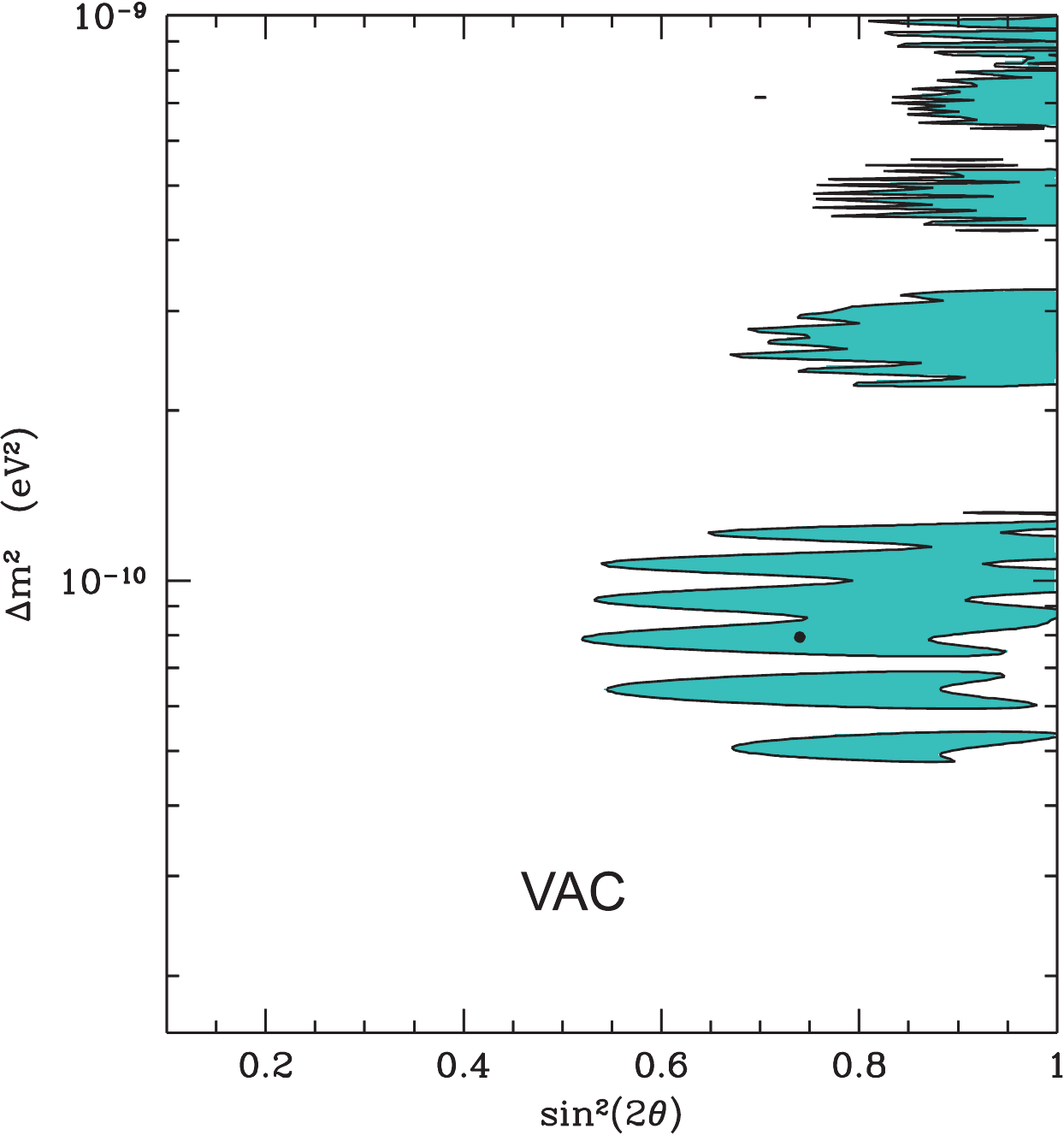}
\caption{Best-fit regions circa 1998 in a two-flavor oscillation interpretation of the
measured rates of Homestake, GALLEX, SAGE and Super-Kamiokande
together with the predictions of the Bahcall and Pinsonneault (1998)
standard solar model. (Adapted from Ref.~\cite{Bahcall:1998jt}.)
\label{fig:solarsolutions}}
\end{figure}

Interpreting the solar neutrino observations of Homestake, GALLEX,
SAGE and Super-Kamiokande in terms of two-flavor oscillations led
around 1998 to the situation shown in fig.~\ref{fig:solarsolutions}.
There were three MSW solutions where the matter effect in the Sun is
important, the small-mixing angle solution (SMA), the large
mixing-angle solution (LMA) and the LOW solution. In addition there
was a solution with large mixing angle and pure vacuum oscillations
(VAC), corresponding to an oscillation length of the Sun-Earth
distance of 150 million km. The SMA solution, where a small mixing
angle gives a large flavor conversion by the MSW mechanism, was
still favored by many.

Then the situation changed quickly with Super-K in 1998 producing
first unambiguous evidence for atmospheric $\nu_\mu\to\nu_\tau$
oscillations with a near-maximal mixing angle~\cite{Fukuda:1998mi},
showing neutrino flavor oscillations with a large mixing angle.
Moreover, when Super-K began including high-statistics spectral and
zenith-angle information for solar neutrinos, the SMA and VAC
solutions became less and less of a good fit~\cite{Fukuda:2001nk}.

\begin{figure}
\centering
\includegraphics[height=0.58\textwidth]{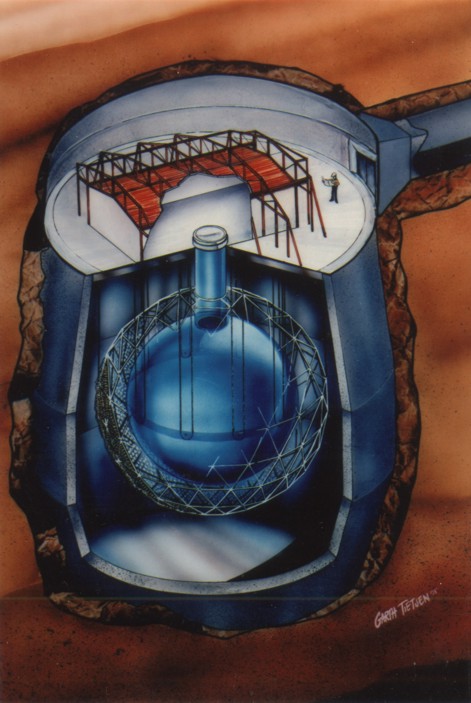}\hfill
\includegraphics[height=0.58\textwidth]{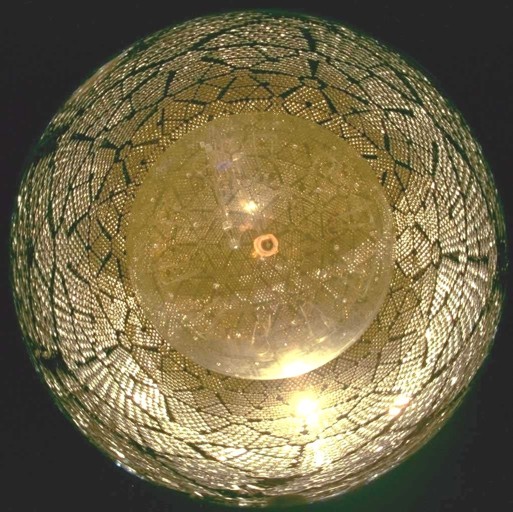}
\caption{Sudbury neutrino observatory (SNO), Cherenkov detector
with 1000 tons of heavy water. Left: Artists rendition of detector.
Right: Fish-eye picture. (Photos courtesy of SNO.)
\label{fig:sno}}
\end{figure}

\begin{figure}
\centering
\includegraphics[height=0.5\textwidth]{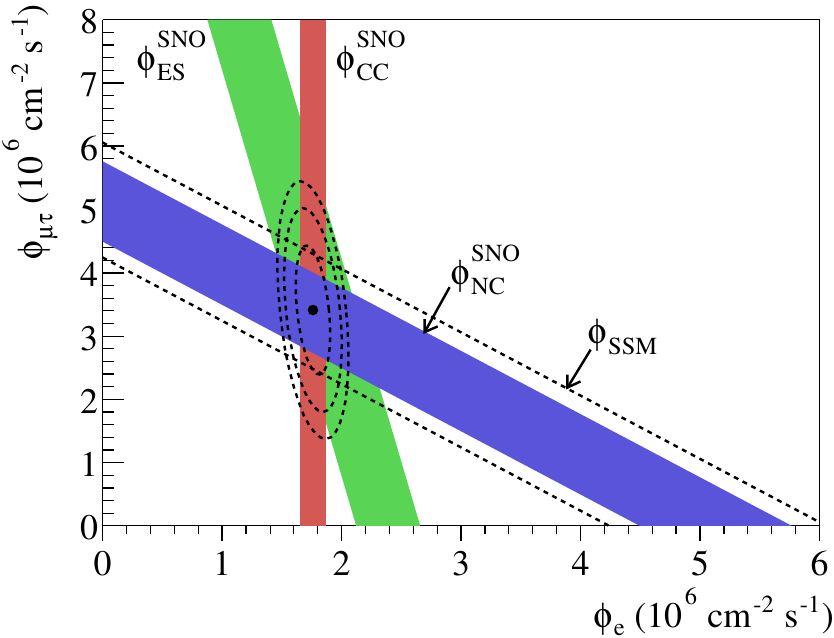}
\caption{SNO solar neutrino measurements (2002)
for charged current (CC) and neutral current (NC) deuterium
disintegration and electron scattering (ES) \cite{Ahmad:2002jz}.
\label{fig:snodata}}
\end{figure}

The solar oscillation story was finally wrapped up by two new
experiments. One was the Sudbury Neutrino Observatory (SNO) in
Canada, a water Cherenkov detector that used 1000~t of heavy water,
${\rm D}_2{\rm O}$, as a target, taking data 1999--2006
(fig.~\ref{fig:sno}). It uses electron scattering (ES) that is
sensitive primarily to $\nu_e$ and also the other flavors. It
further uses a pure $\nu_e$ channel by charged-current (CC) deuteron
disintegration, $\nu_e+d\to p+p+e^-$, and an all-flavor channel by
neutral-current (NC) disintegration, $\nu+d\to p+n+\nu$. When first
results from all three channels became available in 2002, the iconic
picture of fig.~\ref{fig:snodata} revealed a consistent solution
where the all-flavor $^8$B flux was as predicted by solar models and
the $\nu_e$ deficit was clearly explained by flavor
conversion~\cite{Ahmad:2002jz}.

\begin{figure}[b]
\centering
\includegraphics[height=0.5\textwidth]{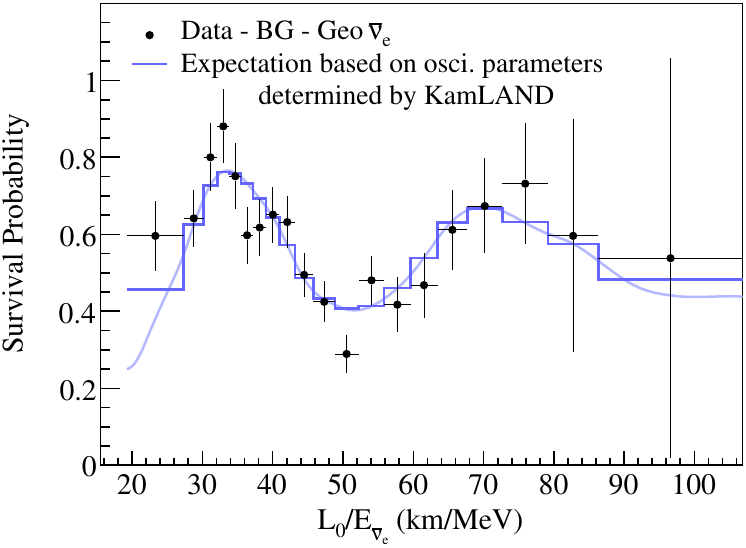}
\caption{Energy variation in terms of $L/E$ of the KamLAND
reactor neutrino measurements \cite{Abe:2008ee}, clearly
showing flavor oscillations.\label{fig:kamlanddata}}
\end{figure}

After Super-K had been built, the old Kamiokande water Cherenkov
detector was replaced with KamLAND, a scintillator detector, with
correspondingly lower energy threshold that could measure the
neutrino flux from the Japanese nuclear power reactors, the dominant
distance being around 180~km. In this way the solar LMA solution
could be tested with a laboratory experiment, of course against
theoretical advice, favoring the SMA solution. The year 2002 became
the {\it annus mirabilis} of neutrino physics in that KamLAND indeed
found $\nu_e$ disappearance corresponding to the solar LMA
solution~\cite{Eguchi:2002dm}. With more statistics, KamLAND later
produced the beautiful $L/E$ plot of fig.~\ref{fig:kamlanddata}. The
flavor oscillation probability of
eq.~(\ref{eq:oscillationprobability}) varies with $L/E$ so that one
can see an oscillation pattern when plotting the measurements as a
function of this variable. This is probably the most convincing
evidence for the reality of flavor oscillations.

\begin{figure}
\centering
\includegraphics[height=0.6\textwidth]{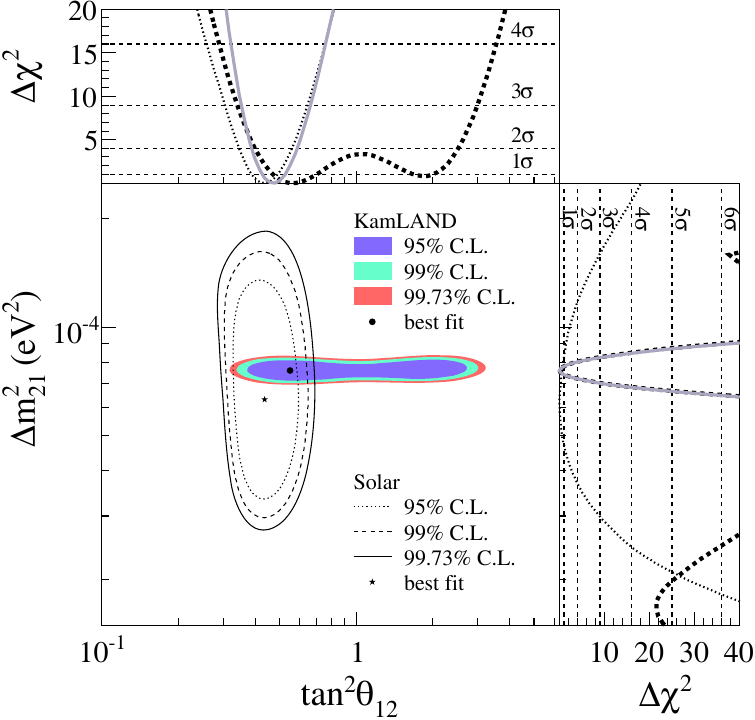}
\caption{Allowed region for neutrino oscillation parameters from
KamLAND and solar neutrino experiments~\cite{Abe:2008ee}. The side-panels show
the $\chi^2$-profiles for KamLAND (dashed) and solar experiments
(dotted) individually, as well as the combination of the two (solid).
\label{fig:solarfit}}
\end{figure}

Combining all solar neutrino measurements and the KamLAND reactor
results in a two-flavor oscillation interpretation yields the
best-fit parameters shown in fig.~\ref{fig:solarfit}. It is
essentially KamLAND that fixes $\Delta m^2$ with high precision,
whereas the solar measurements fix the mixing angle. The values
above and below $45^\circ$ are not symmetric because of the matter
effect in the Sun. In other words, the solar matter effect fixes the
mass ordering to be $m_1<m_2$ and the mixing angle is large but not
maximal.

While the solar neutrino problem has been settled since 2002, this
is not the end of solar neutrino measurements. The task now is
precision and detailed tests. One new contribution in solar neutrino
spectroscopy comes from the Borexino experiment in the Gran Sasso
laboratory. It is an ultrapure scintillator detector (278 tons) and
measures solar neutrinos by electron scattering. It is particularly
sensitive to the monochromatic $^7$Be neutrinos (0.863~MeV) and pep
neutrinos (1.445~MeV) because they produce a distinct shoulder in
the electron recoil spectrum. After many delays, data taking began
in August 2007 and the detector worked beautifully. The most recent
result provides the $\nu_e$ equivalent $^7$Be flux of
$(3.10\pm0.15)\times10^9~{\rm cm}^{-2}~{\rm s}^{-1}$ and under the
assumption of flavor oscillations a $\nu_e$ survival probability of
$0.51\pm0.07$ at 862~keV~\cite{Bellini:2011rx}. Most recently, a
measurement of the much smaller pep flux was also
reported~\cite{Bellini:2011}.

\begin{figure}
\centering
\includegraphics[width=0.7\textwidth]{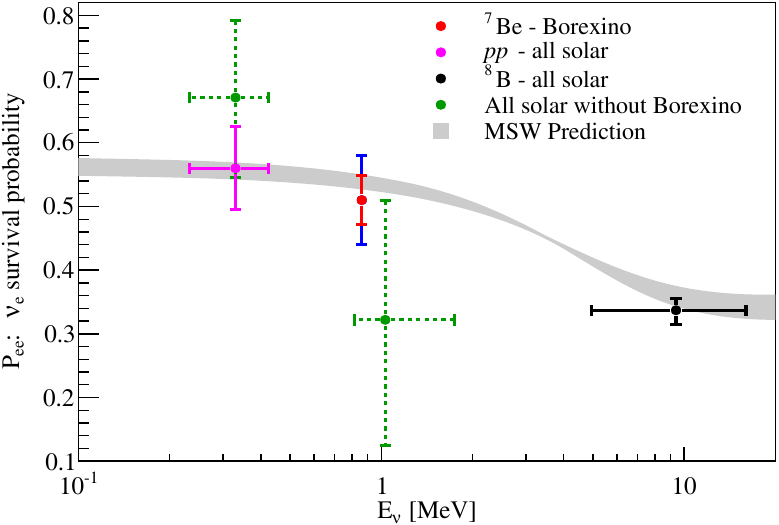}
\caption{Energy-dependent survival probability $P_{ee}$ for solar
$\nu_e$~\cite{Bellini:2011rx}. The grey band indicates the standard
solar model (SSM) expectation together with the best-fit LMA solution.
For the $^7$Be point, measured by Borexino, the inner (red) error
bars show the experimental uncertainty, while the outer (blue) error
bars show the total (experimental + SSM) uncertainty. The remaining
points were obtained from a combined analysis of the results of all
solar neutrino experiments. The green (dashed) points are calculated
without Borexino data.\label{fig:survival}}
\end{figure}

The $\nu_e$ survival probability $P_{ee}$ in the Sun at $E\alt
1$~MeV is essentially given by vacuum oscillations because $\Delta
m^2/2E$ is too large to be much affected by solar matter, so
$P_{ee}\sim 1-\frac{1}{2}\,\sin^22\theta$ (fig.~\ref{fig:survival}).
On the other hand, for $E\agt 6$~MeV it is given by the MSW value
$P_{ee}\sim \sin^2\theta$. The energy-dependent solar measurements
confirm this picture. Borexino has made this crucial test much more
precise (fig.~\ref{fig:survival}).

Solar neutrino oscillations are usually analyzed in a two-flavor
context, but of course we have three active flavors that are
superpositions of three mass eigenstates,
\begin{equation}
\begin{pmatrix}
\nu_e\\ \nu_\mu\\ \nu_\tau
\end{pmatrix}
={\sf U}
\begin{pmatrix}
\nu_1\\ \nu_2\\ \nu_3
\end{pmatrix}\,,
\end{equation}
where the unitary transformation can be parameterized in the form
\begin{equation}
{\sf U}=
\begin{pmatrix}
1&0&0\\
0&c_{23}&s_{23}\\
0&-s_{23}&c_{23}
\end{pmatrix}
\begin{pmatrix}
c_{13}&0&e^{-\I\delta}s_{13}\\
0&1&0\\
-e^{\I\delta}s_{13}&0&c_{13}
\end{pmatrix}
\begin{pmatrix}
c_{12}&s_{12}&0\\
-s_{12}&c_{12}&0\\
0&0&1
\end{pmatrix}
\,,
\end{equation}
where $c_{12}=\cos\theta_{12}$, $s_{12}=\sin\theta_{12}$ and so
forth. Besides two mass differences $m_{21}^2=m_2^2-m_1^2$ and
$m_{31}^2=m_3^2-m_1^2$, flavor oscillations depend on three mixing
angles $\theta_{12}$, $\theta_{23}$, and $\theta_{13}$ as well as a
CP-violating phase (Dirac phase) $\delta$.

\begin{table}
  \caption{Neutrino oscillation parameters from a global fit of
  all solar, reactor, atmospheric and long-baseline
  experiments~\cite{GonzalezGarcia:2010er}. The preliminary value on
  $\theta_{13}$ is based on T2K and first Double Chooz
  data~\cite{Kerret:2011}.
  \label{tab:mixingparameters}}
  \begin{tabular}{lllll}
    \hline
    Parameter&Units&Best-fit&$1\sigma$ range&$3\sigma$ range\\
    \hline
    $\delta m^2=m_2^2-m_1^2$&meV$^2$&$+75.8$&73.2--78.0&69.9--81.8\\
    $\Delta m^2=m_3^2-\frac{1}{2}(m_2^2+m_1^2)$&meV$^2$&$\pm2350$&$\pm$(2260--2470)&$\pm$(2060--2670)\\
    $\sin^2\theta_{12}$&&0.306&0.291--0.324&0.259--0.359\\
    $\sin^2\theta_{23}$&&0.42 &0.39--0.50&0.34--0.64\\
    $\sin^2\theta_{13}$&&\multicolumn{3}{l}{0.085$\pm0.029$(stat)$\pm0.042$(sys)}\\
    $\delta$&&$0^\circ$--$360^\circ$\\
    \hline
  \end{tabular}
\end{table}

The current best-fit values for the oscillation parameters are
summarized in table~\ref{tab:mixingparameters} according to Fogli
et~al.~\cite{Fogli:2011qn} (see Gonzalez-Garcia et
al.~\cite{GonzalezGarcia:2010er} for an alternative analysis). The
third mixing angle $\theta_{13}$ is small so that flavor
oscillations approximately factorize into the two-flavor oscillation
problems of the 12 sector (``solar oscillations'') and the 23 sector
(``atmospheric oscillations''). Until recently, all data were
compatible with a vanishing $\theta_{13}$, although a global
analysis provided first hints for a nonvanishing value at the
$3\sigma$ level~\cite{Fogli:2011qn}. Most recently (Nov.~2011),
additional evidence came from the Double Chooz reactor
experiment~\cite{Kerret:2011}. This question will be convincingly
settled within a few years with more data from the T2K long baseline
experiment and the reactor experiments Double Chooz, Reno, and
Daya~Bay. If indeed $\theta_{13}$ is not very small, then the next
step will be to measure the Dirac phase $\delta$, causing CP
violation in oscillation experiments. The other parameter that
remains to be settled is the mass hierarchy, i.e.\ if $\Delta m^2>0$
(normal hierarchy) or $\Delta m^2<0$ (inverted hierarchy). In the 12
sector, the mass ordering $\delta m^2>0$ has been settled by the
matter effect in the Sun.

\begin{figure}[b]
\centering
\includegraphics[height=0.32\textwidth]{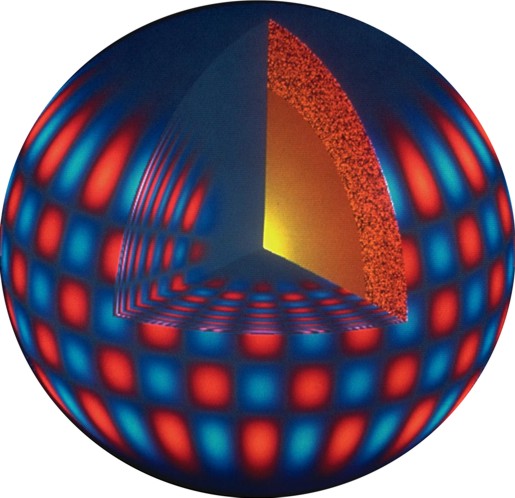}\hfil
\includegraphics[height=0.32\textwidth]{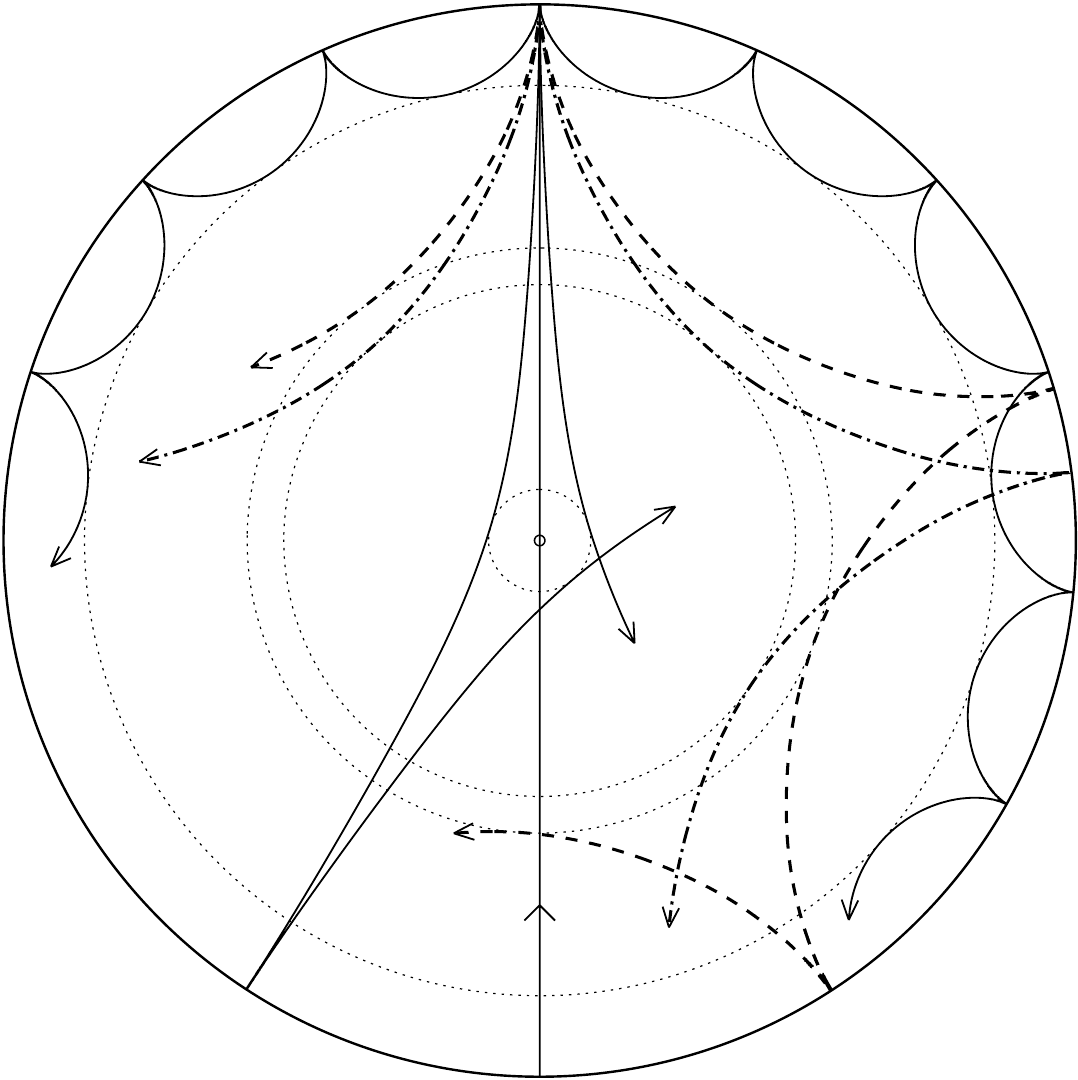}
\caption{Left: One example for solar p-mode oscillations
(Credit: Global Oscillation Network Group/National Solar Observatory/AURA/NSF).
Right: Propagation of p-modes in the Sun~\cite{Christensen:2003}
(Credit: J.~Christensen-Dalsgaard, TAC Aarhus).\label{fig:pmodes}}
\end{figure}

\begin{figure}[b]
\centering
\includegraphics[height=0.60\textwidth]{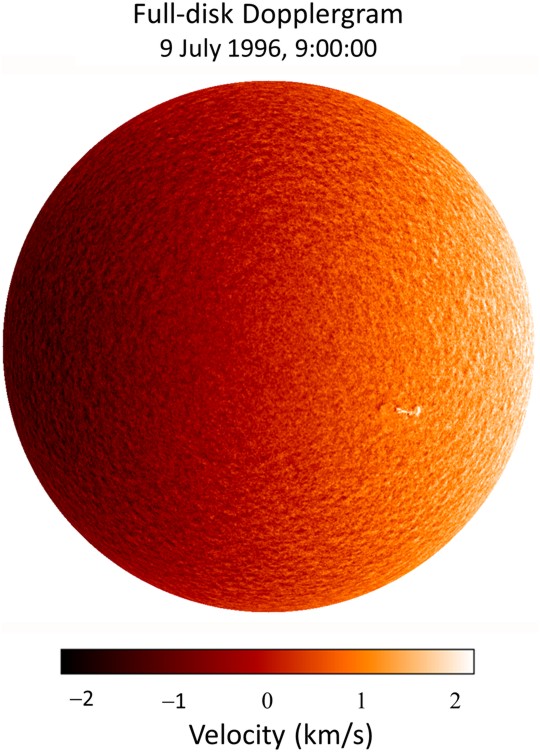}\hfil
\includegraphics[height=0.60\textwidth]{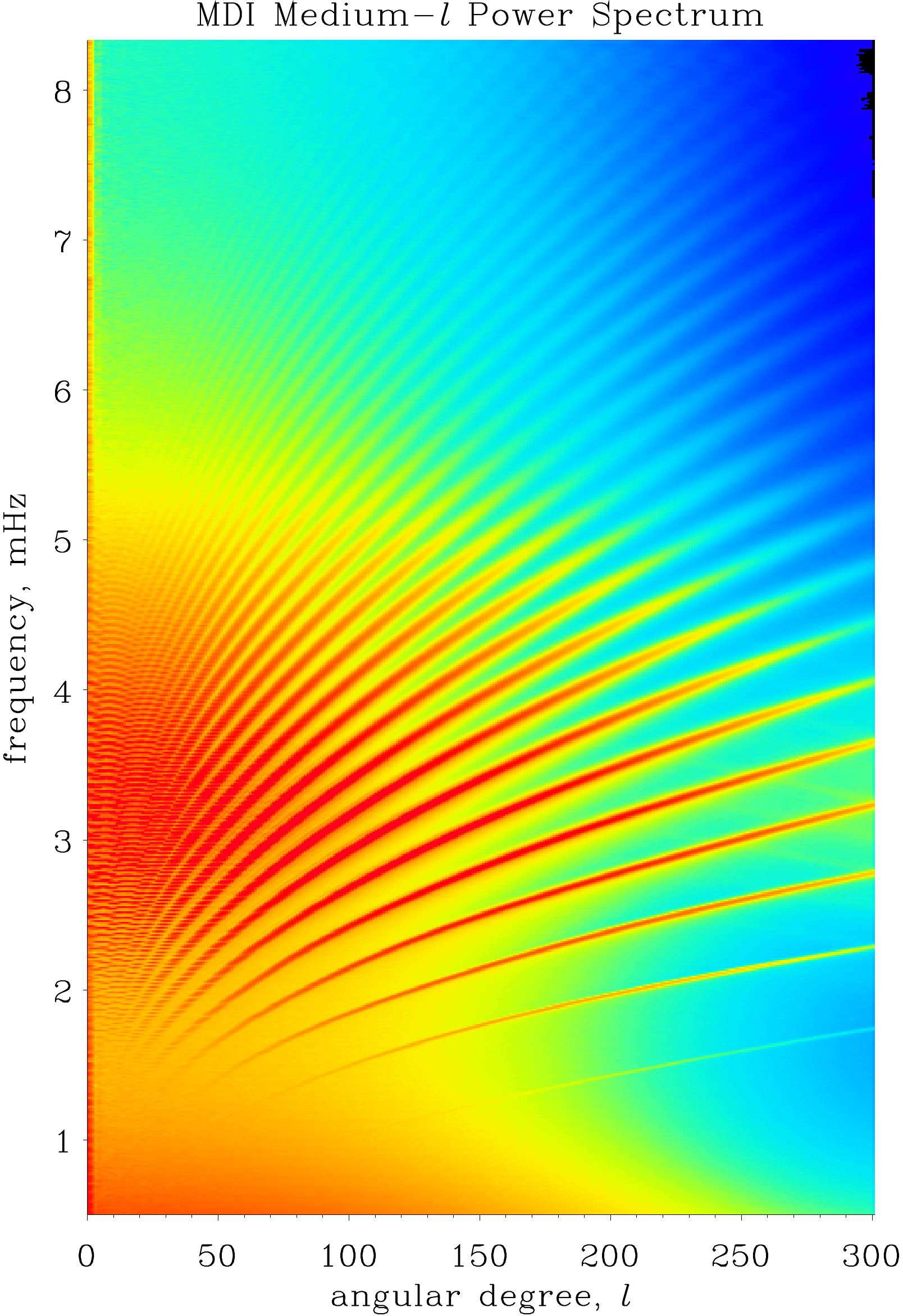}
\caption{Full-disk Dopplergram of the Sun taken with the MDI instrument
on the SOHO satellite (left). Power spectrum of p-modes (right).
Credit: SOHO (ESA \& NASA).\label{fig:dopplergram}}
\end{figure}

\subsection{Helioseismology and the solar opacity problem}
\label{sec:opacity}

The inner properties of the Sun can be studied with neutrinos and
helioseismology. For many years, helioseismology yielded perfect
agreement with standard solar models, whereas the neutrino
measurements were plagued by the mysterious $\nu_e$ deficit that was
finally explained by flavor oscillations. Just as the neutrino
problem got sorted out, the helioseismic agreement began to sour and
today poses a new problem about the Sun.

The solar structure can vibrate around its hydrostatic equilibrium
configuration in different ways. Of main interest are the p-modes
(pressure modes), essentially sound waves with few-minute
frequencies, that get constantly excited by the convective overturns
in the outer layers of the Sun. Depending on their frequency, these
seismic waves probe more or less deep into the solar interior
(fig.~\ref{fig:pmodes}), allowing one to probe the solar sound-speed
profile as a function of radius. One needs to measure the p-mode
frequencies as a function of multipole order $\ell$. To this end one
measures the motion of the solar surface by the Doppler effect and
can produce a ``Dopplergram'' as shown in fig.~\ref{fig:dopplergram}
where one can also see the global rotation of the Sun by the
systematic speed variation across the solar disk. To determine the
frequencies one needs a long uninterrupted time series for Fourier
transformation. This is achieved either by satellite observations
such as the MDI instrument on the SOHO satellite
(http://sohowww.nascom.nasa.gov) or by networks of terrestrial
telescopes that offer 24h vision of the Sun such as BiSON
(http://bison.ph.bham.ac.uk) and GONG (http://gong.nso.edu). A
typical power spectrum derived by this method is also shown in
fig.~\ref{fig:dopplergram}. The theory of how to invert this
information to derive a solar sound speed profile is described, for
example, in the lecture notes of
J.~Christensen-Dalsgaard~\cite{Christensen:2003}. In this way one
can derive a ``seismic model'' of the Sun that allows for comparison
with standard solar models. Besides the sound-speed profile, one
also derives the depth of the convective zone $R_{\rm CZ}$ and the
surface helium mass fraction $Y_{\rm S}$, an adjustable solar-model
parameter that is not directly observable.

\begin{figure}
\centering
\includegraphics[width=0.62\textwidth]{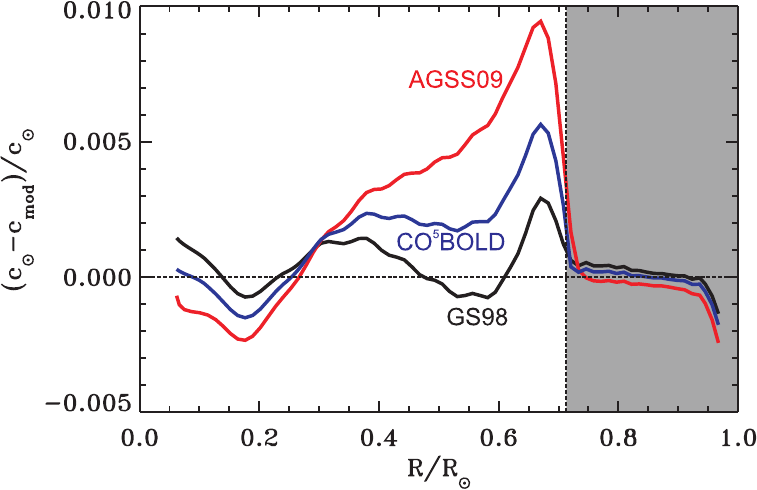}
\caption{Solar soundspeed profile relative to helioseismic model
for the indicated cases of opacities. The grey region
is the convection zone.
(Adapted from Serenelli 2011~\cite{Serenelli:2011ea}.)\label{fig:soundspeed}}
\end{figure}

\begin{table}
  \caption{Properties of solar
  models with different opacities~\cite{Serenelli:2011ea}.\label{tab:solarmodels}}
  \begin{tabular}{llll}
    \hline
    Model&Metallicity&Depth Convection Zone&Surface He Abundance\\
    &$Z/X$&$R_{\rm CZ}/R_\odot$&$Y_{\rm S}$\\
    \hline
    Seismic   &      &$0.713\pm0.001$&$0.2485\pm0.0035$\\
    GS98      &0.0229& 0.713         & 0.243\\
    CO$^5$BOLD&0.0209& 0.717         & 0.237\\
    AGSS09    &0.0178& 0.723         & 0.232\\
    \hline
  \end{tabular}
\end{table}

A traditional solar model compared with helioseismology is shown by
the black line (GS98) in fig.~\ref{fig:soundspeed}. The perfect
agreement, taken for a long time as evidence for our excellent
understanding of the Sun, depends crucially on the solar opacities,
which in turn depend on the abundances of chemical elements.
Traditional models are based on the Grevesse and Sauval 1998 (GS98)
opacities~\cite{Grevesse:1998bj}. Since 2005, however, Martin
Asplund and collaborators have provided new solar element abundances
based on a 3D-hydrodynamics model atmosphere, better selection of
spectral lines (identification of blends) and detailed treatment of
radiative transport in the line-formation modeling. This leads to a
30--40\% reduction of the CNO and Ne abundances. Solar models based
on the Asplund, Grevesse, Sauval and Scott 2009 (AGSS09)
opacities~\cite{Asplund:2009fu} lead to significant modifications of
the sound-speed profile, depth of convection zone and surface helium
abundance (fig.~\ref{fig:soundspeed} and
table~\ref{tab:solarmodels}), in stark conflict with the seismic
model. Caffau and collaborators (CO$^5$BOLD) have embarked on a
similar task, but arrive at different
abundances~\cite{arXiv:1003.1190}. The corresponding solar models
are halfway between GS98 and AGSS09. Either way, the discrepancy
with helioseismology remains unresolved. For example, phases of
accretion during solar evolution do not seem to be
successful~\cite{Serenelli:2011py}.

The solar neutrino flux predictions are also modified as shown in
table~\ref{tab:solarfluxes}. However, the directly measured $^8$B
and $^7$Be fluxes are roughly halfway between the GS98 and AGSS09
models and agree with either within uncertainties. On the other
hand, the predicted CNO-cycle neutrino fluxes naturally are much
smaller, but for the moment only crude experimental upper limits
exist. Sufficiently precise neutrino observations of the CNO
neutrino fluxes could settle the question of the element abundances
in the deep solar interior, but it appears doubtful that Borexino
can measure these fluxes with sufficient precision, even if it
achieves to measure them.

The new question of solar element abundances has opened up a new
frontier for solar neutrino astronomy. Evidently our understanding
of flavor oscillations is crucial for using neutrinos as legitimate
astrophysical probes. Solar neutrino measurements began to prove
that nuclear reactions were the power source of stars. After the
``distraction'' of flavor oscillations, the field is back to its
roots as a probe of the solar interior.

\subsection{Sun as a particle source}

The Sun is a very well understood neutrino source and has provided
invaluable information on neutrino oscillation parameters. Some of
the solar $\nu_e$ fluxes, notably the pp flux, arguably are better
known than the $\bar\nu_e$ flux from a power reactor where a
possible adjustment of several percent has recently caused a lot of
attention~\cite{Mention:2011rk}. The Sun as a $\nu_e$ source can
provide additional information beyond oscillation parameters. For
example, a hypothetical $\nu_e\to\bar\nu_e$ conversion, perhaps by
Majorana transition moments, has been constrained by Borexino to a
probability of less than $1.3\times10^{-4}$ (90\% CL) for
$E_{\bar\nu}>1.8$~MeV, the most restrictive limit of this
kind~\cite{Bellini:2010gn}. One can also constrain radiative
neutrino decays $\nu_2\to\nu_1+\gamma$ by the absence of solar
$\gamma$ rays \cite{Raffelt:1985rj}, but the small neutrino mass
differences render such constraints on the effective transition
moment less interesting than, for example, the globular cluster
limit from plasmon decay given in eq.~(\ref{eq:neutrinoemlimits}).

\begin{figure}
\centering
\includegraphics[width=0.4\textwidth]{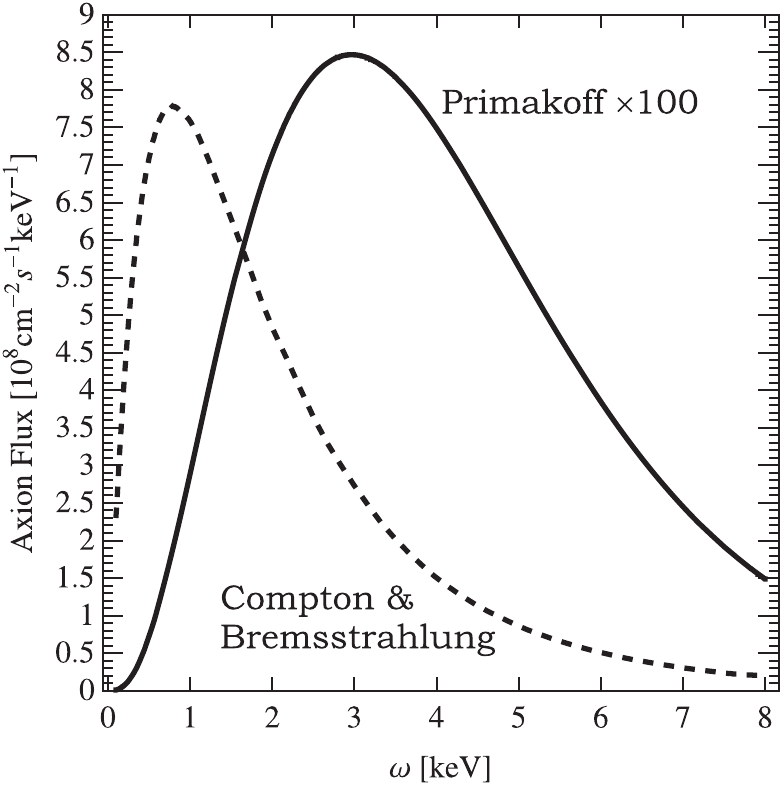}
\caption{Solar axion flux at Earth from electron processes, assuming
$g_{ae}=1\times10^{-13}$ (dashed line) and from the Primakoff process,
assuming an axion-photon coupling of $g_{a\gamma}=10^{-12}~{\rm GeV}^{-1}$,
corresponding to DFSZ axions with $f_a=0.85\times10^9$~GeV, $C_e=1/6$
and $E/N=8/3$ \cite{Irastorza:2011gs}.\label{fig:axionflux}}
\end{figure}

The Sun can also emit hypothetical low-mass particles other than
neutrinos where both nuclear reactions and thermal plasma process
can be the source. For example, in the reaction $d+p\to{}^3{\rm
He}+\gamma$ of the solar pp chains (table~\ref{tab:ppchains}) the
photon can be substituted with an axion that can subsequently decay
outside of the Sun, producing $\gamma$ rays, an argument that has
led to an early constraint on ``standard axions''
\cite{Raffelt:1982dr}. Today, ``invisible axions'' are of much
greater interest with such low masses that they are easily produced
in the thermal processes of fig.~\ref{fig:axionprocesses} that are
based either on the axion-electron or the axion-photon coupling. In
the so-called DFSZ axion model, the axion-photon interaction
strength is given by $E/N=8/3$ in
eq.~(\ref{eq:axiontwophotoncoupling}). If we assume $C_e=1/6$, the
solar axion flux prediction at Earth is shown in
fig.~\ref{fig:axionflux}, based on the white-dwarf inspired
axion-electron coupling of $g_{ae}=10^{-13}$
(section~\ref{sec:whitedwarfcooling}).

Solar axions can be searched with the ``helioscope''
technique~\cite{PRINT-83-0597 (FLORIDA)}. Particles with a
two-photon vertex can transform into photons and vice versa in an
external electromagnetic field. For a microscopic target this is a
scattering process with photon exchange, the Primakoff process shown
in fig.~\ref{fig:axionprocesses}. In a macroscopic field, the
conversion $a\to\gamma$ is more akin to a flavor
oscillations~\cite{MPI-PAE/PTh-54/87}. The ``flavor variation''
along a beam in $z$ direction is then given, in full analogy to
neutrino flavor oscillations, by the Schr\"odinger-like equation
\begin{equation}
{\rm i}\,\frac{\partial}{\partial z}
\begin{pmatrix}\gamma\\ a\end{pmatrix}=\frac{1}{2\omega}
\begin{pmatrix}\omega_{\rm pl}^2&g_{a\gamma} B\omega\\
g_{a\gamma} B\omega&m_a^2
\end{pmatrix}
\begin{pmatrix} \gamma\\ a\end{pmatrix}\,,
\end{equation}
where $\omega$ is axion or photon energy, $B$ is the transverse
magnetic field, and we have included an effective photon mass in
terms of the plasma frequency if the process does not take place in
vacuum. The conversion probability after a distance $L$ is
\begin{equation}
P_{a\to\gamma}=\left(\frac{g_{a\gamma}BL}{2}\right)^2\,
\frac{\sin^2(qL/2)}{(qL/2)^2}\,,
\end{equation}
where the required momentum transfer is
\begin{equation}
q=\sqrt{\left(\frac{\omega_{\rm pl}^2-m_a^2}{2\omega}\right)^2+(g_{a\gamma}B)^2}\,.
\end{equation}

\begin{figure}
\centering
\includegraphics[width=1.0\textwidth]{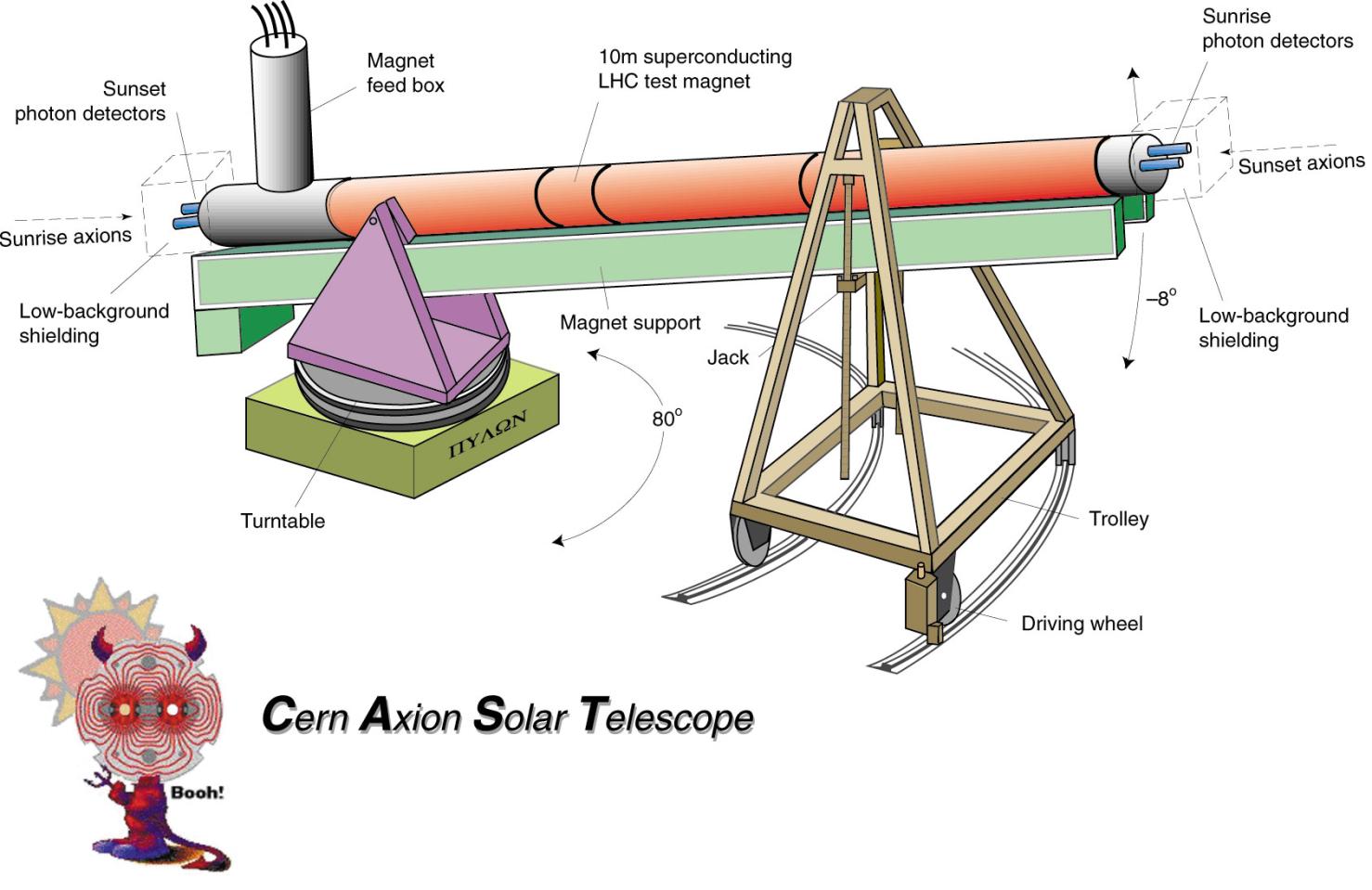}
\caption{CAST experiment at CERN to search for solar axions.\label{fig:cast}}
\end{figure}

\begin{figure}
\centering
\includegraphics[width=0.5\textwidth]{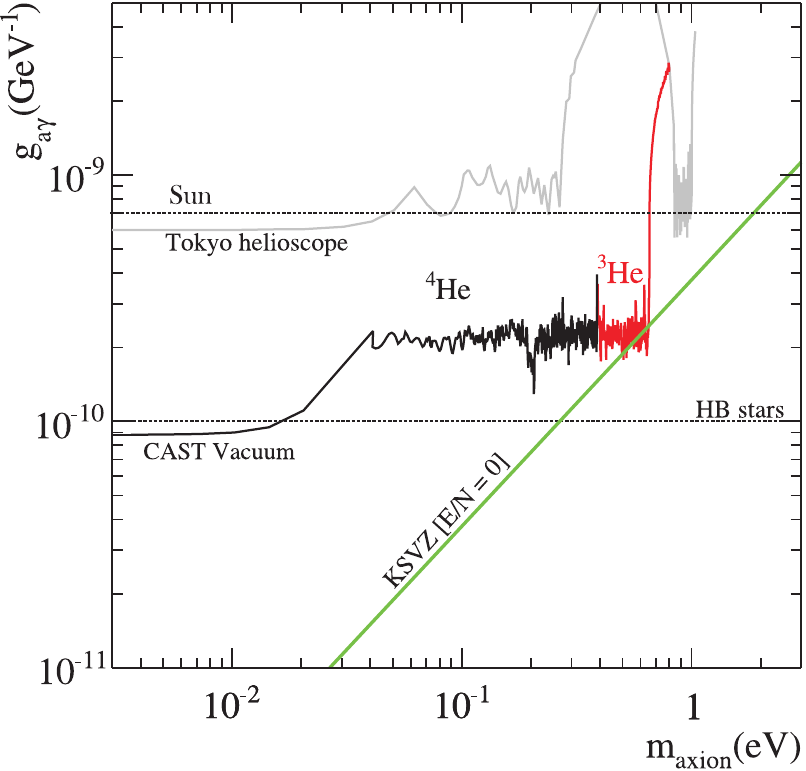}
\caption{Axion $g_{a\gamma}$-$m_a$ exclusion range by the CAST solar
axion search at CERN. (Adapted from Ref.~\cite{Aune:2011rx}.)\label{fig:castexclusion}}
\end{figure}

To detect solar axions one would thus orient a dipole magnet toward
the Sun and search for keV-range x-rays at the far end of the
conversion pipe. After a pioneering effort in
Brookhaven~\cite{Lazarus:1992ry}, a fully steerable instrument was
built in Tokyo \cite{Moriyama:1998kd, Inoue:2002qy, Inoue:2008zp}.
The largest helioscope yet is the CERN Axion Solar Telescope (CAST),
using a refurbished LHC test magnet ($L=9.26$~m, $B\sim 9.0$~T)
mounted to follow the Sun for about 1.5~h both at dawn and dusk
\cite{Zioutas:2004hi, Andriamonje:2007ew, Arik:2008mq, Aune:2011rx},
see fig.~\ref{fig:cast}. CAST began operation in 2003 and after two
years of data taking achieved a limit of
$g_{a\gamma}<0.88\times10^{-10}~{\rm GeV}^{-1}$ at 95\% CL for
$m_a\alt0.02$~eV. For these parameters, the conversion probability
is $P_{a\to\gamma}\sim1.3\times10^{-17}$. The limit on $g_{a\gamma}$
is comparable to the globular cluster limit from the energy loss in
horizontal-branch stars (fig.~\ref{fig:castexclusion}). Of course,
it is only interesting for those axion models where they do not
interact with electrons (hadronic axion models) because otherwise
the white-dwarf limit is more restrictive. For axion-like particles
with a two-photon vertex and small masses, CAST provides the most
restrictive limit on $g_{a\gamma}$.

\begin{figure}
\centering
\includegraphics[height=0.32\textwidth]{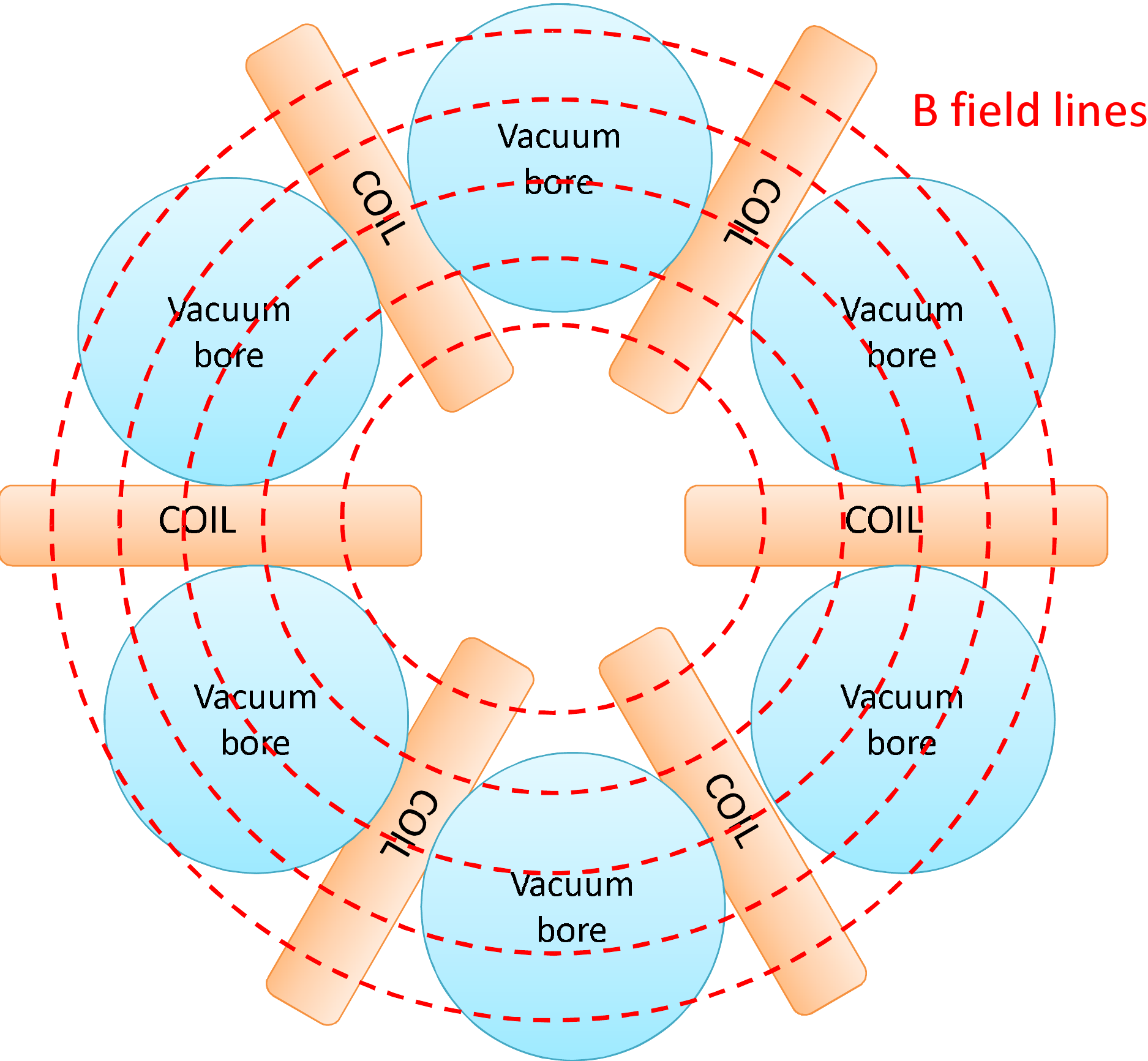}\hfill
\includegraphics[height=0.32\textwidth]{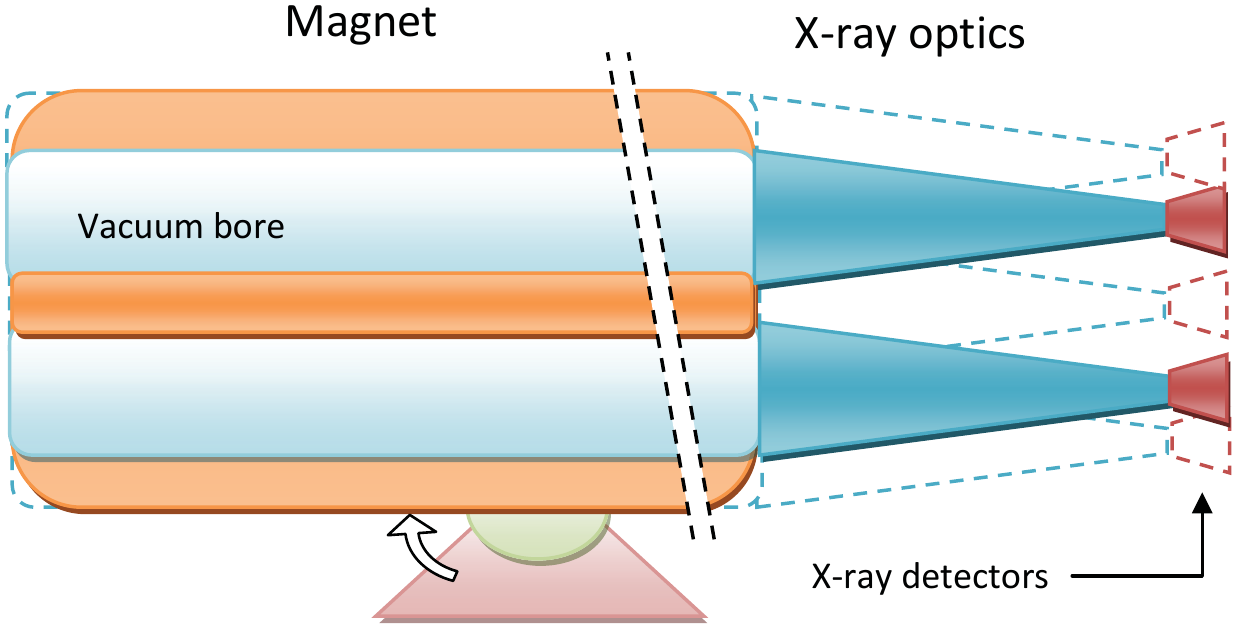}
\caption{Possible design of a next-generation axion helioscope
(NGAH)~\cite{Irastorza:2011gs}. Each vacuum bore could have a cross section
of 1~m$^2$.\label{fig:ngah}}
\end{figure}

For $m_a\agt0.02$~eV the inverse momentum transfer becomes of order
$L$ and the oscillating term  $\sin^2(qL/2)/(qL/2)^2$, which is 1
for small $m_a$, reduces the maximum transition probability. In
other words, the axion-photon oscillation length becomes smaller
than $L$, the conversion probability saturates and the CAST limits
on $g_{a\gamma}$ degrade with increasing mass. To extend the search
to larger masses one can fill the conversion pipe with helium as
buffer gas to provide the photons with a refractive mass
$\omega_{\rm pl}$. For an axion masses around $m_a\sim\omega_{\rm
pl}$ one can thus restore the full conversion
efficiency~\cite{LBL-25908}. This effect is rather comparable to the
matter effect in neutrino flavor oscillations.  Varying the gas
pressure allows one to step through many search masses and extend
the sensitivity to larger masses. This method was applied both in
the Tokyo axion helioscope and CAST using $^4$He as buffer gas,
extending the limits as shown in fig.~\ref{fig:castexclusion}. For
CAST, the maximum possible $^4$He pressure, the vapor pressure at
the liquid helium temperature of the superconducting magnet,
corresponds to $\omega_{\rm pl}\sim0.4$~eV. To reach yet larger
masses, CAST used $^3$He as buffer gas; first results are shown in
fig.~\ref{fig:castexclusion}. For the first time, the mass-coupling
relation for KSVZ axions was crossed, the prototype hadronic axion
model. Meanwhile, a search mass of 1.17~eV has been reached,
essentially the largest achievable with this method because for
larger gas densities absorption is becoming a serious problem. In
any event, the CAST constraints now connect seamlessly to
cosmological hot dark matter bounds, $m_a\alt0.7$~eV, that apply
because axions with the relevant parameters would have been
thermally produced in the early universe~\cite{arXiv:1004.0695}. To
cover more realistic model space one needs to push towards smaller
$g_{a\gamma}$ values. This may be achieved with a next generation
axion helioscope (NGAH) \cite{Irastorza:2011gs} with comparable $L$
and $B$, but much larger magnetic-field cross section
(fig.~\ref{fig:ngah}).

We should finally make sure that using the Sun as an axion source in
this way is self consistent. Axion emission represents a new
energy-loss channel for the Sun and would require increased fuel
consumption and thus an increased central temperature~$T_{\rm c}$.
This effect, in turn, would show up as increased neutrino fluxes,
notably an increased $^8$B flux that varies approximately as $T_{\rm
c}^{18}$. Based on numerical solar models with axion losses by the
Primakoff process~\cite{Schlattl:1998fz} one finds that the $^8$B
neutrino flux increases with axion luminosity $L_a$ relative to the
unperturbed flux as~\cite{Gondolo:2008dd}
\begin{equation}
\Phi_{\rm B8}^a=\Phi_{\rm B8}^0\,\left(\frac{L_\odot+L_a}{L_\odot}\right)^{4.6}\,.
\end{equation}
After accounting for neutrino flavor oscillations, the measured
$\Phi_{\rm B8}$ agrees well with standard solar model predictions
within errors, although the dominant uncertainty of the calculated
fluxes evidently comes from the assumed element abundances and
concomitant opacity. It appears reasonably conservative to assume
the true neutrino flux does not exceed the prediction by more than
50\% so that
\begin{equation}
L_a<0.1\,L_\odot\,.
\end{equation}
This limit implies the conservative bound
\begin{equation}
g_{a\gamma}<7\times10^{-10}~{\rm GeV}^{-1}\,,
\end{equation}
shown as a horizontal line ``Sun'' in fig.~\ref{fig:castexclusion}.
The Tokyo limits are just barely self-consistent whereas CAST probes
to much lower $g_{a\gamma}$ values than are already excluded by the
measured solar neutrino flux.

\section{Supernova neutrinos}                           \label{sec:SN}

\subsection{Classification of supernovae}

\begin{figure}
\centering
\includegraphics[width=0.48\textwidth]{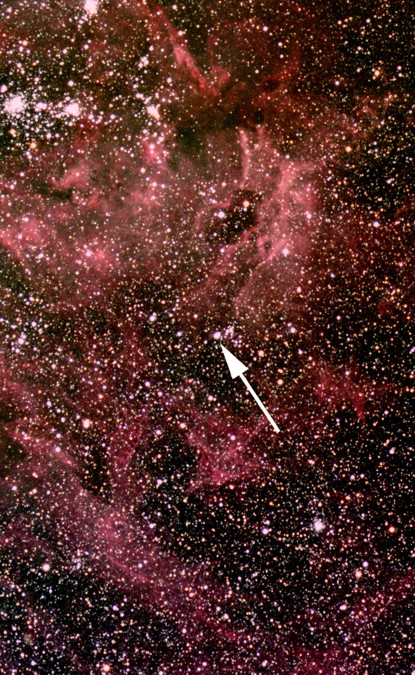}\hfill
\includegraphics[width=0.48\textwidth]{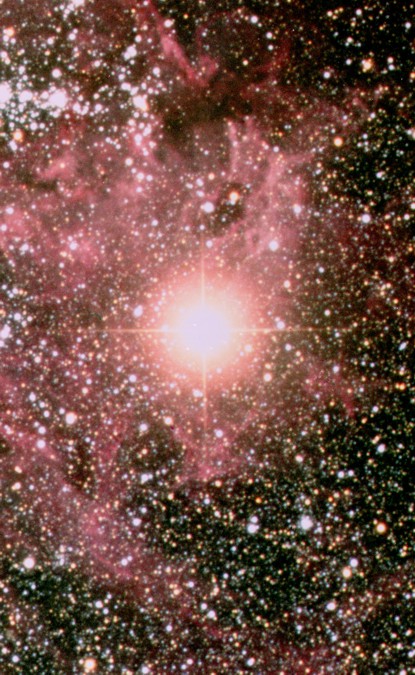}
\caption{The blue supergiant star Sanduleak $-69\,202$ in the Large Magellanic Cloud, before
and after it exploded on 23 February 1987 (SN~1987A). This was the closest observed
SN since Kepler's SN of 1603 and was the first example of a SN where the progenitor star could
be identified. \copyright~Australian Astronomical Observatory.\label{fig:sanduleak}}
\end{figure}

Supernova (SN) explosions are the most energetic astrophysical
events since the big bang~\cite{Burrows:1990ts, Burrows:2000mk,
Janka:2006fh}. A star suddenly brightens and at the peak of its
light curve shines as bright as the host galaxy
(fig.~\ref{fig:sanduleak}). Baade and Zwicky identified SNe as a new
class of objects in the late 1920s and in 1934 speculated that a SN
may be the end state of stellar evolution and that the energy source
was provided by the gravitational binding energy from the collapse
to a neutron star~\cite{Baade:1934}. They also speculated that SNe
were the energy source for cosmic rays. A few years later, Gamow and
Schoenberg (1941) developed first ideas about the connection between
core collapse and neutrinos~\cite{Gamow1941}, fifteen years before
neutrinos were experimentally detected.

\begin{figure}
\centering
\includegraphics[height=0.48\textwidth]{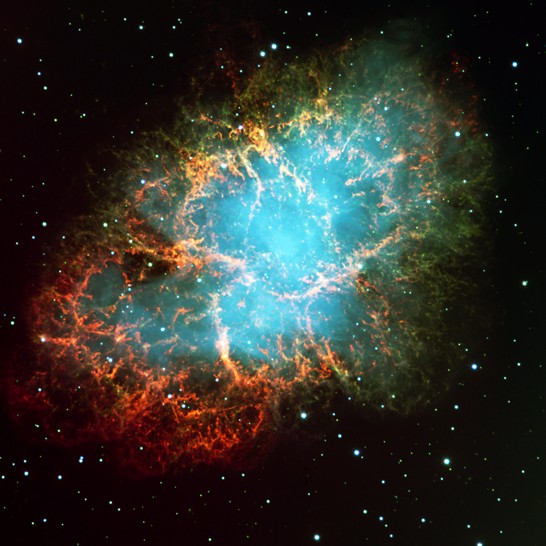}\hfill
\includegraphics[height=0.48\textwidth]{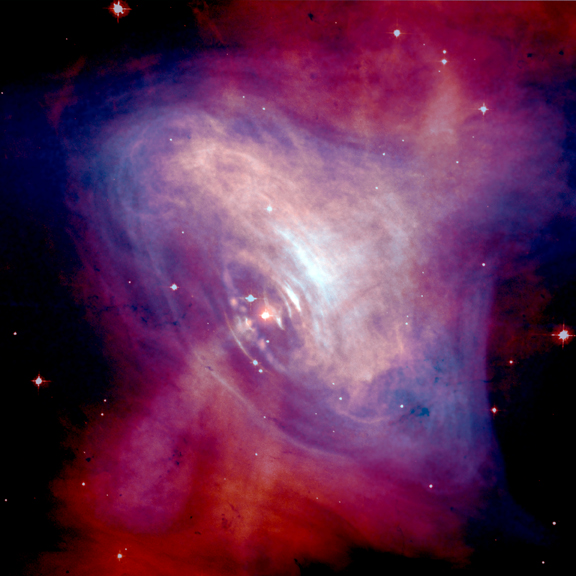}
\caption{Remnant of the historical supernova of 1054. Left: Crab Nebula,
the dispersed ejecta from the explosion.
Credit: ESO (see also http://apod.nasa.gov/apod/ap991122.html).
Right: Crab Pulsar in the center of the Crab Nebula,
the compact neutron star remaining from the
collapse, as a superposition of an HST optical image (red) and a
false-color Chandra x-ray image (blue).
Credit: J.~Hester (ASU) et al., CXC, HST, NASA (see also
http://apod.nasa.gov/apod/ap050326.html). \label{fig:crab}}
\end{figure}

Today we believe indeed that a star with mass exceeding
6--$8\,M_\odot$, after going through all nuclear burning stages
(fig.~\ref{fig:onionshell}), ends its life when its degenerate core
has reached the Chandrasekhar limit and collapses, in the process
ejecting the stellar mantle and envelope. When the core, a mass of
about $1.5\,M_\odot$, collapses to a compact star with nuclear
density and a radius of around 12~km, almost the complete
gravitational binding energy of about $3\times10^{53}~{\rm erg}$ is
released in neutrinos of all flavors in a burst lasting a few
seconds. For that period, the neutrino luminosity of a core-collapse
SN is comparable to the combined photon luminosity of all stars in
the visible universe. About one core collapse takes place per second
in the visible universe, so on average stars liberate as much energy
in neutrinos (from core collapse) as they release in photons (from
nuclear binding energy). The diffuse SN neutrino background (DSNB)
in the universe from all past SNe thus provides an energy density
comparable to that of the extra-galactic background light. Detecting
the DSNB is the next milestone of low-energy neutrino astronomy.

What remains of a SN explosion is the dispersed ejected gas, as for
example the Crab Nebula (fig.~\ref{fig:crab}), the remnant of the
historical SN of 1054 that was reported in Chinese records. While
99\% of the liberated energy appears as neutrinos, about 1\% goes
into the kinetic energy of the explosion, and only about 0.01\% into
the optical SN outburst. The remaining neutron star usually appears
as a fast-spinning pulsar, the Crab Pulsar being a prime example
(fig.~\ref{fig:crab}). Many pulsars receive a ``kick'' at birth,
moving with velocities of up to 2000~km~s$^{-1}$ relative to the
ejecta, implying that they even can be shot out of their host
galaxy. Modern multi-dimensional SN simulations seem to be able to
explain pulsar kicks by the asymmetry of the hydrodynamical
explosion~\cite{Scheck:2006rw, Wongwathanarat:2010ip}.

\begin{figure}
\centering
\includegraphics[width=0.75\textwidth]{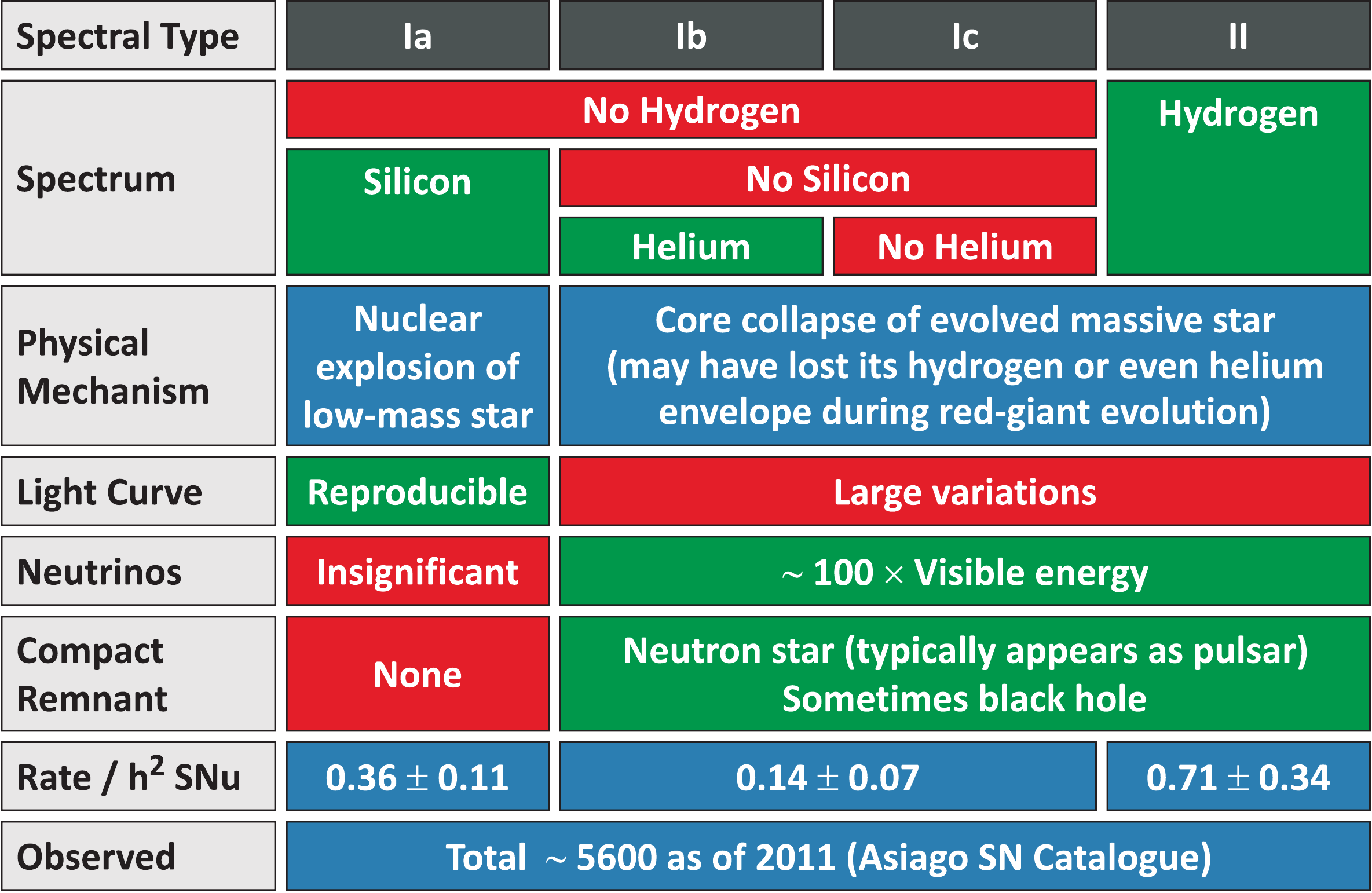}
\caption{Spectral classification of supernovas. The rate is measured in the
supernova unit, $1~{\rm SNu}=1$ SN per century per $10^{10}\,L_{\odot,B}$ (B-band
solar luminosity).
\label{fig:classification}}
\end{figure}

The astronomically observed SNe correspond to two entirely different
classes of physical phenomena~\cite{astro-ph/0012455}, i.e.\
core-collapse and thermonuclear SNe, the latter appearing as
spectral type Ia (fig.~\ref{fig:classification}). Astronomically, SN
types differ in their spectra and shape of the light curves. A
thermonuclear SN is thought to arise from a white dwarf that
accretes matter from a companion star in a binary system. When the
companion enters its giant phase, it inflates and matter can be
transferred to the white dwarf. Its mass increases until it reaches
its Chandrasekhar limit and collapses. However, the white dwarf
consists of carbon and oxygen and the collapse triggers explosive
nuclear burning, leading to complete disruption of the star. Nuclear
burning beyond helium formation releases around 1~MeV energy per
nucleon. A core-collapse SN, on the other hand, releases
gravitational binding energy of 100--200~MeV per nucleon, of which
99\% emerge as neutrinos. So both types of SN release around 1~MeV
visible energy per nucleon, explaining the superficial similarity.
Of course, a thermonuclear SN does not leave a pulsar behind. The
spectral type Ia corresponds to a thermonuclear SN, whereas the
spectral types Ib, Ic and II correspond to core collapse
(fig.~\ref{fig:classification}). The spectral types Ib and Ic are
core-collapse events where the progenitor star has shed its hydrogen
envelope before collapse.

Thermonuclear SNe are surprisingly reproducible. Their light curves
form a one-parameter class of functions that can be made uniform
with an empirical transformation, the Phillips relationship, that
connects the peak luminosity with the duration of the light curve.
In this way, SNe Ia can be used as cosmic standard candles and
because they can be seen across the entire visible universe, they
have been systematically used to study the expansion of the
universe~\cite{Goobar:2011iv}. The 1998 detection of accelerated
cosmic expansion by this method~\cite{Perlmutter:1998np,
Riess:1998cb} was awarded with the 2011 physics noble prize to Saul
Perlmutter, Brian Schmidt and Adam Riess. Core-collapse SNe, on the
other hand, show diverse light curves, depending on the mass and
envelope structure of the progenitor star, and typically are dimmer
than SNe~Ia. At the time of this writing, a total of around 5600 SNe
have been detected, primarily by the automated searches used for
cosmology. A table of all detected SNe is maintained by the Padova
Astronomical Observatory, the Asiago Supernova Catalogue
(http://graspa.oapd.inaf.it). Note that the first observed SN in a
given year, for example 2011, is denoted as SN 2011A, counting until
2011Z, and then continuing with small letters as SN 2011aa, 2011ab,
and so forth. The simple alphabet was exhausted for the first time
in 1988. For historical SNe, the type is clear when a pulsar or
neutron star is seen in the remnant, or by the historical record of
the peak luminosity and light curve. For Tycho's SN of 1572, a
spectrum could be taken in 2008 by virtue of a light echo,
confirming the suspected type Ia~\cite{Krause:2008bn}.

\subsection{Explosion mechanism}

While a thermonuclear SN explosion is intuitively easy to understand
as a ``fusion bomb,'' core collapse is primarily an implosion and
how to turn this into an explosion of the stellar mantle and
envelope is far from trivial and indeed not yet fully resolved. The
explosion could be a purely hydrodynamic event in form of the
``bounce and shock'' scenario, first proposed in 1961 by Colgate,
Grasberger and White~\cite{colgate:1961}. As the core collapses it
will finally reach nuclear density where the equation of state (EoS)
stiffens---essentially nucleon degeneracy provides a new source of
pressure. When the collapse suddenly halts (core bounce), a shock
wave forms at its edge and travels outward, expelling the overlying
layers of the star. Alternatively, Colgate and White (1966) appealed
to the large neutrino luminosity that carries away the gravitational
binding energy of the collapsed core~\cite{Colgate:1966ax}.
Neutrinos stream through the overlying star and, by occasional
interactions, transfer momentum and expel matter.

\begin{figure}
\centering
\includegraphics[width=1.0\textwidth]{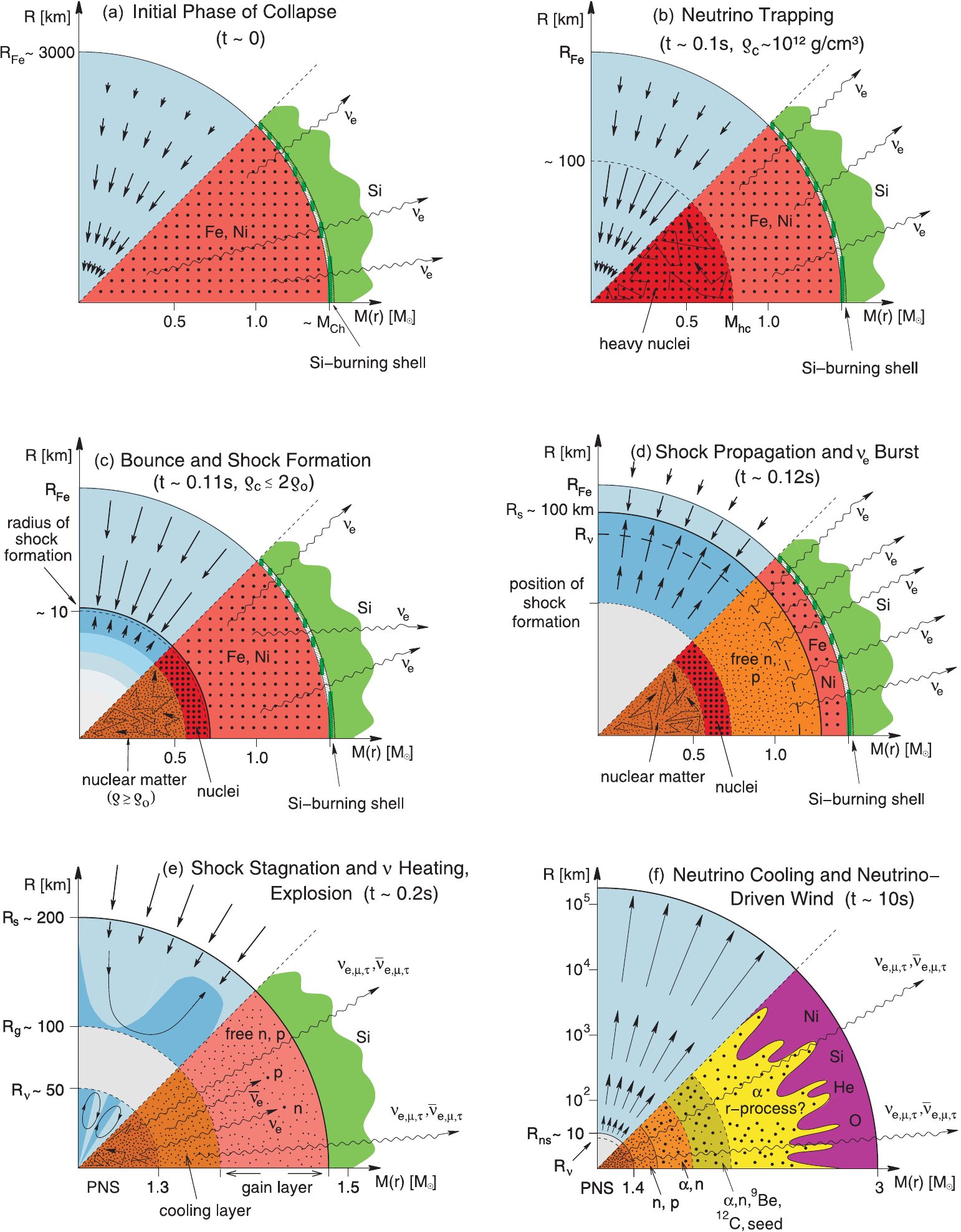}
\caption{Stages of core collapse and supernova explosion
as described in the text~\cite{Janka:2006fh}.\label{fig:collapsescheme}}
\end{figure}

The modern picture of the ``delayed explosion scenario,'' or
``neutrino mechanism,'' incorporates elements of both ideas. It was
first found by Wilson (1982) in a numerical
simulation~\cite{Wilson:1982} and spelled out in 1984 by Bethe and
Wilson~\cite{Bethe:1984ux}. In a series of cartoons
(fig.~\ref{fig:collapsescheme}), the events from collapse to
explosion are:
\begin{itemize}
\item[{\bf (a)}] {\bf Initial phase of collapse.} A
    Chandrasekhar-mass iron-nickel core of an evolved massive
    star becomes unstable. Electrons squeezed into high-energy
    states begin to dissociate the heavy nuclei, convert to
    neutrinos, escape, and in this way accelerate the loss of
    pressure. Photo dissociation of heavy nuclei is also
    important.

\item[{\bf (b)}] {\bf Neutrino trapping.} The core collapses,
    separated into a nearly homologous inner core that remains
    in hydrodynamic contact with itself, and the outer core with
    supersonic collapse. When densities of about $10^{12}~{\rm
    g}~{\rm cm}^{-3}$ are reached, neutrinos are trapped by
    coherently enhanced elastic scattering on large nuclei.

\item[{\bf (c)}] {\bf Bounce and shock formation.} The inner
    core reaches nuclear density of about $3\times10^{14}~{\rm
    g}~{\rm cm}^{-3}$, the EoS stiffens, the collapse halts, and
    the supersonic infall rams into a ``solid wall'' and gets
    reflected, forming a shock wave. Across the outward moving
    shock wave, the velocity field jumps discontinuously from
    supersonic inward to outward motion. The density also jumps
    discontinuously across the shock wave.

\item[{\bf (d)}] {\bf Shock propagation and \boldmath$\nu_e$
    burst.} The shock propagates outward and eventually reaches
    the edge of the iron core. The dissociation of this layer
    allows for electron capture, $e^-+p\to n+\nu_e$, producing
    the ``prompt $\nu_e$ burst'' or ``prompt deleptonization
    burst.'' Only the outer $\sim0.1\,M_\odot$ of the former
    iron core deleptonizes in this way, deeper layers
    deleptonize slowly on the diffusion time scale of seconds.

\item[{\bf (e)}] {\bf Shock stagnation, neutrino heating,
    explosion.} The shock wave runs out of pressure and
    stagnates at a radius of 150--200~km. Matter keeps falling
    in (``accretion shock''), i.e.\ the shock wave surfs on the
    infalling material that deposits energy near the nascent
    neutron star and powers a strong neutrino luminosity that is
    dominated by $\nu_e\bar\nu_e$ pairs. Convection sets in.
    Neutrino streaming continues to heat the material behind the
    shock wave, building up renewed pressure. After several
    hundred ms the shock wave takes off, expelling the overlying
    material.

\item[{\bf (f)}] {\bf Neutrino cooling and neutrino-driven
    wind.} The neutron star settles to about 12~km radius and
    cools by diffusive neutrino emission over seconds. A
     wind of matter is blown off with chemical composition
    governed by neutrino processes. Nucleosynthesis takes place
    in this ``hot bubble'' region, conceivably including the
    r-process production of heavy neutron-rich elements.
\end{itemize}

Some of these events deserve additional comments, notably the effect
of neutrino trapping. In the final hot nuclear-density core,
neutrinos are trapped by elastic scattering on nucleons and in
addition by beta processes for the electron flavor, a typical mean
free path after collapse being of order meters. However, for the SN
dynamics, the early trapping at around $10^{12}~{\rm g}~{\rm
cm}^{-3}$ is crucial because the electron lepton number, initially
in the form of electrons, cannot escape during infall in the form of
$\nu_e$. Therefore, the collapsed core will  have essentially the
same number of electrons per baryon, $Y_e\sim 0.42$, that was
present in the pre-collapse nickel-iron core. In other words,
radiation and thus entropy (in the form of neutrinos) cannot escape
and the collapse is essentially isentropic with crucial impact on
the hydrodynamics.

This ``low-density'' trapping occurs because of coherent enhancement
of the elastic scattering cross section first pointed out in 1973 by
Daniel Freedman~\cite{Freedman:1973yd} immediately after the
discovery of neutral-current neutrino
interactions~\cite{Hasert:1973cr}. Whenever some particle or
radiation scatters on a collection of $N$ targets, and when the
momentum transfer in the collision is so small that the target is
not ``resolved'' (the inverse momentum transfer exceeds the
geometric size of the target), the targets will act as one coherent
scatterer. The scattering amplitudes then add up in phase, implying
that the scattering cross section is $N^2$ times the individual
cross section. Elastic low-energy neutrino-nucleus scattering by
$Z^0$ exchange sees $N$ neutrons and $Z$ protons with the effective
coupling constants given in table~\ref{tab:couplings}. The
axial-current interaction is essentially proportional to the overall
nuclear spin which is small because nucleon spins tend to pair off
and coherent scattering leads to a reduced overall axial-current
cross section. For the vector current, the ``weak charges'' add
coherently, but are very small for protons, $C_V\sim 0$, because
$\sin^2\Theta_{\rm W}=0.23\sim 1/4$. So essentially only the
neutrons contribute and the scattering cross section scales as $N^2$
for neutrino energies up to a few ten~MeV. The collapsing core of an
evolved star consists of iron-group elements with $N\sim 30$ so that
coherently enhanced cross sections will be
important~\cite{Freedman:1977xn}. Measuring coherent
neutrino-nucleus scattering in the laboratory remains an open task.

\begin{figure}
\centering
\includegraphics[width=0.60\textwidth]{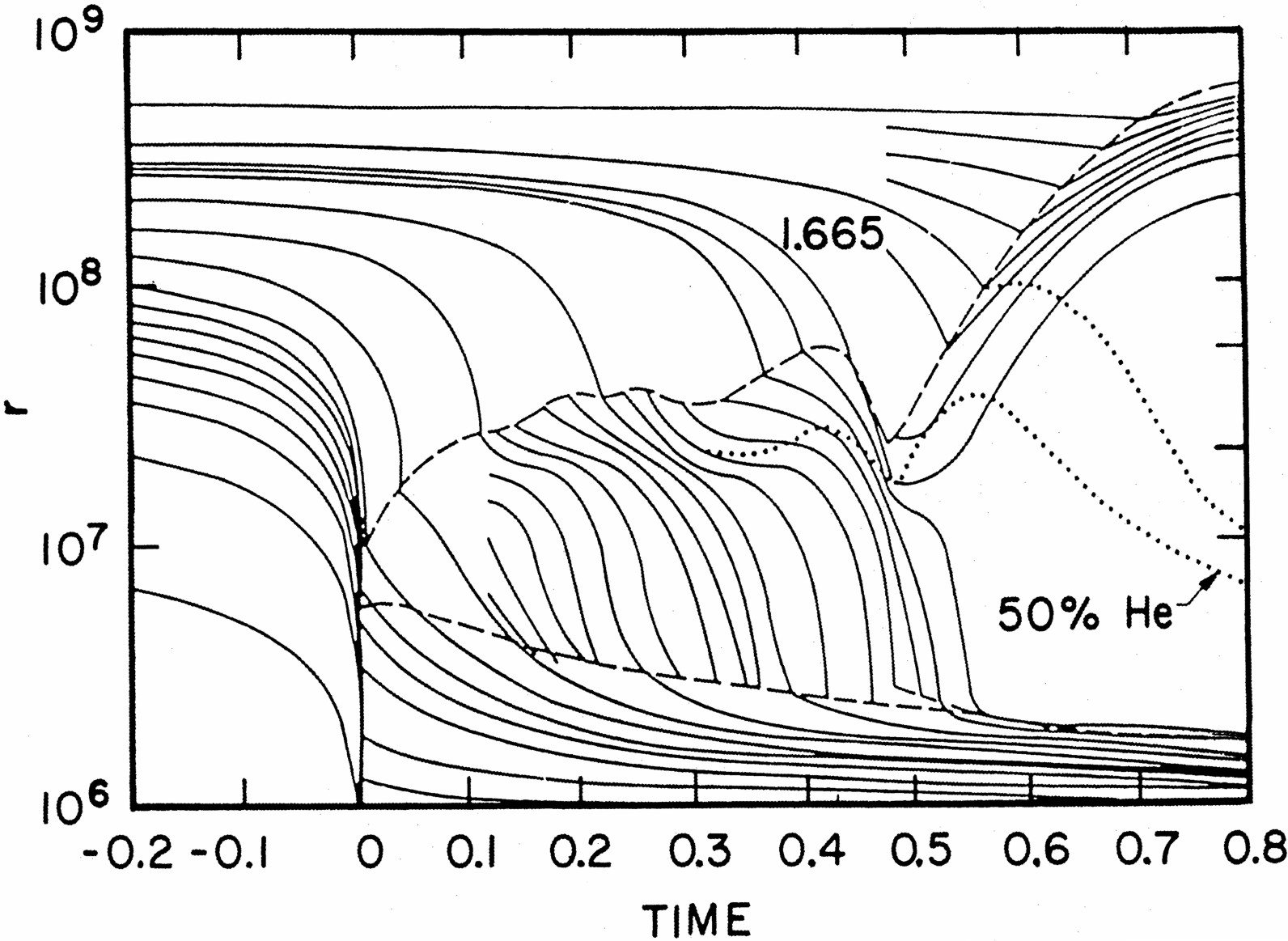}
\caption{Delayed explosion scenario in Wilson's numerical simulation
(1982) \cite{Wilson:1982} and explained by Bethe and
Wilson~\cite{Bethe:1984ux}. Shown are the trajectories of various mass points
(radius in cm, time in s). The lower dashed curve is the position of the neutrino
sphere, the upper one is the shock. At $t=0.48$~s, two neighboring trajectories
begin to diverge. The region between them is the matter-depleted
hot bubble region.\label{fig:wilson}}
\end{figure}

\begin{figure}
\centering
\includegraphics[height=0.42\textwidth]{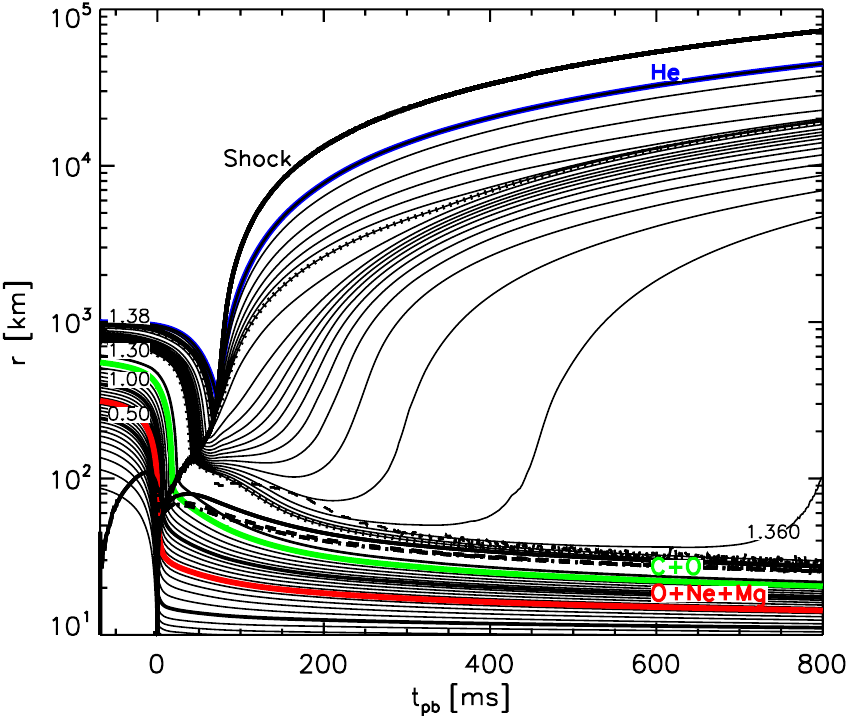}\hfill
\includegraphics[height=0.42\textwidth]{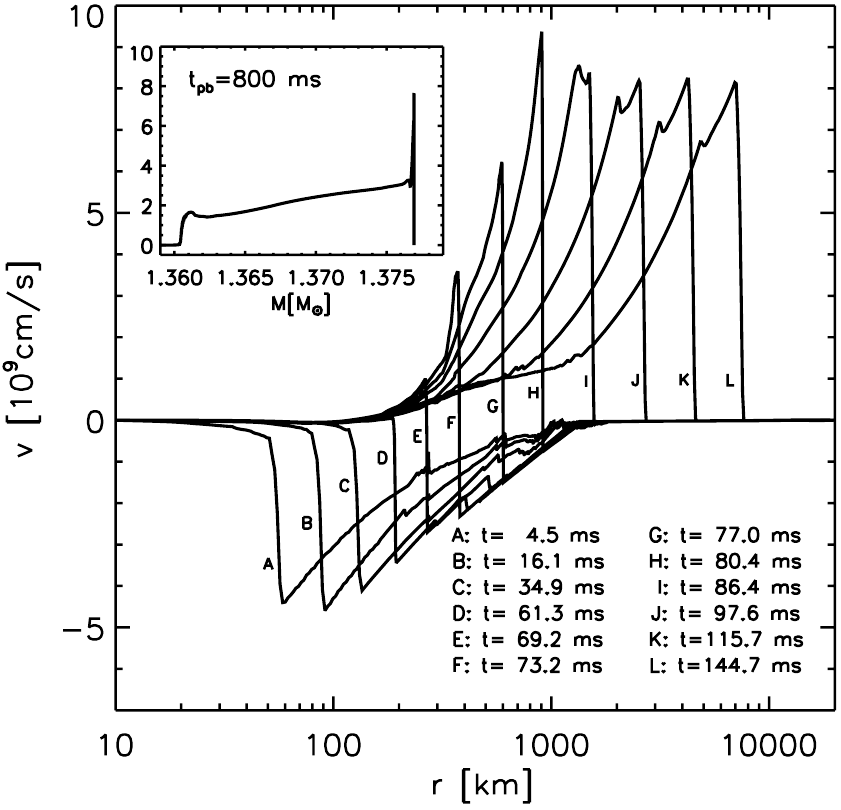}
\caption{Explosion in spherical symmetry of an
O-Ne-Mg-core SN, characteristic for progenitor masses 8--$10\,M_\odot$,
where the accretion phase is very short~\cite{Kitaura:2005bt}.
Left: Trajectories of various mass shells. Right: Velocity profiles
at different times.\label{fig:kitaura}}
\end{figure}

Concerning the bounce-and-shock delayed explosion mechanism, a
crucial point is that the edge of the inner homologous core is
inside the iron core, i.e.\ the shock wave dissociates iron on its
way out. Behind the shock wave, matter is composed of free protons,
neutrons, electrons and neutrinos. Dissociating $0.1\,M_\odot$ of
iron requires an energy of $1.7\times10^{51}~{\rm erg}$, comparable
to the explosion energy.\footnote{$10^{51}~{\rm erg}$ is sometimes
denoted 1~foe for ``(ten to) fifty one ergs'' or more lately as
1~Bethe.} This effect robs the shock wave of the energy to explode
the star, and without neutrino heating, it re-collapses and the end
state would be a black hole. Pressure can build up again by neutrino
energy deposition behind the shock wave that can lead to a delayed
explosion. This was first observed by Jim Wilson, a pioneer of
numerical SN modeling, in 1982 with the result shown in
fig.~\ref{fig:wilson}. On balance, the hot material above the SN
core loses energy by neutrino emission, whereas the colder material
behind the shock wave gains energy. The ``gain radius'' between the
SN core and the shock wave separates the two regimes.

However, modern simulations do not produce explosions in spherical
symmetry except for very low-mass progenitor stars
(fig.~\ref{fig:kitaura}). The Livermore simulations of the Wilson
group used simplified neutrino transport methods and the effect of
neutron-finger convection, no longer considered realistic, was used
to increase the early neutrino luminosity. Sometimes it has been
speculated that the explosion is aided by new channels of energy
transfer, for example by axion-like particles~\cite{Schramm:1981mk,
Berezhiani:1999qh}, neutrino flavor
oscillations~\cite{Fuller:1992,Suwa:2011ac,Dasgupta:2011jf}, or
sterile neutrinos~\cite{Hidaka:2007se,Fuller:2009zz}, but the
required particle parameters are either now excluded or not
necessarily well motivated.

\begin{figure}
\centering
\includegraphics[width=0.75\textwidth]{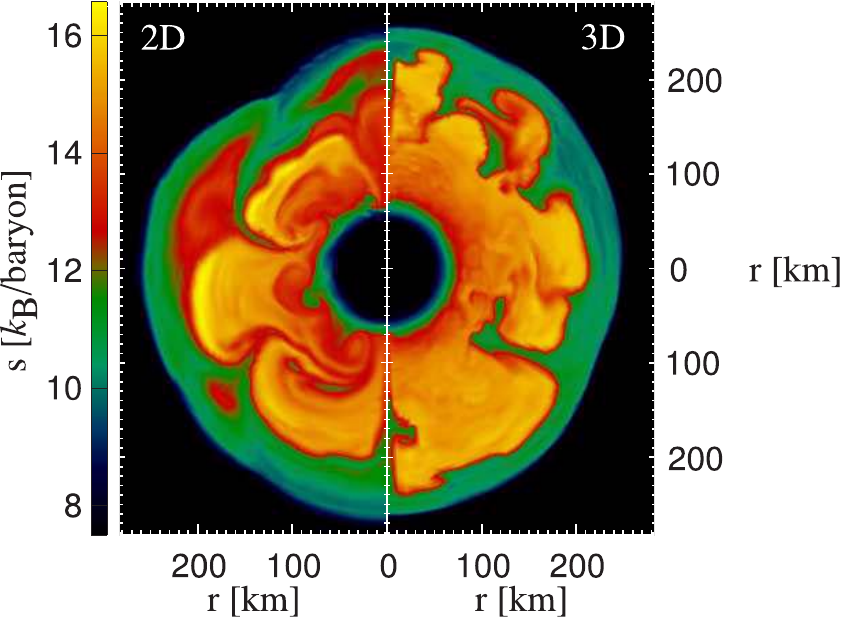}
\caption{Convection between proto neutron star and stagnating shock
wave~\cite{Hanke:2011jf}. Shown are entropy contours at 400~ms post bounce (pb)
of an $11.2\,M_\odot$ model from a 2D and 3D simulation which both
explode at about 550~ms pb.\label{fig:convection}}
\end{figure}

\begin{figure}
\centering
\includegraphics[width=0.6\textwidth]{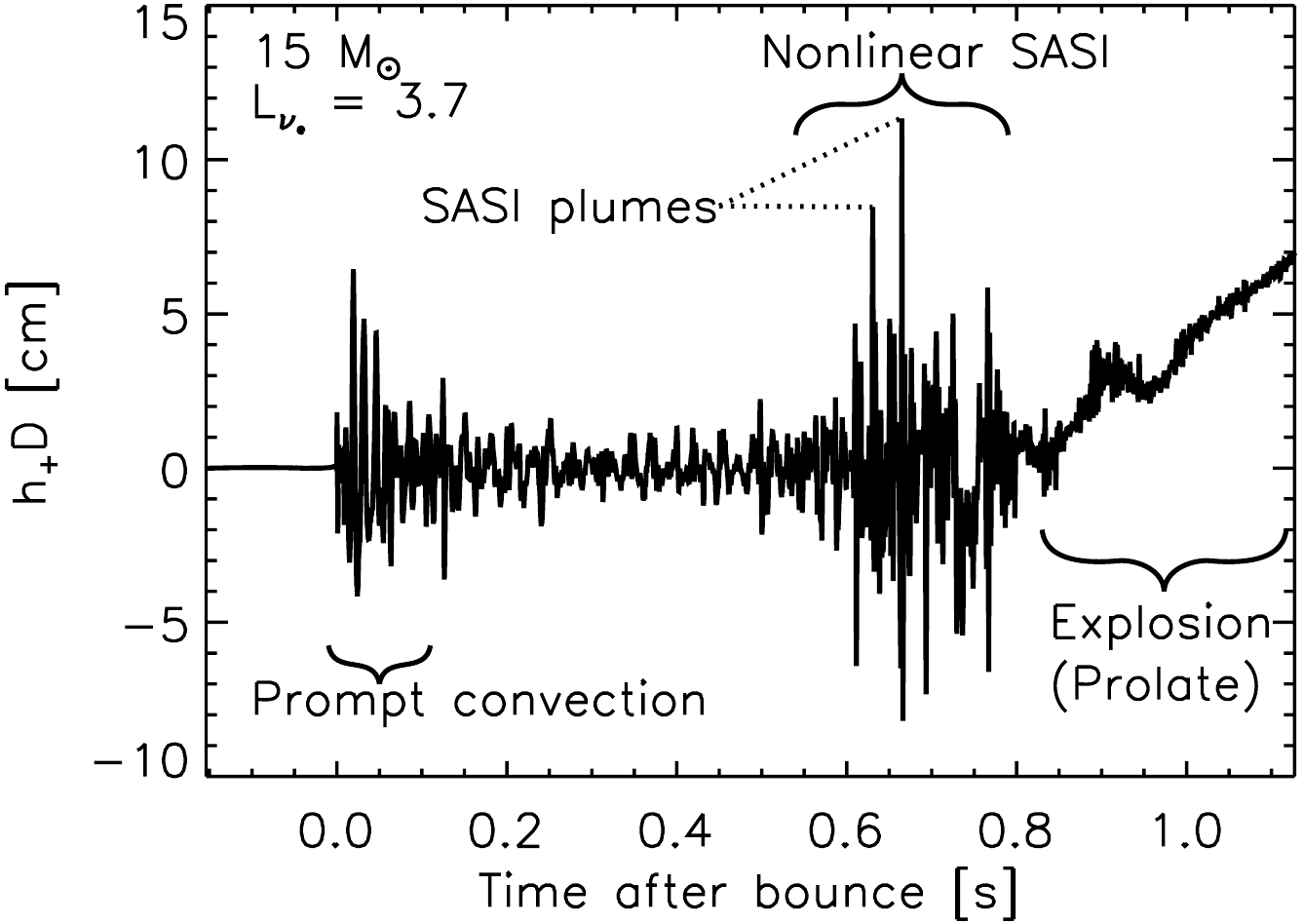}
\caption{Schematic gravitational wave signal (gravitational strain $h_+$ times
distance $D$) from a core-collapse
SN~\cite{Murphy:2009dx}.
Prompt convection, which results from a negative entropy
gradient left by the stalling shock, is the first distinctive feature
from approximately 0--50~ms post bounce (pb). For about 50--550~ms pb,
the signal is dominated by proto neutron star (PNS) and post shock convection.
Afterward and until the onset of explosion (800~ms),
strong nonlinear SASI motions dominate.
The most distinctive features are spikes that correlate with dense and narrow
down-flowing plumes striking the PNS surface ($\sim50$~km). The
aspherical (predominantly prolate) explosion manifests in a monotonic
rise in $h_+D$ that is similar to the ``memory signature''
of asymmetric neutrino emission.\label{fig:gw}}
\end{figure}

The most probable solution of the SN explosion problem is of more
mundane origin. The assumption of approximate spherical symmetry is
poorly satisfied because the region between SN core and standing
accretion shock is convectively unstable. Already the first 2D
numerical simulations (axial symmetry) and later 3D simulations
revealed the development of large-scale convective overturns
(fig.~\ref{fig:convection}). In addition, the standing accretion
shock instability (SASI) leads to spectacular dipolar oscillations
of the SN core against the ``cavity'' formed by the standing shock
wave~\cite{astro-ph/0210634, astro-ph/0509765, astro-ph/0606640}.
The strong deviation from spherical evolution leads to powerful
gravitational wave emission (fig.~\ref{fig:gw}) that can be observed
from the next nearby SN with the upcoming generation of
gravitational wave observatories~\cite{Marek:2008qi, Murphy:2009dx}.

Convection and SASI activity can help with shock reheating in
several ways. Hot material is dredged up from deeper layers to the
region behind the shock wave. Moreover, the material is exposed to
the neutrino flux for a longer time and absorbs more energy. 2D
simulations lead to successful explosions for some range of
progenitor masses~\cite{Marek:2007gr}. Self-consistent 3D
simulations do not yet exist because of the numerical challenge of
implementing neutrino transport without simplifying assumptions in
the most general case~\cite{Cardall:2011zz}. Parametric studies are
not yet conclusive whether going from 2D to 3D will further enhance
or perhaps even diminish the impact of non-sphericity on the final
explosion~\cite{Hanke:2011jf, Nordhaus:2010uk}. It appears unlikely
that fast rotation is crucial for the explosion because most
progenitor stars do not seem to rotate fast enough. Likewise,
magnetic fields would have to be exceedingly strong to have a major
impact on the explosion dynamics. Transferring energy to the shock
by acoustic waves~\cite{Burrows:2006uh}, generated by neutron-star
ringing, is probably too slow to trigger the explosion before the
neutrino mechanism does the job. The final verdict on the delayed
neutrino driven explosion mechanism will depend on careful numerical
3D modeling and observational input from gravitational wave and
neutrino observations from the next nearby SN.

\subsection{Characteristics of neutrino signal}

Observing a high-statistics neutrino signal from the next nearby SN
is a major goal of low-energy neutrino astronomy and interpreting
the SN~1987A signal is a crucial test for SN theory, so we first
discuss what to expect for different flavors. Usually one
distinguishes between three species $\nu_e$, $\bar\nu_e$ and
$\nu_x$, where the latter refers to any of $\nu_{\mu,\tau}$ or
$\bar\nu_{\mu,\tau}$. The dominant source of opacity is
$\nu_en\leftrightarrow p e^-$ and $\bar\nu_ep\leftrightarrow n e^+$
for the electron flavor and elastic neutral-current scattering
$\nu_x N\leftrightarrow N \nu_x$ for the others. The absence of
muons (mass 106~MeV) and \hbox{$\tau$-leptons} (mass 1777~MeV)
prevents charged-current reactions for the heavy-lepton neutrinos,
although some thermal muons may exist in the innermost core if $T$
becomes large enough. Note that $\nu N$ scattering differs somewhat
between $\nu_{\mu,\tau}$ and $\bar\nu_{\mu,\tau}$ due to weak
magnetism~\cite{Horowitz:2001xf}, but the small difference is often
ignored.

\begin{figure}
\centering
\includegraphics[width=1.0\textwidth]{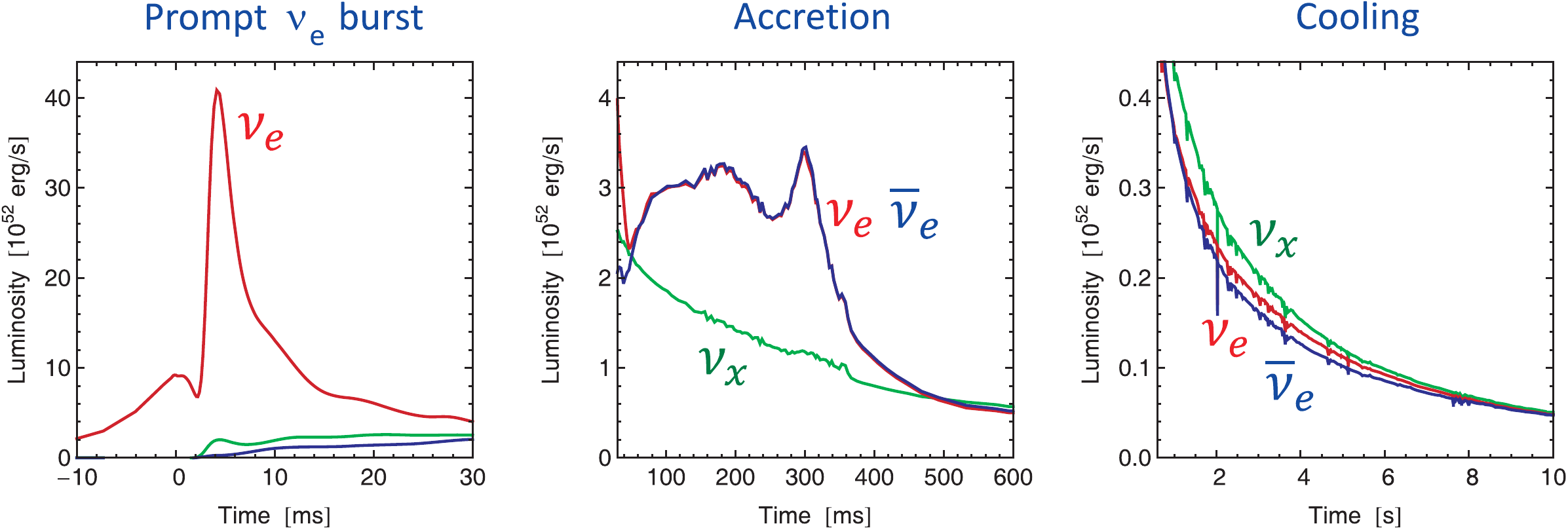}
\caption{Neutrino signal using data from a spherically symmetric
$10.8\,M_\odot$ simulation of the Basel group \cite{Fischer:2009af}.
The explosion was manually triggered.\label{fig:basel}}
\end{figure}

The detectable neutrino signal has three main phases shown in
fig.~\ref{fig:basel} from a numerical simulation of a
$10.8\,M_\odot$ spherically symmetric simulation of the Basel group.
The explosion was triggered manually by increasing the numerical
energy absorption rate in the gain region behind the shock wave. The
neutrino signal has three distinct phases, corresponding to three
phases of the collapse and explosion dynamics.
\begin{itemize}
\item[{\bf (1)}]{\bf Prompt \boldmath$\nu_e$ burst.} The shock
    wave breaks through the edge of the core, allowing for fast
    electron capture on free protons. A $\nu_e$ burst (5--10~ms)
    from deleptonization of the outer core layer emerges, the
    emission of $\bar\nu_e$ and $\nu_x$ is slowly beginning.
    This phase should not depend much on the progenitor mass.
\item[{\bf (2)}]{\bf Accretion phase.} The shock wave stagnates
    and matter falls in, releasing gravitational energy that
    powers neutrino emission. The $\nu_e$ and $\bar\nu_e$
    luminosities are similar, but the $\nu_e$ number flux is
    larger, carrying away the lepton number of the infalling
    material. The heavy-lepton flavors are emitted closer to the
    SN core, and their flux is smaller, but their energies
    larger. So we typically have a hierarchy $L_{\nu_e}\sim
    L_{\bar\nu_e}>L_{\nu_x}$ and $\langle
    E_{\nu_e}\rangle<\langle E_{\bar\nu_e}\rangle <\langle
    E_{\nu_x}\rangle$, with $\langle
    E_{\bar\nu_e}\rangle\sim12$--13~MeV. The duration of the
    accretion phase, typically a few hundred ms, and the
    detailed neutrino signal depend on the mass profile of the
    accreted matter.
\item[{\bf (3)}]{\bf Cooling phase.} The shock wave takes off,
    accretion stops, the SN core settles to become a neutron
    star, and cools by neutrino emission. The energy stored deep
    in its interior, largely in the form of $e$ and $\nu_e$
    degeneracy energy, emerges on a diffusion time scale of
    seconds. The luminosities of all species are similar
    $L_{\nu_e}\sim L_{\bar\nu_e}\sim L_{\nu_x}$ and decrease
    roughly exponentially with time. The $\nu_e$ number flux is
    larger because of deleptonization. The average energies
    follow the hierarchy $\langle E_{\nu_e}\rangle<\langle
    E_{\bar\nu_e}\rangle \sim\langle E_{\nu_x}\rangle$ and
    decrease with time. The characteristics of the cooling phase
    probably do not depend strongly on the progenitor mass.
\end{itemize}
Overall, a total energy of 2--$4\times10^{53}~{\rm erg}$ is emitted,
depending on the progenitor mass and equation of state, very roughly
equipartitioned among all flavors.

\begin{figure}
\centering
\includegraphics[width=1.0\textwidth]{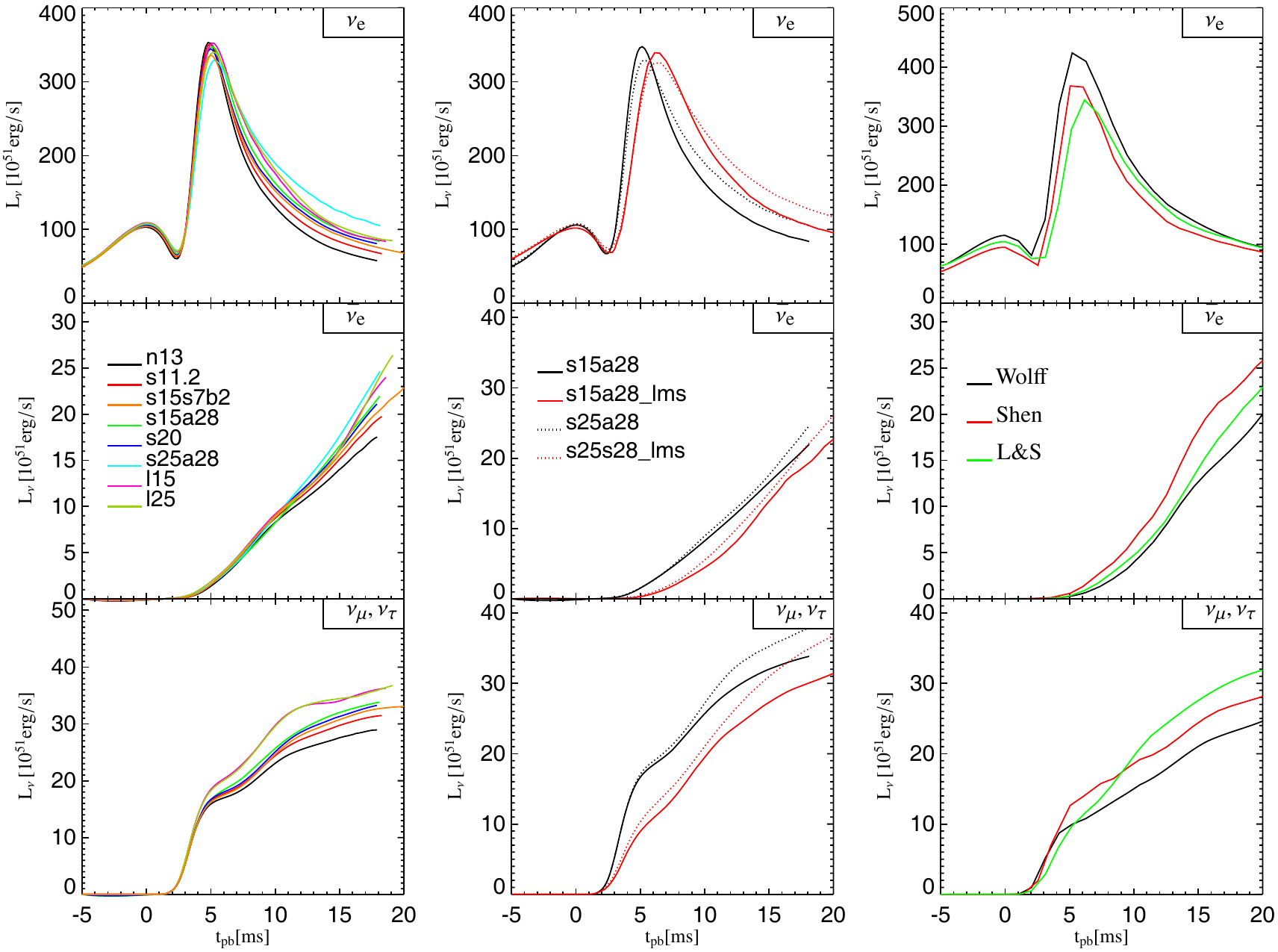}
\caption{Onset of neutrino luminosity and prompt $\nu_e$ burst for
a broad range of model assumptions~\cite{Kachelriess:2004ds}. Rows from
top to bottom for the indicated flavors.
Left column: Progenitor masses 11.2--$25\,M_\odot$ where
the mass is indicated by the number after the first letter of the shown model
name. Center column:
New treatment of electron captures by nuclei~\cite{nucl-th/0101039} (red lines)
compared to the traditional description (black lines) for a
$15\,M_\odot$ and $25\,M_\odot$ star. Right column: Three
different nuclear equations of state applied to a
$15\,M_\odot$ progenitor.\label{fig:promptburst}}
\end{figure}

The most generic of these phases is the prompt $\nu_e$ burst that
does not seem to depend much on the progenitor mass, assumed
equation of state (EoS), or details of neutrino opacities
(fig.~\ref{fig:promptburst}). When the $\nu_e$ burst is released,
the associated large chemical potential suppresses $\bar\nu_e$
emission, showing a slow start compared with the heavy-lepton
flavors. A possible observation of the prompt $\nu_e$ burst from the
next nearby SN requires a sensitive $\nu_e$ detector, in contrast to
the existing large-scale $\bar\nu_e$ experiments that are primarily
sensitive to the inverse beta reaction $\bar\nu_e+p\to n+e^+$.
Moreover, flavor oscillations will lead to large $\nu_e\to\nu_x$
flavor conversion, depending on the value of the neutrino mixing
angle $\theta_{13}$ and the atmospheric mass hierarchy.

It is only recently that SN neutrino signals have been simulated all
the way to the cooling phase with modern Boltzmann solvers of
neutrino transport~\cite{Fischer:2009af, Huedepohl:2009wh}.
Previously expectations were often gauged after the long-term
neutrino signal published by the Livermore
group~\cite{Totani:1997vj}. This pioneering work combined
relativistic hydrodynamics with multigroup three-flavor neutrino
diffusion in spherical symmetry (1D), simulating the entire
evolution self-consistently. The spectra were hard over a period of
at least 10~s with increasing hierarchy $\langle
E_{\nu_e}\rangle<\langle E_{\bar\nu_e}\rangle<\langle
E_{\nu_x}\rangle$. These models, however, included significant
numerical approximations and omitted neutrino reactions that were
later recognized to be important \cite{Keil:2002in}. Relativistic
calculations of proto neutron star (PNS) cooling with a flux-limited
equilibrium~\cite{Burrows:1986me, Keil:1995hw} or multigroup
diffusion treatment~\cite{Suzuki:1991} found monotonically
decreasing neutrino energies after no more than a short
($\sim$100~ms) period of increase. Pons~et~al.~\cite{Pons:1998mm}
studied PNS cooling for different EoS and masses, using flux limited
equilibrium transport with diffusion coefficients adapted to the
underlying EoS. They always found spectral hardening over 2--5~s
before turning over to cooling.

\begin{figure}[b]
\centering
\includegraphics[width=0.8\textwidth]{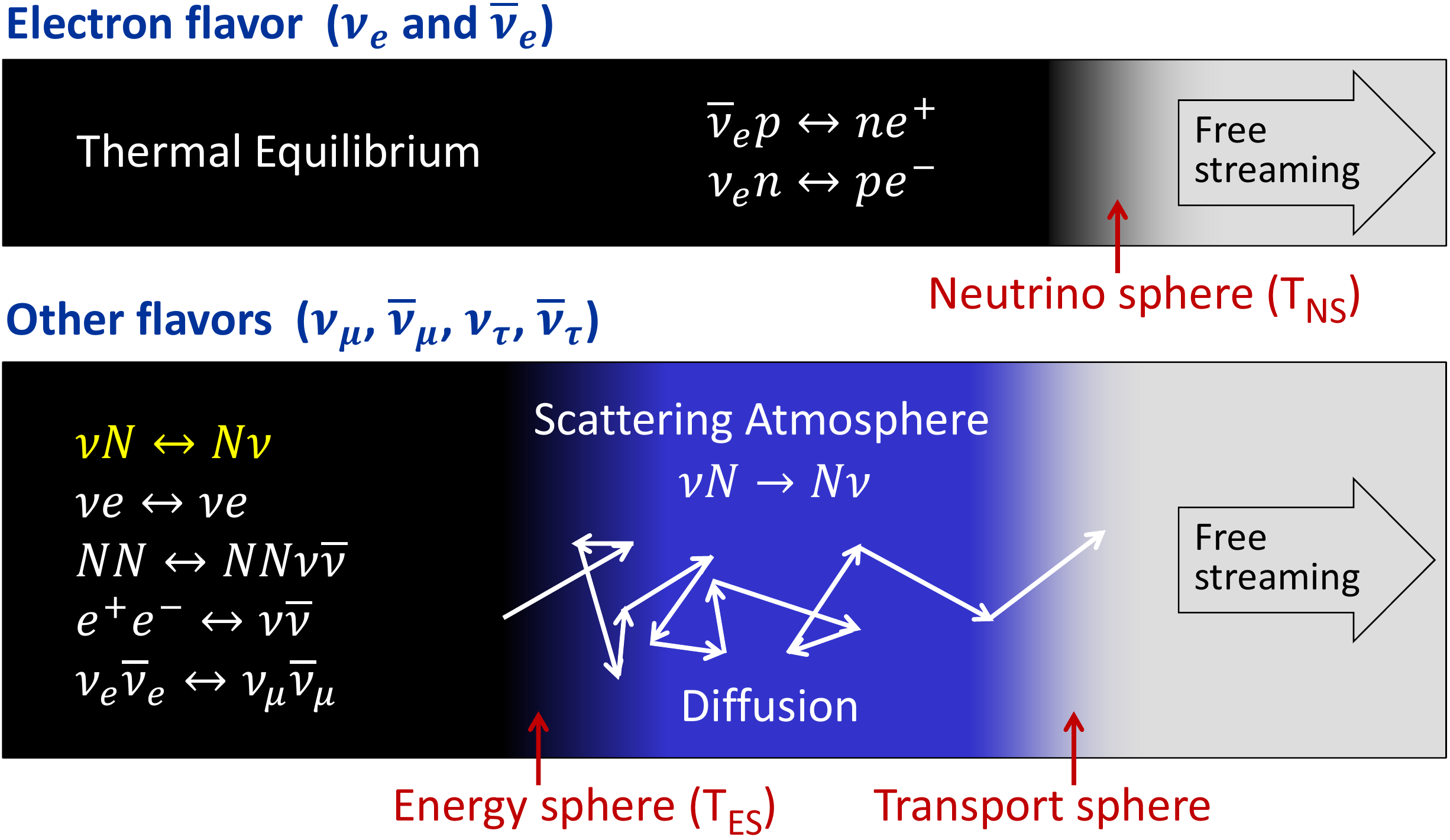}
\caption{Spectra formation for neutrinos of different flavors
as they stream from a SN core~\cite{Raffelt:2001kv}.\label{fig:transport}}
\end{figure}

However, the strong hierarchy of average energies, especially during
the cooling signal, that was often discussed in the context of
flavor oscillations, is certainly unrealistic. For the electron
flavor, neutrinos are trapped by charged-current reactions and begin
to stream freely at a radius where these reactions become
inefficient (fig.~\ref{fig:transport}). The energy-dependent
decoupling radius is called ``neutrino sphere'' and the spectra of
$\nu_e$ and $\bar\nu_e$ are determined by the temperature of the
matter in that region. The excess of neutrons over protons implies
that $\nu_e$ decouple at a larger radius and thus lower $T$,
explaining the traditional hierarchy $\langle
E_{\nu_e}\rangle<\langle E_{\bar\nu_e}\rangle$.

\begin{figure}[b]
\centering
\includegraphics[width=0.68\textwidth]{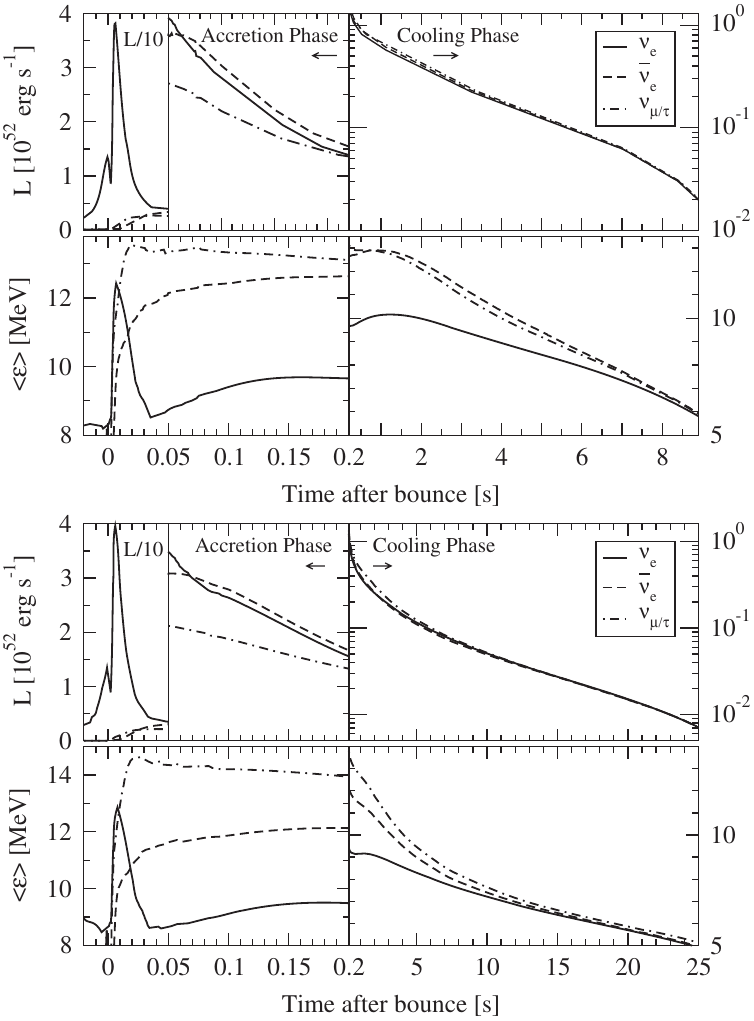}
\caption{Neutrino signal from an electron capture SN (progenitor
mass $8.8\,M_\odot$) that explodes in a
spherically symmetric simulation of the Garching
group~\cite{Huedepohl:2009wh}. Top: Full set of neutrino opacities,
including $NN$ correlations that reduce the opacities.
Bottom: Reduced set of opacities, no $NN$ correlations and no nucleon
recoil in $\nu N$ collisions. \label{fig:ecsn}}
\end{figure}

For the other species, decoupling is a two-step process, although
the main opacity always arises from neutrino-nucleon scattering
(fig.~\ref{fig:transport}). Deep inside, other processes are
important that produce $\nu_x\bar\nu_x$ pairs and exchange energy,
notably $\nu e$ and $\nu\nu$ scattering, nucleon-nucleon
bremsstrahlung, and $e^-e^+$ and $\nu_e\bar\nu_e$ annihilation. The
textbook wisdom that heavy-lepton neutrinos primarily emerge from
$e^-e^+$ annihilation is incorrect. Older simulations only used $\nu
N$ scattering and $e^-e^+$ annihilation, missing some of the crucial
processes. The energy-exchanging processes decouple at the ``energy
sphere,'' but the matter temperature in this region does not
directly fix the spectrum of the $\nu_x$ that stream from the
``transport sphere'' where $\nu N$ scattering has become
ineffective. The ``scattering atmosphere'' between these regions, by
the $E^2$ dependence of the $\nu N$ cross section, acts as a ``low
pass filter,'' skewing the emerging spectrum to lower energies and
leading to a flux spectrum with an effective $T$ as low as 60\% of
the matter $T$ at the energy sphere~\cite{Raffelt:2001kv}. Moreover,
nucleon recoils, often neglected in numerical simulations, further
soften the emerging spectrum. Even though the $\nu_x$ energy sphere
is at much larger $T$ than the $\nu_e$ and $\bar\nu_e$ neutrino
sphere, the emerging spectrum at late cooling times need not be
harder, and actually can be softer.

The impact of opacity details was studied by the Garching group for
a low-mass progenitor ($8.8\,M_\odot$) that collapses after
developing a degenerate O-Ne-Mg core and explodes in a spherically
symmetric simulation~\cite{Huedepohl:2009wh}. In fig.~\ref{fig:ecsn}
(upper panel) we show the neutrino signal for the full set of
opacities described in the Appendix of Ref.~\cite{Buras:2005rp} that
includes all processes indicated in fig.~\ref{fig:transport}. In
addition, nucleon-nucleon correlations in dense nuclear matter are
included that significantly reduce the neutrino scattering rate. In
the lower panel, these correlations and nucleon recoils are switched
off, corresponding roughly to the opacities used, for example, in
the Basel simulations. As a result, the cooling time increases (no
$NN$ correlations) and the emerging $\langle E_{\nu_x}\rangle$
increases (no $N$~recoils). So the $\langle E_{\nu_x}\rangle$ values
found in the long-term Basel simulations~\cite{Fischer:2009af}
probably should be reduced by 1--2~MeV to account for $N$ recoils.

\subsection{Supernova~1987A and its neutrino signal}

\begin{figure}[b]
\centering
\includegraphics[height=0.42\textwidth]{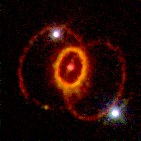}\hfill
\includegraphics[height=0.42\textwidth]{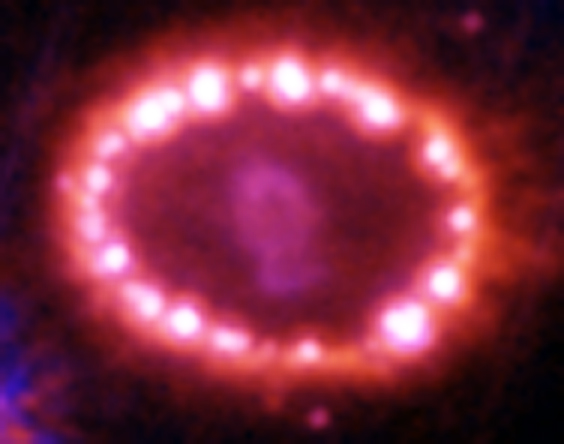}
\caption{Rings of SN~1987A illuminated by the explosion.
 Left: Hubble Space Telescope image, taken in Feb.\ 1994.
 Credit: C.~Burrows, ESA/STScI and NASA.
 Right: Image of inner ring, taken 28 Nov.\ 2003, showing
 bright spots caused by the supernova shock wave hitting the gas.
 The elongated ``nebula'' inside the ring is the supernova remnant.
 Credit: NASA, P.~Challis, R.~Kirshner (Harvard-Smithsonian Center
 for Astrophysics) and B.~Sugerman~(STScI).
  \label{fig:rings}}
\end{figure}

One of the most important events in the history of neutrino
astronomy was the observation of the neutrino signal of SN~1987A
that exploded on 23 February 1987 in the Large Magellanic Cloud, a
satellite galaxy of our Milky Way at a distance of about 50~kpc
(160,000 light years). The exploding star was the blue supergiant
Sanduleak $-69\,202$ (fig.~\ref{fig:sanduleak}), this being the
first SN that could be associated with an observed progenitor star.
SN~1987A was the closest visible SN in modern times. Previous
historical SNe in our galaxy of the second millennium occurred in
1006 (the brightest ever observed SN), 1054 (leading to the crab
nebula), Tycho's SN of 1572, Kepler's of 1604 and one around 1680
(Cas~A). While it is believed that a few SNe occur in our galaxy per
century, most are obscured by dust in the galactic plane, so one
expects only about 15\% of all galactic SNe to become directly
visible.

\begin{figure}
\centering
\includegraphics[width=0.8\textwidth]{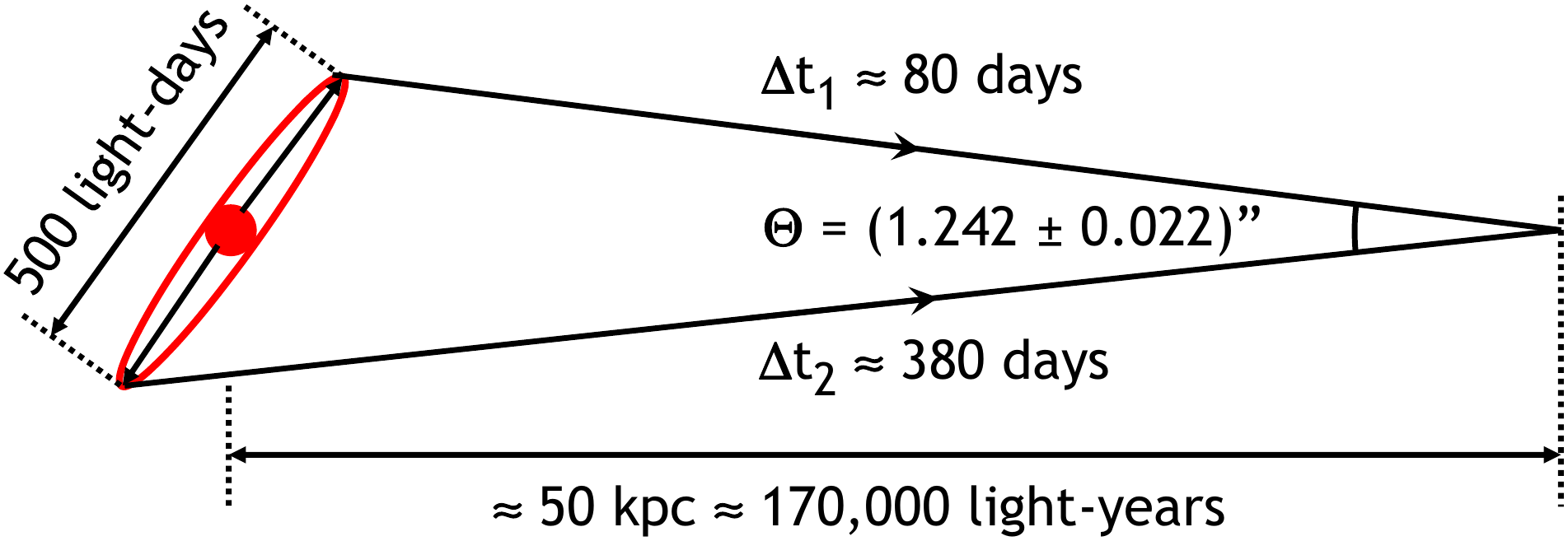}
\caption{SN~1987A distance determination by the arrival time difference
between the first light from the near and far side of the inner ring. The
implied distance is $51.4\pm1.2$~kpc according to Panagia~\cite{Panagia:1999}
or $47.2\pm0.9$~kpc according to
Gould and Uza~\cite{Gould:1997cc}.\label{fig:distance}}
\end{figure}

One of the most spectacular SN~1987A images (fig.~\ref{fig:rings})
was provided by the Hubble Space Telescope after its repair,
revealing a complicated ring system consisting of one inner ring and
two symmetrically located outer rings, all of which derive from
material ejected by the progenitor star and have nothing to do with
the SN itself. The rings were illuminated by the UV flash from the
SN~1987A explosion. The diameter of the inner ring is about
500~light days, so it turned on significantly after the SN
explosion, and the outer rings even later. The inner ring is tilted
relative to the line of sight, so the arrival time at Earth of light
from different parts of the ring allows one to determine the SN
distance in a purely geometric way (fig.~\ref{fig:distance}). Once
the shock wave reaches the inner ring years after the SN, it lights
up again with knot-like structures showing up (right panel in
fig.~\ref{fig:rings}). Within the inner ring one sees an elongated
nebula, representing the SN ejecta, providing direct evidence for
the lack of spherical symmetry of the explosion. SN~1987A has
provided a host of crucial astronomical information on the core
collapse phenomenon and nucleosynthesis in the SN environment.

Turning to the SN~1987A neutrino detection, in the late 1970s and
early 1980s, dedicated detectors were built to search for neutrinos
from galactic core-collapse events. The core-collapse rate was
thought to be fairly large, perhaps one every decade. The Baksan
Scintillator Telescope (BST) in the Caucasus Mountains (200 tons)
took up continuous operation on 30 June 1980 and has watched the
neutrino sky ever since. The smaller 90~ton Liquid Scintillator
Detector (LSD) took up operation in a side cavern of the Mont Blanc
tunnel in October 1984 and operated until the catastrophic tunnel
fire (24 March 1999). LSD was equipped with a real-time SN alert
system. Moreover, in the early 1980s the search for proton decay,
predicted in grand unified theories, led to the construction of the
Irvine-Michigan-Brookhaven (IMB) water Cherenkov detector (6800
tons) in the USA, reporting first results in 1982 and operating
until 1991. Likewise, Kamiokande (2140 tons of water) in Japan took
up operation in April 1983. In order to search for solar neutrinos
it was refurbished to lower the energy threshold. It began operation
as Kamkiokande-II in January 1987, only weeks before SN~1987A, and
took solar neutrino data until February 1995.

\begin{figure}
\centering
\includegraphics[width=0.6\textwidth]{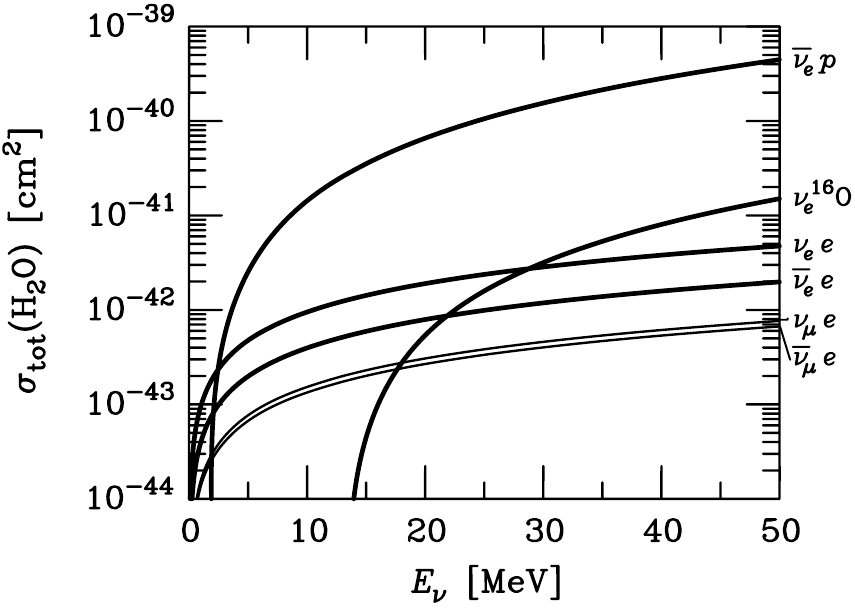}
\caption{\label{fig:sigma}
Total cross section per water molecule
for the measurement of neutrinos in a water Cherenkov detector. A
factor of~2 for protons and~10 for electrons is already included.
A SN neutrino signal is primarily detected by inverse beta decay
$\bar\nu_e+p\to n+e^+$.}
\end{figure}

\begin{figure}
\centering
\includegraphics[width=0.5\textwidth]{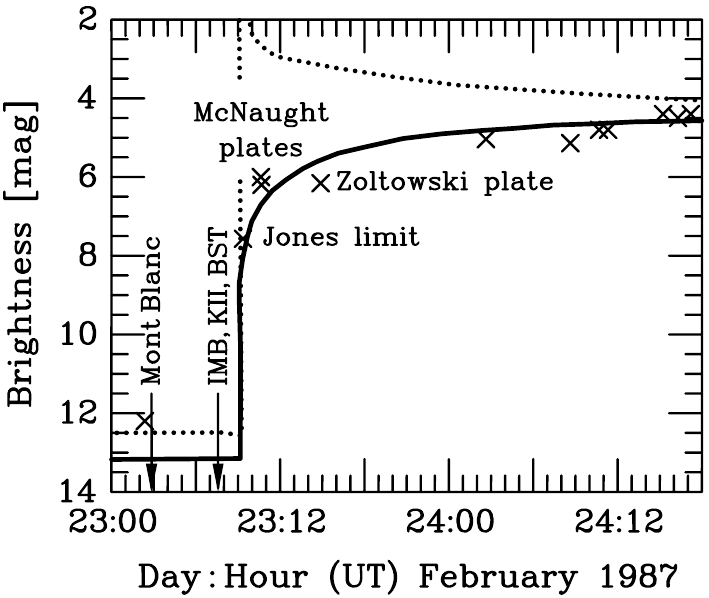}
\caption{\label{fig:earlylightcurve} Early optical observations of
SN~1987A according to the IAU
Circulars, notably No.~4316 of February 24, 1987. The times of the IMB,
Kamiokande~II (KII) and Baksan (BST) neutrino observations (23:07:35) and of
the Mont Blanc events (23:02:53) are also indicated. The solid line is the
expected visual brightness, the dotted line the bolometric brightness
according to model calculations. (Adapted, with permission, from Arnett
et~al.\ 1989 \cite{Arnett:1989}, Annual Review of Astronomy and Astrophysics, Volume 27,
\copyright~1989, by Annual Reviews Inc.)}
\end{figure}

\begin{figure}
\centering
\includegraphics[width=0.52\textwidth]{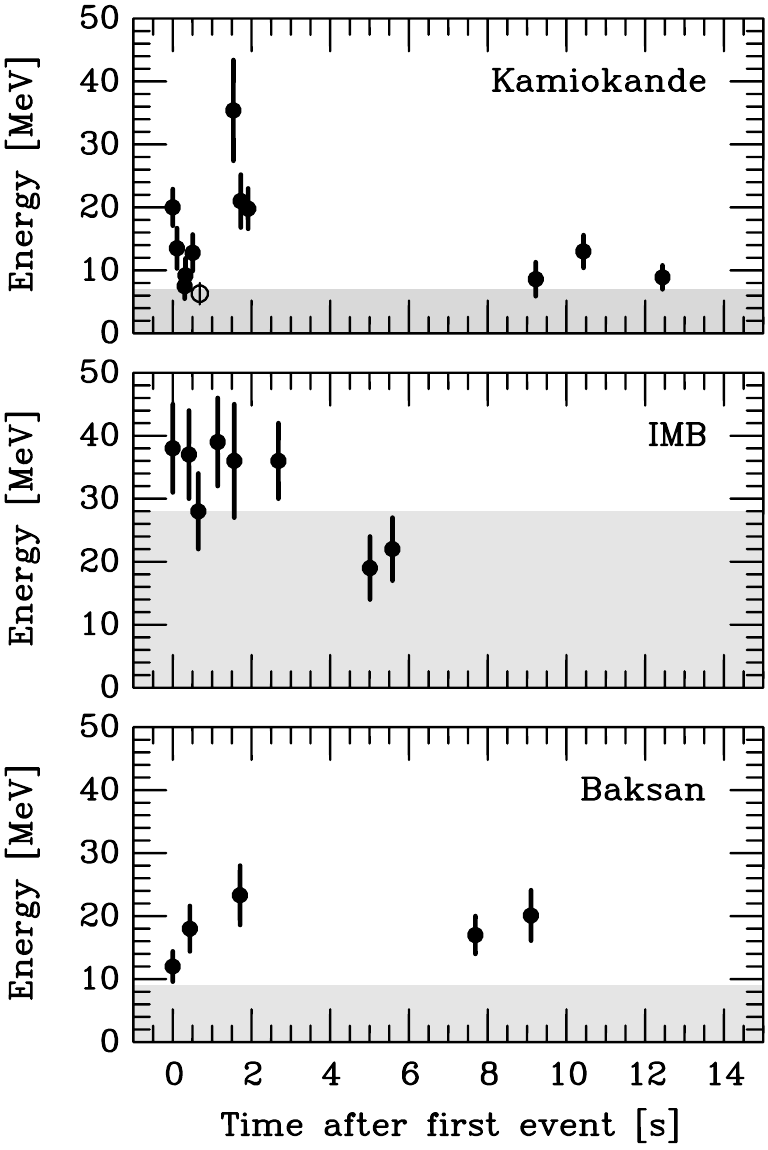}
\caption{\label{fig:sndat} SN~1987A neutrino observations at
Kamiokande~\cite{Hirata:1987hu, Hirata:1988ad},
IMB~\cite{Bionta:1987qt, Bratton:1988ww} and
Baksan \cite{Alekseev:1987ej, Alekseev:1988gp}. The energies refer
to the secondary positrons from $\bar\nu_e p\to n e^+$.
In the shaded area the trigger efficiency is less than 30\%. The clock
uncertainties are reported to be $\pm 1$~min in Kamiokande,
$\pm 50$~ms in IMB, and $+2/{-}54$~s in BST;
in each case the first event was shifted to $t=0$. In Kamiokande,
the event marked as an open circle is attributed to background.}
\end{figure}

These detectors see SN neutrinos primarily in the $\bar\nu_e$
channel from inverse beta decay (fig.~\ref{fig:sigma}). All of them
reported events associated with SN~1987A arriving a few hours before
the optical SN explosion as expected
(fig.~\ref{fig:earlylightcurve}). The
Kamiokande~\cite{Hirata:1987hu, Hirata:1988ad},
IMB~\cite{Bionta:1987qt, Bratton:1988ww} and
Baksan~\cite{Alekseev:1987ej, Alekseev:1988gp} observations
(fig.~\ref{fig:sndat}) are contemporaneous within clock
uncertainties. A 5-event cluster in the LSD
experiment~\cite{Dadykin:1987ek, Aglietta:1987it} was observed
4.72~h earlier and had no counterpart in the other detectors and
vice versa. Moreover, the LSD detector was too small to expect a
signal from as far away as the Large Magellanic Cloud. It can be
associated with SN~1987A only if one invokes very non-standard
double-bang scenarios of stellar collapse~\cite{Imshennik:2004iy}.
Still, no similar event cluster was ever observed again in LSD over
its 15 years of operation and its origin remains unresolved. A
lively account of the exciting and somewhat confusing history of the
SN~1987A neutrino detection was given by
M.~Koshiba~\cite{Koshiba:1992yb} and A.~Mann~\cite{Mann:1997}.

\begin{figure}
\centering
\includegraphics[width=0.8\textwidth]{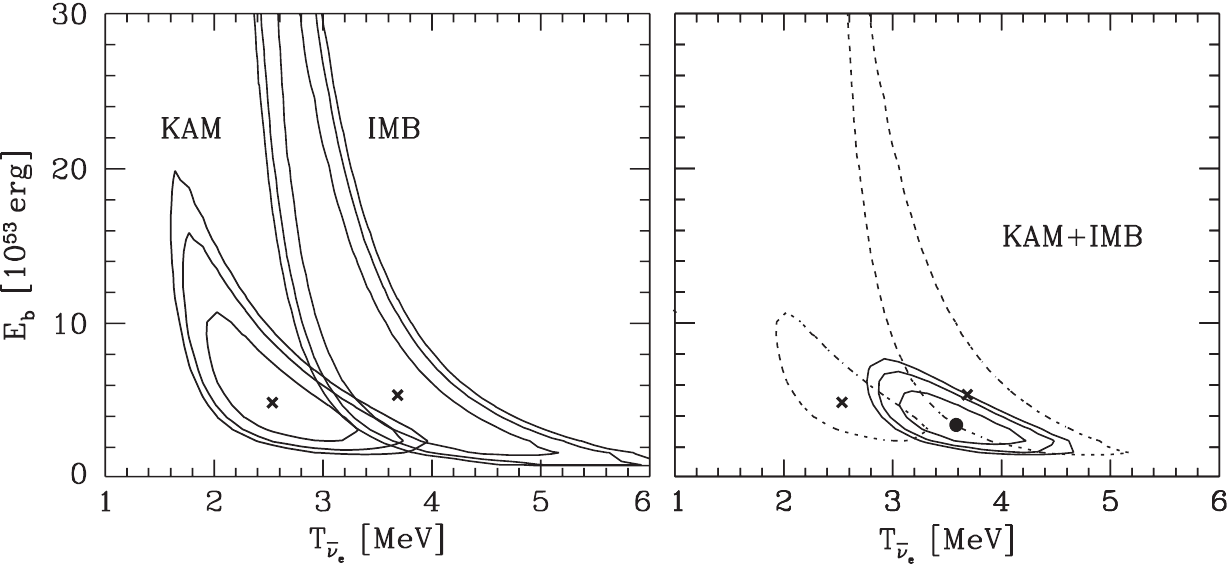}
\caption{\label{fig:SNfit}Confidence contours for the signal fit parameters
$E_{\rm b}$ (total released binding energy) and $\bar\nu_e$ spectral
temperature $T_{\bar\nu_e}$ based on the Kamiokande and IMB data and a
combined fit~\cite{Jegerlehner:1996kx}.
The confidence contours are for 68.3\%, 90\% and 95.4\%.}
\end{figure}

The event energies and signal duration roughly agree with
theoretical expectations. The IMB event energies are larger than
those in Kamiokande, in part because IMB had a higher energy
threshold---it had not been optimized for low-energy neutrino
detection. While the instantaneous neutrino spectra tend to be
``pinched,'' i.e.\ a bit narrower than a simple thermal spectrum,
the time-integrated flux probably can be reasonably well
approximated by the Maxwell-Boltzmann form $F_{\bar\nu_e}(E)\propto
E^2 e^{-E/T_{\bar\nu_e}}$. With this assumption one can derive the
fit parameters $T_{\bar\nu_e}$ and total emitted energy $E_{\rm b}$,
assuming 1/6 of the total energy arrived in the $\bar\nu_e$ channel.
Confidence contours for the fit parameters $E_{\rm b}$ and
$T_{\bar\nu_e}$ are shown in fig.~\ref{fig:SNfit}; other authors
have found similar results. The Kamiokande data alone imply a rather
soft spectrum, so there is tension between the data sets,  but they
are statistically compatible. Theoretically one expects $E_{\rm
b}=2$--$4\times 10^{53}$~erg and $T_{\bar\nu_e}=\frac{1}{3}\,\langle
E_{\bar\nu_e}\rangle\sim 4$~MeV if one ignores the possibility of
flavor oscillations. Flavor oscillations are unavoidable, so if the
$\langle E_{\bar\nu_e}\rangle$ predictions are roughly correct,
$\langle E_{\bar\nu_x}\rangle$ at the source cannot be much larger
than $\langle E_{\bar\nu_e}\rangle$, in contrast to the older
simulations, but in agreement with the more recent picture.

Much more sophisticated analyses have been
performed~\cite{Loredo:2001rx, Mirizzi:2005tg, Pagliaroli:2008ur},
but in the end the information contained in a sparse signal is
limited. The SN~1987A neutrino observations have provided a general
confirmation of the neutrino emission scenario with appropriate
energies over a diffusion time scale of seconds. A serious
quantitative test of the core collapse paradigm, however, requires a
high-statistics observation, ideally in several complementary
detectors, including gravitational wave observatories.

\subsection{Neutrinos from the next nearby supernova}

Galactic SNe are rare, perhaps a few per century
(table~\ref{tab:rate}), so measuring a high-statistics neutrino
signal from the next nearby SN is a once-in-a-lifetime opportunity
that should not be missed. Many currently operating detectors with a
primary physics focus on other topics have good SN sensitivity
(table~\ref{tab:SNdetectors}), providing for an optimistic outlook
that a high-statistics SN neutrino light curve will be measured
eventually~\cite{Scholberg:2010zz}. When it occurs, because neutrinos
arrive a few hours before the visual SN explosion, an early warning
can be issued. To this end, several detectors together form the
Supernova Early Warning System (SNEWS), issuing an alert if they
measure candidate signals in coincidence~\cite{Antonioli:2004zb,
Scholberg:2008fa, SNEWS}.

The workhorse process remains inverse beta decay, $\bar\nu_e+p\to
n+e^+$, either in water Cherenkov detectors consisting of ${\rm
H}_2{\rm O}$ as target, or in scintillator detectors, consisting
primarily of mineral oil with an approximate chemical composition
${\rm C}_n{\rm H}_{2n}$. Therefore, 1~kt of water contains about
$6.7\times10^{31}$ protons, whereas 1~kt of mineral oil about
$8.6\times10^{31}$ protons. The total inverse beta cross section is
at lowest order~\cite{Vogel:1999zy, Strumia:2003zx}
\begin{equation}
\sigma_{\bar\nu_ep}=9.42\times10^{-44}~{\rm cm}^2(E_\nu/{\rm MeV}-1.3)^2\,.
\end{equation}
To estimate the expected event rate we assume a fiducial SN at a
distance of 10~kpc that emits a total of $3\times10^{53}~{\rm erg}$
in the form of neutrinos, and 1/6 of that in the form of $\bar\nu_e$
with an average energy $E_{\rm av}=\langle
E_{\bar\nu_e}\rangle=12$~MeV as suggested by recent numerical work
and compatible with SN~1987A. These assumptions provide for a total
number of $2.6\times10^{57}$ emitted $\bar\nu_e$ and a fluence
(time-integrated flux) at Earth of
\begin{equation}
F_{\bar\nu_e}=2.18\times10^{11}~{\rm cm}^{-2}\,\frac{L_{\bar\nu_e}}{5\times10^{52}~{\rm erg}}
\,\frac{12~{\rm MeV}}{E_{\rm av}}\,
\left(\frac{10~{\rm kpc}}{D}\right)^2\,.
\end{equation}
We assume that the time-integrated spectrum follows a
Maxwell-Boltzmann distribution
\begin{equation}
f(E_\nu)=\frac{27}{2}\,\frac{E_\nu^2}{E_{\rm av}^3}\, e^{-3E_\nu/E_{\rm av}}
\end{equation}
that could also be written in terms of the spectral temperature
$T=E_{\rm av}/3$. We then expect 223 produced positrons per kiloton
water, the exact event rate depending on the detector threshold and
efficiency, and about 287 positrons per kiloton mineral oil.

\begin{table}
 \caption{\label{tab:SNdetectors}Existing and near-future SN neutrino
 detectors and event rates for a SN at 10~kpc, emission
 of $5\times10^{52}~{\rm erg}$ in $\bar\nu_e$, average energy 12~MeV,
 and thermal energy distribution. For HALO and ICARUS, the event
 rates depend on assumptions about the other species.
 For references and details see Ref.~\cite{Scholberg:2010zz}~.}
 \begin{tabular}{llllll}
 \hline
 Detector&Type&Location&Mass [kt]&Events&Status\\
 \hline
 IceCube&Ice Cherenkov&South Pole&0.6/OM&$10^6$&Running\\
 Super-K IV&Water&Japan&32&7000&Running\\
 LVD&Scintillator&Italy&1&300&Running\\
 KamLAND&Scintillator&Japan&1&300&Running\\
 SNO+&Scintillator&Canada&1&300&Commissioning 2013\\
 MiniBOONE&Scintillator&USA&0.7&200&Running\\
 Borexino&Scintillator&Italy&0.3&80&Running\\
 BST&Scintillator&Russia&0.2&50&Running\\
 HALO&Lead&Canada&0.079&tens&Almost ready\\
 ICARUS&Liquid argon&Italy&0.6&200&Running\\
 \hline
 \end{tabular}
\end{table}

\begin{figure}
\centering
\includegraphics[width=0.8\textwidth]{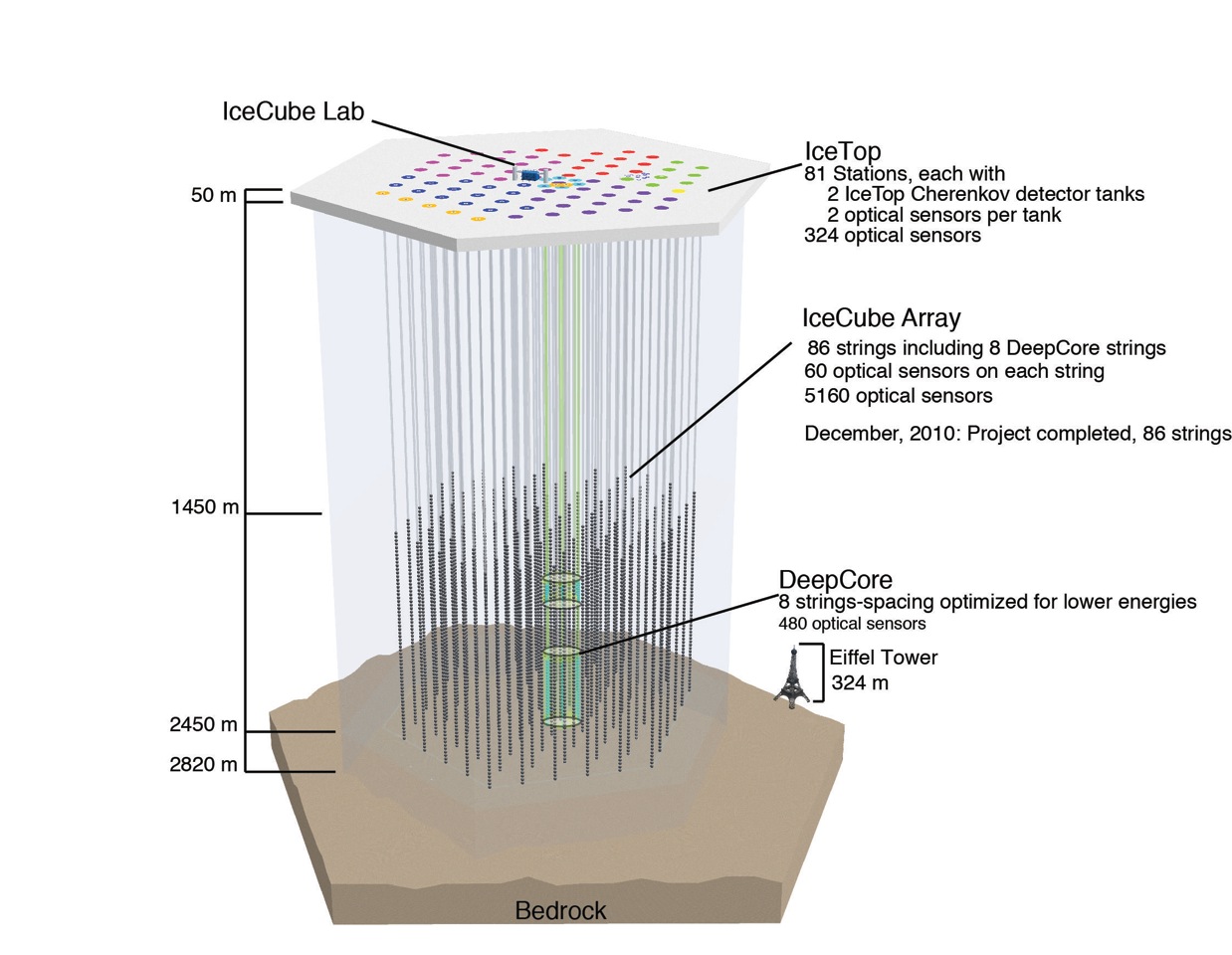}
\caption{IceCube neutrino observatory at the South Pole.
Credit: IceCube Collaboration.\label{fig:icecube}}
\end{figure}

Somewhat surprisingly, the largest SN neutrino detector to date is
the high-energy neutrino telescope IceCube at the South Pole
(fig.~\ref{fig:icecube}), where 1~km$^3$ of ice is instrumented with
a total of 5160 optical modules (OMs). It consists of 78 sparsely
instrumented strings (17~m vertical distance between OMs, 125~m
horizontal string distance) and 8~densely instrumented strings
(7--10~m vertical distance, 60~m horizontal distance), forming the
deep core sub-detector that is optimized for lower-energy neutrinos
in the range 10--300~GeV. When a SN neutrino burst passes through
the ice, the inverse beta reaction produces positrons which in turn
produce Cherenkov light, but typically at most one photon from any
one $\bar\nu_e$ is picked up, no Cherenkov rings can be
reconstructed, and the SN burst simply adds to the noise in the OMs.
For our fiducial SN at 10~kpc, each OM picks up a total of around
300 Cherenkov photons over a few seconds, compared with an internal
singles noise rate of 286~Hz. The correlated noise among all OMs
therefore provides a highly significant signal~\cite{Pryor:1987tz,
Halzen:1994xe, Dighe:2003be, Abbasi:2011ss}, even though there is no
spectral information. For neutrino telescopes in water, this method
is strongly constrained by the high level of radioactive
backgrounds, notably potassium, that is dissolved in sea water.

\begin{figure}
\centering
\includegraphics[width=0.45\textwidth]{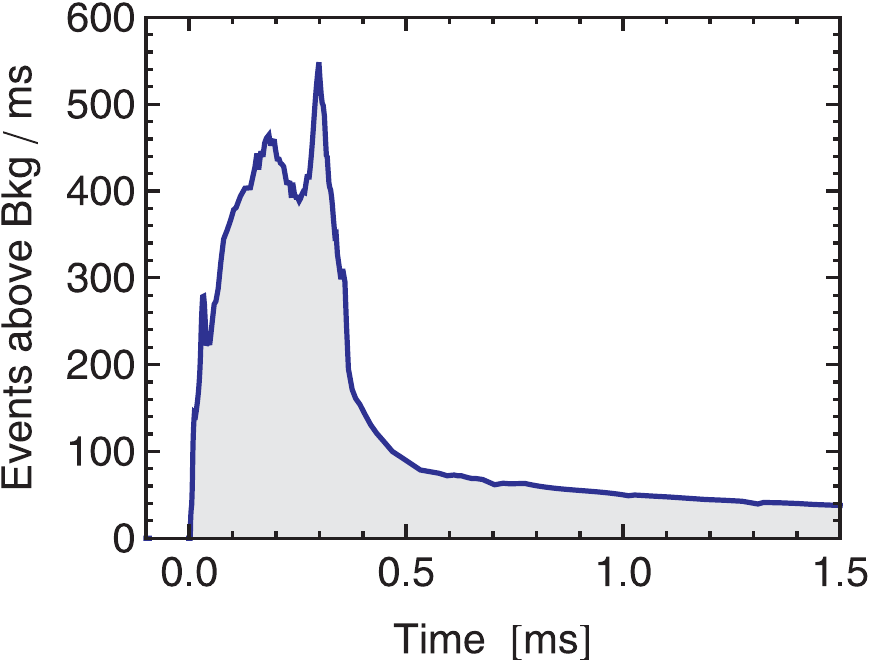}
\caption{Neutrino signal above background in Icecube for a fiducial SN at 10~kpc,
based on the $10.8\,M_\odot$ model of the Basel group shown in
figure~\ref{fig:basel}.\label{fig:icecubesignal}}
\end{figure}

Assuming for simplicity an exact $E_\nu^2$ dependence of the inverse
beta cross section, an approximate expression for the count rate
above background in IceCube is~\cite{Lund:2010kh}
\begin{equation}
R_{\bar\nu_e}=114~{\rm ms}^{-1}\,\frac{L_{\bar\nu_e}}{10^{52}~{\rm erg}~{\rm s}^{-1}}\,
\left(\frac{10~{\rm kpc}}{D}\right)^2\,
\left(\frac{E_{\rm rms}}{15~{\rm MeV}}\right)^2
\quad\hbox{where}\quad
E_{\rm rms}^2=\frac{\langle E_{\bar\nu_e}^3\rangle}{\langle E_{\bar\nu_e}\rangle}\,.
\end{equation}
Note that for a Maxwell-Boltzmann spectrum one finds $E_{\rm
rms}=\sqrt{20/9}\,E_{\rm av}\sim 1.49\,E_{\rm av}$. However, the
instantaneous spectra tend to be pinched and so a realistic $E_{\rm
rms}$ would be smaller. Based on the Basel SN model of
fig.~\ref{fig:basel} we show the expected counting rate above
background in fig.~\ref{fig:icecubesignal}. This is to be compared
with a typical IceCube background rate of $1300~{\rm ms}^{-1}$,
larger than the signal and thus dominating the shot noise. If one
were to use 5~ms bins, the $1\,\sigma$ shot noise would be $\pm
16~{\rm ms}^{-1}$ or about 5\% during the accretion phase in
fig.~\ref{fig:icecubesignal}.

\begin{figure}
\centering
\includegraphics[height=0.35\textwidth]{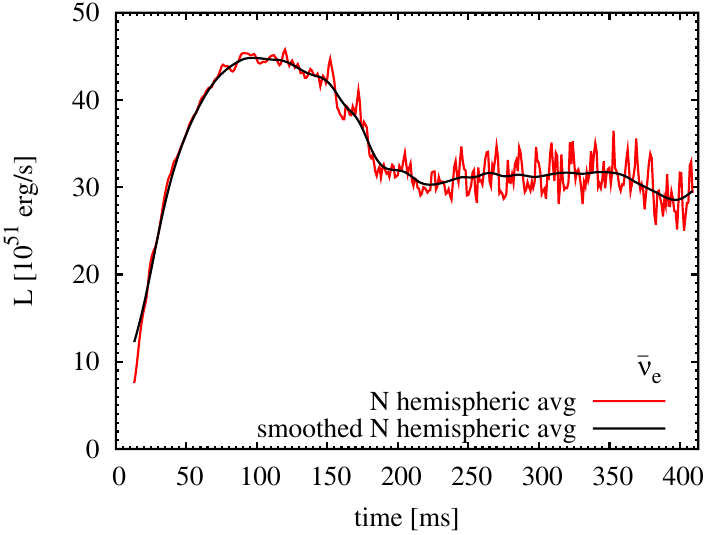}\hfill
\includegraphics[height=0.35\textwidth]{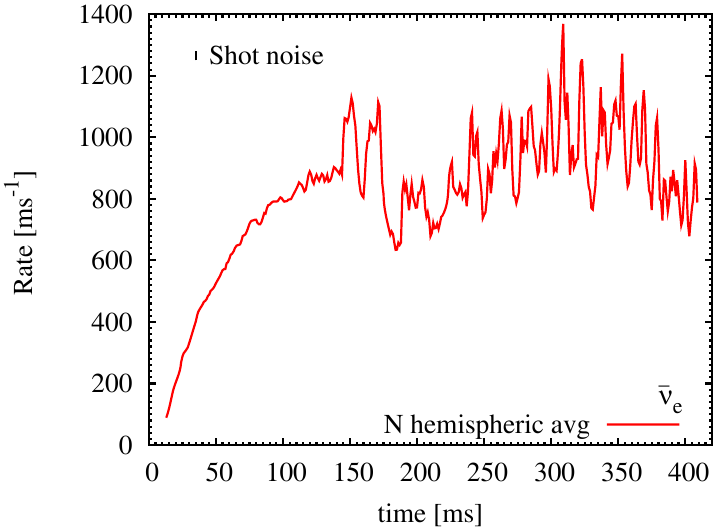}
\caption{Supernova $\bar\nu_e$ signal~\cite{Lund:2010kh} from a 2D Garching
simulation~\cite{Marek:2008qi}. Left: Luminosity and an approximate
time average in the north polar direction. Right: Corresponding
IceCube detection rate and $1\,\sigma$ shot noise for an
assumed 1~ms bin width.\label{fig:lund}}
\end{figure}

The strength of IceCube as a SN neutrino detector is the large rate
of uncorrelated Cherenkov photons that minimizes the shot noise
relative to the number of events and thus offers superior resolution
for the signal time variation. One application is to determine the
signal onset to within a few ms that would be particularly useful in
combination with gravitational wave detection of the bounce
time~\cite{Halzen:2009sm, Pagliaroli:2009qy}. Another application is
to resolve fast time variations caused by convective overturns and
strong SASI activity, leading to significant signal modulations on
time scales of tens of ms (fig.~\ref{fig:lund}). The shown example
is based on a 2D simulation where the SASI activity may be stronger
than in 3D. It depends both on the strength of the modulations and
the distance of the SN whether these features can be resolved.

\begin{figure}[b]
\centering
\includegraphics[height=0.36\textwidth]{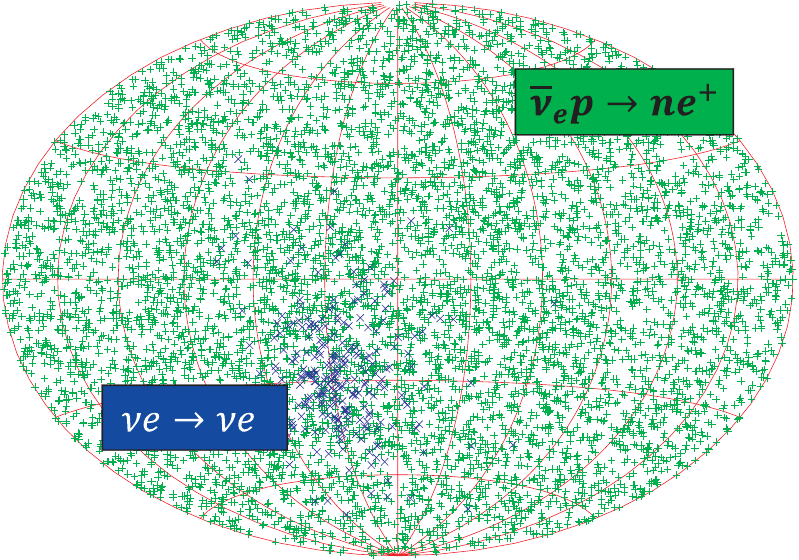}
\caption{Angular distribution of $\bar\nu_ep\to ne^+$
events (green) and elastic scattering events
$\nu e\to e\nu$ (blue) of a simulated SN \cite{Tomas:2003xn}.\label{fig:pointing}}
\end{figure}

The other existing large detector is Super-Kamiokande, after
refurbished electronics in its incarnation IV, with a lowered energy
threshold. Its main detection channel is once more inverse beta
decay, but of course it obtains event-by-event energy and
directional information. Like IceCube, it will provide a superb
neutrino light curve, except with less power to resolve fast time
variations. As a sub-dominant channel, Super-K can statistically
identify electron recoil events by their angular distribution,
$\nu+e\to e+\nu$, that is primarily sensitive to $\nu_e$ and
$\bar\nu_e$ (fig.~\ref{fig:pointing}). In this way, the SN can be
located in the sky by neutrinos alone~\cite{Beacom:1998fj,
Tomas:2003xn}, a possibility that is of particular interest if the
SN is visually obscured.

Telling $\nu e\to e\nu$ from $\bar\nu_ep\to ne^+$ on an
event-by-event basis requires to identify the final-state neutron.
It recombines with a proton to form a deuteron, emitting a 2.2~MeV
$\gamma$-ray that is below threshold in a water Cherenkov detector.
If a sufficient amount of gadolinium, one of the most efficient
neutron catchers, is dissolved in the tank, the subsequent 8~MeV
$\gamma$ cascade could be measured, tagging the inverse beta
reaction \cite{Beacom:2003nk, Watanabe:2008ru}. A dedicated R\&D
program, the ongoing EGADs project, evaluates the full-scale
realistic feasibility of this approach. Without neutron tagging, the
SN pointing accuracy is $7.8^\circ$ for the 95\% CL half-cone
opening angle, whereas for a 90\% tagging efficiency this would
improve to $3.2^\circ$. For a megaton water Cherenkov detector
(30$\times$Super-K), these numbers improve to $1.4^\circ$ and
$0.6^\circ$, respectively~\cite{Tomas:2003xn}.

The ongoing long-baseline neutrino oscillation programs worldwide
suggest that at least one megaton-class water Cherenkov detector
will be built in the foreseeable future \cite{Scholberg:2010zz}.
Such projects are discussed in Japan (``Hyper-Kamiokande''), in
Europe (``Memphys'') and the US (``LBNE''). Such developments will
boost the SN detection capabilities even further and provide yet
more statistics for a SN neutrino light curve.

Scintillator detectors are another class of $\bar\nu_e$ detectors
that can be scaled to large volume. One advantage is the low energy
threshold and concomitant native neutron-tagging capability as well
as stronger light output implying superior energy resolution. Of
course, there is hardly any directional information except in a weak
statistical sense by the displacement of the positron annihilation
and neutron capture vertices~\cite{Apollonio:1999jg,
Hochmuth:2007gv}. Each of the existing detectors
(table~\ref{tab:SNdetectors}) would provide a significant SN neutrino
light curve and energy information, and taken together they provide
formidable statistics. A 50~kt scintillator detector, Low Energy
Neutrino Astronomy (LENA), is under discussion~\cite{Wurm:2011zn}
that combines the advantage of the scintillator technique with the
size of Super-Kamiokande. For the SN parameters assumed earlier, it
would register about $1.1\times10^4$ inverse beta events, somewhat
more than Super-K, with better energy resolution and about 600
electron scattering events. One may also measure proton
recoil~\cite{Beacom:2002hs, Dasgupta:2011wg}, $\nu p\to p\nu$, with
around 1300 events in LENA. Other subdominant channels that may
become detectable with a few hundred events each are the carbon
reactions \hbox{(i)~$\nu+{}^{12}{\rm C}\to {}^{12}{\rm C}^*+\nu$}
followed by ${}^{12}{\rm C}^*\to{}^{12}{\rm C}+\gamma$,
(ii)~$\bar\nu_e+{}^{12}{\rm C}\to {}^{12}{\rm B}+e^+$ followed by
${}^{12}{\rm B}\to{}^{12}{\rm C}+e^-+\bar\nu_e$, and
(iii)~$\nu_e+{}^{12}{\rm C}\to {}^{12}{\rm N}+e^-$ followed by
${}^{12}{\rm N}\to{}^{12}{\rm C}+e^++\nu_e$.

A new type of SN detector, HALO, is being realized in SNO Lab, using
79~tons of existing lead as a target. The relevant processes are the
dominant charged-current reactions $\nu_e+{}^{208}{\rm Pb}\to
{}^{207}{\rm Bi}+n+e^-$ and $\nu_e+{}^{208}{\rm Pb}\to {}^{206}{\rm
Bi}+2n+e^-$ as well as the neutral-current reactions
$\nu+{}^{208}{\rm Pb}\to {}^{207}{\rm Pb}+n$ and $\nu+{}^{208}{\rm
Pb}\to {}^{206}{\rm Pb}+2n$. In all cases, one measures the produced
neutrons with $^3$He detectors remaining from the decommissioned SNO
solar neutrino experiment. HALO provides complementary information on
the spectrum because its high threshold makes it especially sensitive
to the high-energy tail of the neutrino
distribution~\cite{Engel:2002hg, Duba:2008zz, Vaananen:2011bf}.

In the SN neutrino signal, the spectral differences between different
flavors are much larger in the $\nu$ channel than the $\bar\nu$
channel, and in particular the prompt $\nu_e$ burst is a dramatic
feature, yet the existing large detectors are all primarily sensitive
to $\bar\nu_e$. A large $\nu_e$ detector could be based on the liquid
argon time projection chamber technique, with the recently
commissioned 600~t ICARUS module in Gran Sasso being an operational
prototype~\cite{Sala:2011zz}. SN neutrinos are detected by the main
reaction $\nu_e+{}^{40}{\rm Ar}\to{}^{40}{\rm K}+e^-$ plus some
subdominant channels, so one has an excellent $\nu_e$
detector~\cite{GilBotella:2004bv}. While ICARUS would measure a few
hundred events from a SN at 10~kpc, a much bigger detector, perhaps
up to 100~kt, is discussed in Europe under the name of GLACIER
\cite{Autiero:2007zj}.

How often can we expect a signal from any of these detectors? Even
the largest of the existing instruments can only cover our own
galaxy and its satellites such as the Large Magellanic Clouds.
Reaching the Andromeda galaxy, the Milky Way's large partner galaxy
at a distance of around 760~kpc, requires bigger detectors such as a
megaton class water Cherenkov instrument that could then get a few
tens of events. For a high-statistics observation we remain
constrained to our own galaxy and its satellites. The estimated SN
rates by various techniques are summarized in table~\ref{tab:rate},
i.e.\ we can expect a few core collapses per century. Except for
SN~1987A in the Large Magellanic Cloud, no core collapse was
observed over more than 30~years of neutrino observations, already
implying a nontrivial upper limit on the rate of possible failed
SNe.

\begin{table}
 \caption{\label{tab:rate}Estimated rate of galactic core-collapse
 SNe per century.}
 \begin{tabular}{llll}
 \hline
 Method&Rate&Authors&Refs.\\
 \hline
 Scaling from external galaxies&$2.5\pm0.9$
 &van den Bergh&
 \cite{vandenBergh:1994, Diehl:2006cf}\\
 &&\& McClure (1994)\\
 &$1.8\pm1.2$&Cappellaro \&\ Turatto&
 \cite{Cappellaro:1999qy, astro-ph/0012455}\\
 &&(2000)\\
 Gamma-rays from galactic $^{26}$Al&$1.9\pm1.1$
 &Diehl et al.\ (2006)&
 \cite{Diehl:2006cf}\\
 Historical galactic SNe (all types)&$5.7\pm1.7$&
 Strom (1994)&\cite{Strom:1994}\\
 &$3.9\pm1.7$&
 Tammann et al.\ (1994)&\cite{Tammann:1994ev}\\
 No neutrino burst in 30 years$^{a}$&${}<7.7$ (90\% CL)&
 Alekseev \& Alekseeva&\cite{Alekseev:2002ji}\\
 &&(2002)\\
 \hline
 \multicolumn{4}{l}{$^a$We have scaled the limit of
 Ref.~\cite{Alekseev:2002ji} to 30~years of neutrino sky coverage.}
 \end{tabular}
\end{table}

\begin{figure}
\centering
\includegraphics[height=0.38\textwidth]{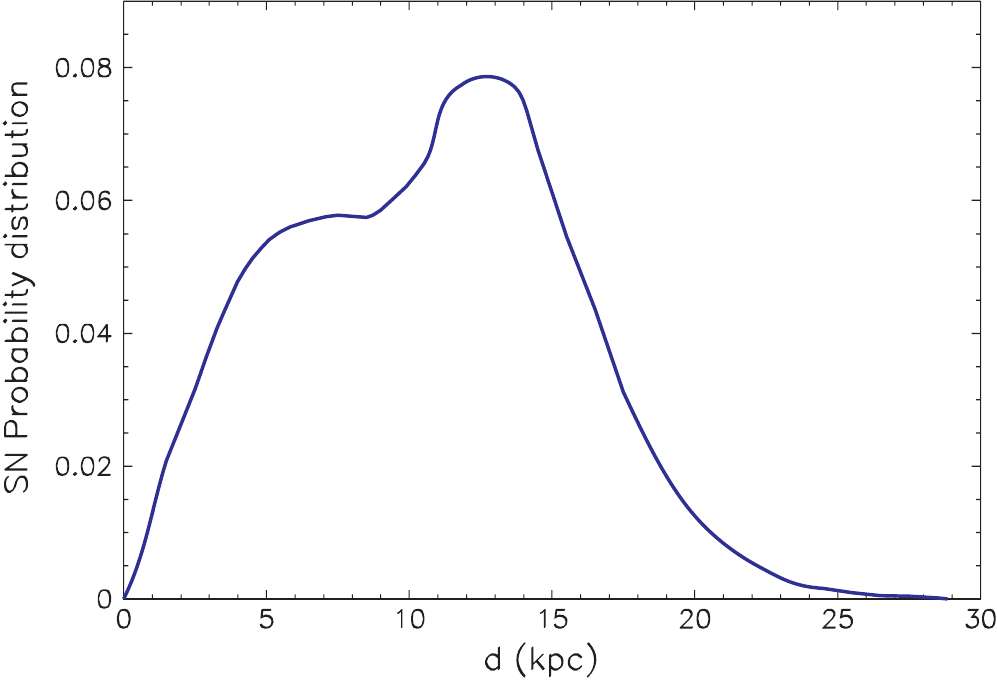}\hfill
\includegraphics[height=0.38\textwidth]{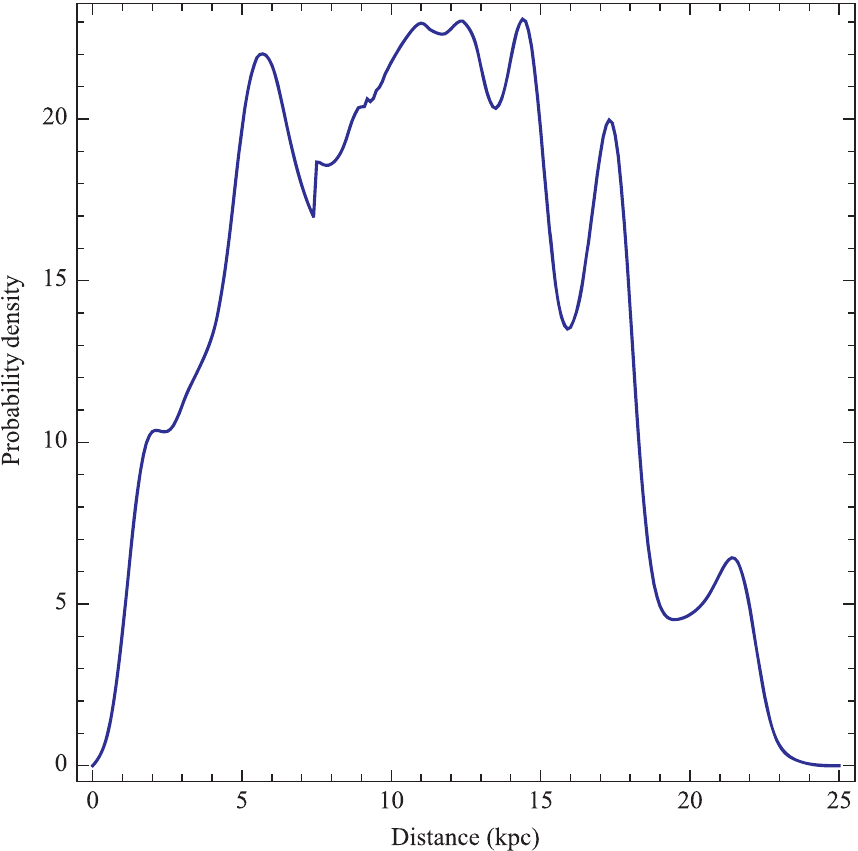}
\caption{SN distance distribution relative to the Earth for a simple
model of progenitor distribution \cite{Mirizzi:2006xx} (left) and one
taking account of the spiral arm structure \cite{Ahlers:2009ae}
(right).\label{fig:SNdistribution}}
\end{figure}

The possible distribution of core-collapse SNe in the galaxy must
follow the regions of star formation, notably in the spiral arms.
The expected distance distribution for two simple models are shown
in fig.~\ref{fig:SNdistribution}. While being different in detail,
the main point is that the distributions are very broad and that
10~kpc is probably a reasonable benchmark value. Sometimes our
distance to the galactic center of 8.5~kpc is used for this purpose,
but SNe are not especially likely in the galactic center region.
However, the expected distribution is so broad that any specific
distance is unlikely to be ``typical'' for the next nearby SN. In
this sense, any forecast of what can be learnt should, in principle,
cover a broad range of cases. Since any distance, say, between 2 and
20~kpc is almost equally likely, the dynamical range of plausible
event statistics is about a factor of 100.

\begin{figure}
\centering
\includegraphics[width=1.0\textwidth]{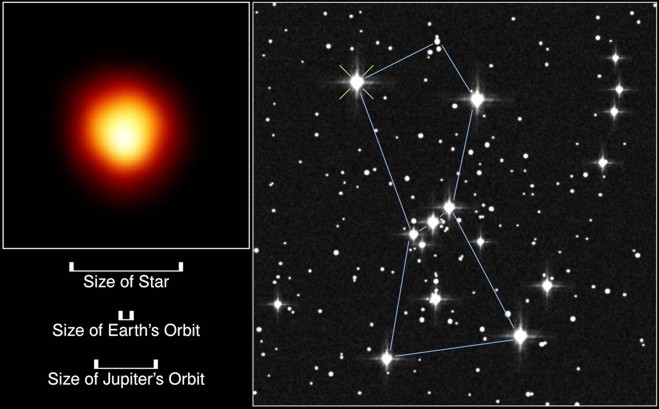}
\caption{The star Betelgeuze (Alpha Orionis) at a
distance of 130 pc (425 lyr)
is the first resolved image of a star other than the Sun.
It is a candidate for the next nearby SN explosion.
HST image taken in ultraviolet on 3 March 1995.
Credit: A.~Dupree (Harvard-Smithsonian CfA), R.~Gilliland (STScI),
NASA and ESA.
  \label{fig:betelgeuze}}
\end{figure}

Of course, we may be especially lucky and the next galactic SN
happens very nearby in that the red supergiant Betelgeuze in the
constellation Orion could explode (fig.~\ref{fig:betelgeuze}),
causing around $4\times10^7$ events in Super-Kamiokande. To handle
the possibility of such a large data flow, special measures for the
data acquisition system have to be taken. This is the closest
conceivable SN among the known stars in the solar neighborhood, but
would still be at a safe distance regarding life on Earth. For such
a close SN, one may be able to pick up the neutrino signal of the
pre-supernova evolution when silicon burning produces a huge flux of
thermal neutrinos with enough energy for the inverse beta reaction.
The increased neutron production rate for a few weeks before the
explosion could provide early warning of the imminent Betelgeuze
explosion~\cite{Odrzywolek:2003vn}.

Reaching beyond the galaxy and its satellites requires new
strategies. Even megaton class detectors will only reach to the
Andromeda galaxy and get only a few tens of events from that
distance. A different strategy would be a multi-megaton detector
such as the proposed 5~megaton Deep-TITAND, that could pick up mini
bursts of a few events from all SNe out to few-Mpc
distances~\cite{Kistler:2008us}. In this way one could build up an
average SN neutrino spectrum from many different SNe over a few
years. Another way to realize the same idea is with an upgraded
deep-core detector in IceCube that could be instrumented with an
ever denser grid of optical modules (PINGU project) such as to reach
eventually the 10~MeV range threshold~\cite{Kowalski:2011}.
Conceivably one could construct a 10~megaton detector in this way,
providing for a novel perspective for low-energy neutrino astronomy.

\subsection{Diffuse supernova neutrino background (DSNB)}

Another way to reach beyond the galaxy is to search for the diffuse
SN neutrino background (DSNB) from all past SNe in the
universe~\cite{Beacom:2010kk}. While SNe in any given galaxy are
rare, the emitted energy in each core collapse is so large that the
long-term average of total neutrino energy emitted is almost exactly
the same as the total photon energy. The cosmic average light
emitted by all stars adds up to the extra galactic background light
(EBL) with an intensity of 50--$100~{\rm nW}~{\rm m}^{-2}~{\rm
ster}^{-1}$, corresponding to an energy density of 13--$26~{\rm
meV}~{\rm cm}^{-3}$, i.e.\ about 10\% of the energy density provided
by the cosmic microwave background. In this sense stellar
populations emit about as much gravitational binding energy (in the
form of neutrinos) as they emit nuclear binding energy (mostly in
the form of photons and some thermal neutrinos).

\begin{figure}[b]
\centering
\includegraphics[width=0.6\textwidth]{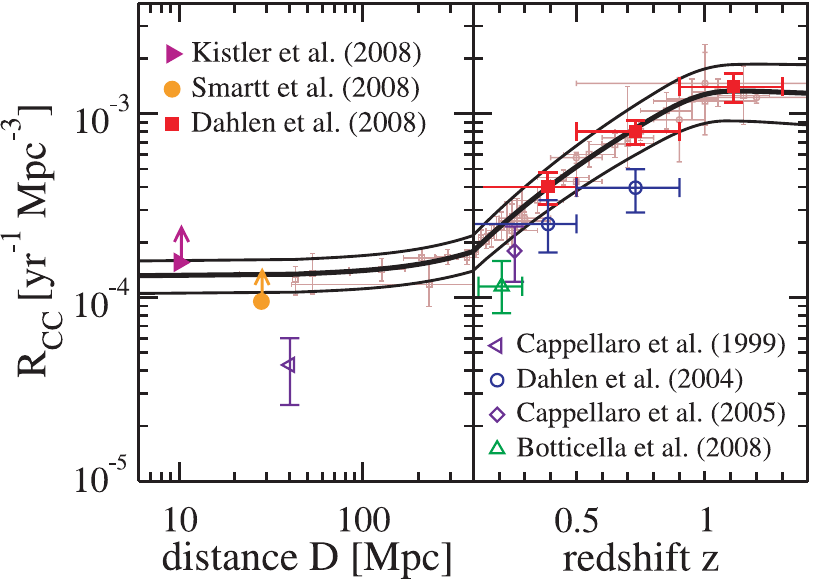}
\caption{Core collapse rate as a function of redshift according
to different measurements of the star formation
rate~\cite{Horiuchi:2008jz}.\label{fig:ccrate}}
\end{figure}

The DSNB signal depends on three ingredients. First, the cosmic core
collapse rate $R_{\rm cc}$, about 10 per second in the causal
horizon; this is determined by astronomical measurements that are
already precise and quickly improving (fig.~\ref{fig:ccrate}).
Second, the average SN neutrino emission, which is expected to be
comparable for all core collapses, including those that fail and
produce black holes; this is the quantity of fundamental interest.
Third, the detector capabilities, including the energy dependence of
the cross section and detector backgrounds.

Detecting the DSNB is important even if a Milky Way burst is
observed. DSNB $\bar\nu_e$ will provide a unique measurement of the
average neutrino emission spectrum to test SN simulations.
Comparison to SN 1987A and an eventual Milky Way SN will test the
variation between core collapses. While the statistics of DSNB
events will be low with foreseeable detectors, comparable to those
of SN 1987A, this data will more effectively measure the
exponentially falling tail of the spectrum at high energies. The
DSNB is also a new probe of stellar birth and death: its energy
density is comparable to that of photons produced by stars, but the
DSNB is unobscured and has no known competition from astrophysical
sources.

\begin{figure}[b]
\centering
\includegraphics[width=0.5\textwidth]{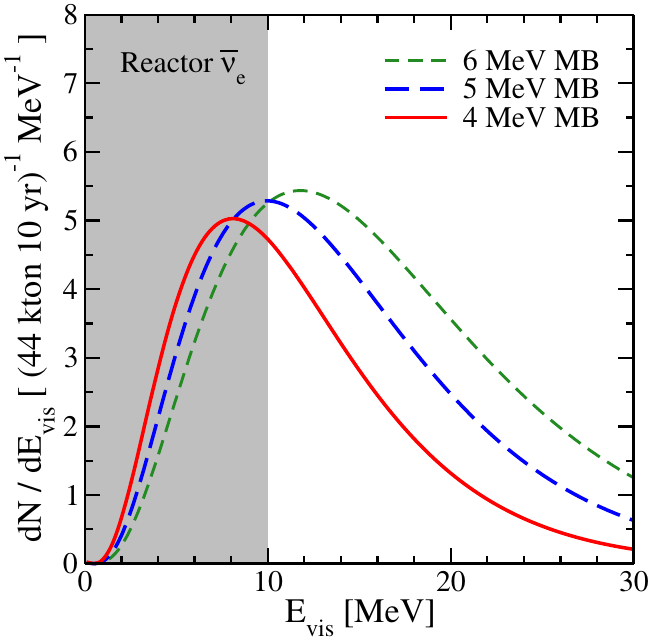}
\caption{Detection positron spectrum in the possible 50~kt
LENA scintillator detector for different values of the assumed
$T$ of the average SN $\bar\nu_e$ emission
spectrum~\cite{Wurm:2011zn}.\label{fig:dsnbspectrum} Below about
10~MeV, the background $\bar\nu_e$ flux from power reactors completely
masks the DSNB. At higher energies, backgrounds from cosmic rays
kick in, but should be controllable for 10--30~MeV.}
\end{figure}

The DSNB event rate spectrum follows from a line of sight integral
for the radiation intensity from a distribution of distant sources.
After integrating over all angles due to the isotropy of the DSNB
and the transparency of Earth, it is, in units ${\rm s}^{-1}~{\rm
MeV}^{-1}$,
\begin{equation}
\frac{dN_{\rm vis}}{d E_{\rm vis}}=\,N_p\sigma(E_\nu)\int_0^\infty R_{\rm cc}(z)
\Bigl\{(1+z)\phi[E_\nu(1+z)]\Bigr\}
\,\left|\frac{dt}{dz}\right|\,dz\,,
\end{equation}
where $E_{\rm vis}$ is the detected positron energy. On the right
hand side, before the integral is the number of targets (protons)
times the detection cross section. Under the integral, the first
ingredient is the comoving cosmic core-collapse rate, in units ${\rm
Mpc}^{-3}~{\rm yr}^{-1}$; it evolves with redshift
(fig.~\ref{fig:ccrate}). The second is the average time-integrated
emission per SN, in units ${\rm MeV}^{-1}$; redshift reduces emitted
energies and compresses spectra. The last term is the differential
distance, where $|dt/dz|^{-1} = H_0(1 + z)[\Omega_\Lambda
+\Omega_{\rm m}(1 + z)^3]^{1/2}$; the cosmological parameters are
taken as $H_0 = 70~{\rm km}~{\rm s}^{-1}~{\rm Mpc}^{-1}$,
$\Omega_\Lambda= 0.7$, and $\Omega_{\rm m} = 0.3$. The cosmological
factor and the SN rate derived from star formation rate data are
really one combined factor proportional to the ratio of the average
luminosity per galaxy in SN neutrinos relative to stellar photons.
For the example of the foreseen 50~kt LENA scintillator detector,
with a fiducial mass of 44~kt, one then finds the detection spectrum
shown in fig.~\ref{fig:dsnbspectrum}. Over a measurement time of
10~years it would collect a significant data set, depending on the
emission spectrum of SN neutrinos.

\begin{figure}
\centering
\includegraphics[width=0.6\textwidth]{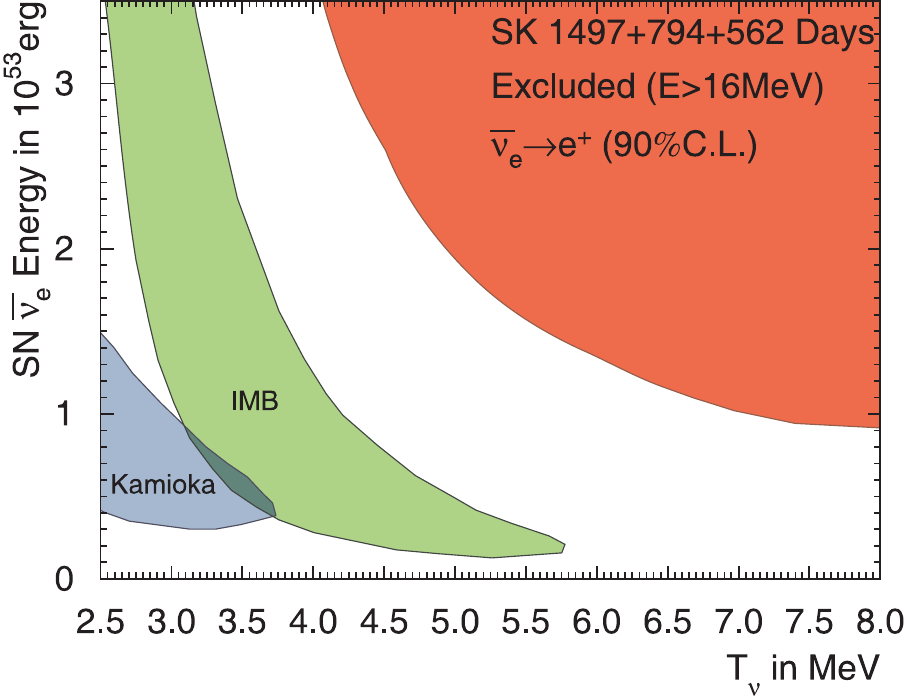}
\caption{DSNB exclusion limits (90\% CL), assuming the
average SN $\bar\nu_e$ emission spectrum is described by a
thermal Maxwell-Boltzmann spectrum~\cite{Bays:2011si}.
For comparison, the best-fit regions for the SN~1987A signal
of fig.~\ref{fig:SNfit} are also shown.\label{fig:dsnbsklimits}}
\end{figure}

The detection is more difficult for a water Cherenkov detector
because it lacks the native neutron tagging capability due to its
larger energy threshold. Therefore, it is not possible to reject
irreducible backgrounds caused by cosmic ray events. The
Super-Kamiokande detector places an upper $\bar\nu_e$ flux limit of
2.8--$3.0~{\rm cm}^{-2}~{\rm s}^{-1}$ for neutrino energies above
17.3~MeV, the exact value depending on the assumed
spectrum~\cite{Bays:2011si}. Depending on the assumed total energy
emitted in $\bar\nu_e$ by any given SN and the spectral shape
assumed to be thermal, they find the exclusion range shown in
fig.~\ref{fig:dsnbsklimits}. To achieve a detection in Super-K one
needs neutron-tagging capability that is currently being developed
in terms of loading the detector with gadolinium as explained
earlier.

\subsection{Particle physics constraints and future possibilities}

The neutrino observations from core collapse and the SN dynamics
itself provide formidable laboratories for particle
physics~\cite{Raffelt:1996wa, Raffelt:1990yz, Raffelt:1999tx,
Raffelt:2006cw, Schramm:1987ra}. It was Georgiy Zatsepin who first
pointed out that the neutrino burst from SN collapse offers an
opportunity to measure the neutrino mass by the energy-dependent
time-of-flight delay~\cite{Zatsepin:1968kt}
\begin{equation}\nonumber
\Delta t=5.1~{\rm ms}\left(\frac{D}{10~{\rm kpc}}\right)
\left(\frac{10~{\rm MeV}}{E_\nu}\right)^2
\left(\frac{m_\nu}{1~{\rm eV}}\right)^2\,.
\end{equation}
However, when the SN~1987A burst was measured, it provided a mass
limit of about 20~eV \cite{Loredo:2001rx, Loredo:1988mk,
Kernan:1994kt}, which even at that time was only marginally
interesting and was soon superseded by laboratory limits. The
neutrino signal of the next nearby SN could improve this at best to
the eV range~\cite{Beacom:1998ya, Nardi:2004zg}. It is more
interesting to note that the restrictive sub-eV cosmological
neutrino mass limits~\cite{Abazajian:2011dt} assure that fast time
variations at the source will not be washed out by time-of-flight
effects and thus are, in principle, detectable at
IceCube~\cite{Abbasi:2011ss, Lund:2010kh}.

A time-of-flight argument can also be used to constraint a putative
neutrino electric charge. It would lead to deflection in the
galactic magnetic field and thus to an energy-dependent pulse
dispersion in analogy to $m_\nu$, providing
$e_\nu\alt3\times10^{-17}\,e$ \cite{Barbiellini:1987zz,
Bahcall:1989ks}.

From a present-day perspective, the most interesting time-of-flight
constraint, however, is the one between neutrinos and photons,
testing the equality of the relativistic limiting propagation speed
between the two species. SN physics dictates that the neutrino burst
should arrive a few hours earlier than the optical brightening, in
agreement with SN~1987A. Given the distance of about 160,000 light
years one finds~\cite{Longo:1987ub,Stodolsky:1987vd}
\begin{equation}\nonumber
\left|\frac{c_\nu-c_\gamma}{c_\gamma}\right|\alt2\times10^{-9}\,.
\end{equation}
At the time of this writing, this result plays a crucial role for
possible interpretations of the apparent superluminal neutrino speed
reported by the OPERA experiment~\cite{Adam:2011zb},
$(c_\nu-c_\gamma)/c_\gamma = (2.37 \pm 0.32_{\rm stat}
+0.34/{-}0.24_{\rm sys})\times 10^{-5}$. No plausible interpretation
for this measurement is available at present.

Both neutrinos and photons should be delayed by their propagation
through the gravitational potential of the galaxy (Shapiro time
delay) which is estimated to be a few months toward the Large
Magellanic Cloud. The agreement between the arrival times within a
few hours confirms a common time delay within about
0.7--$4\times10^{-3}$, i.e.\ neutrinos and photons respond to
gravity in the same way~\cite{Longo:1987gc,Krauss:1987me}. This is
the only experimental proof that neutrinos respond to gravity in the
usual way. These results could be extended to include the
propagation speed of gravitational waves if the next nearby SN is
observed both in neutrinos and with gravitational wave detectors.
The onset of both bursts would coincide with the SN bounce time to
within a few ms and the coincidence could be measured with this
precision~\cite{Halzen:2009sm, Pagliaroli:2009qy}. In view of the
current discussion of superluminal neutrino propagation, such a
measurement would provide important additional constraints on
possible interpretations.

After core collapse, neutrinos are trapped in the SN core and energy
is emitted on a neutrino diffusion time scale of a few
seconds~\cite{Sawyer:1979xs}. This basic picture was confirmed by
the SN~1987A neutrino burst, indicating that the gravitational
binding energy was not carried away in the form of some other
radiation, more weakly coupled than neutrinos, that would escape
directly without diffusion~\cite{Raffelt:1987yt, Turner:1987by,
Mayle:1987as}. This ``energy-loss argument'' has been applied to a
large number of cases, notably axions, Majorons, and right-handed
neutrinos, often providing the most restrictive limits on the
underlying particle-physics model; extensive reviews are
Refs.~\cite{Raffelt:1996wa, Raffelt:1990yz, Raffelt:1999tx,
Raffelt:2006cw, Schramm:1987ra}. More recently, the argument was
applied to Kaluza-Klein gravitons~\cite{Cullen:1999hc,
Hanhart:2000er, Hanhart:2001fx, Hannestad:2003yd}, light
neutralinos~\cite{Dreiner:2003wh}, light dark matter
particles~\cite{Fayet:2006sa}, and
unparticles~\cite{Davoudiasl:2007jr, Hannestad:2007ys,
Dutta:2007tz}. While there is no good reason to doubt the validity
of this widely used argument, it is based on very sparse data.
Measuring a high-statistics neutrino signal from the next nearby SN
would put these crucial results on much firmer experimental ground.

\begin{figure}
\centering
\includegraphics[width=1.0\textwidth]{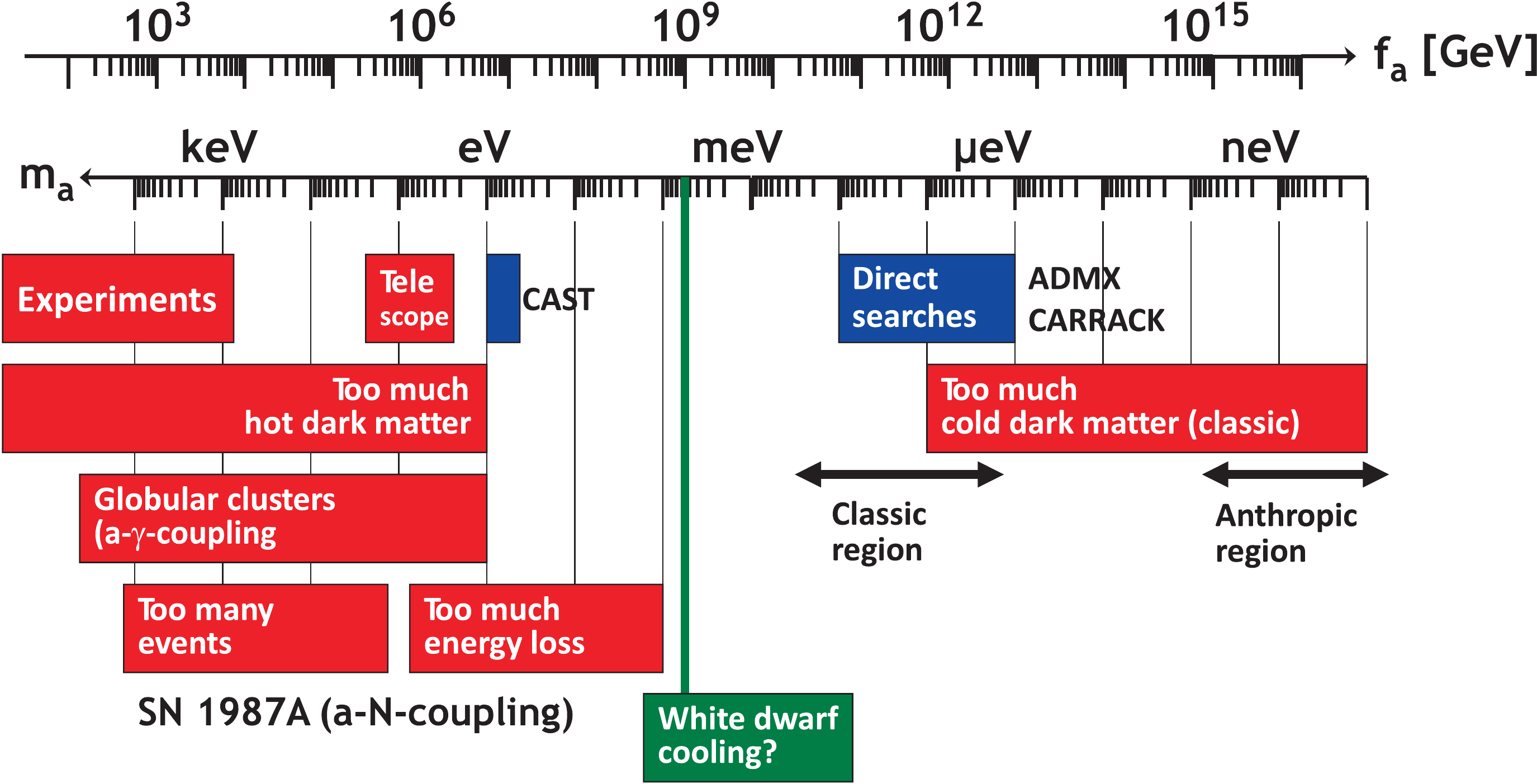}
\caption{Summary of axion bounds, where red bars imply exclusion,
green a tentative signature, and blue experimental search
ranges~\cite{Raffelt:2006cw}.\label{fig:axionbounds}}
\end{figure}

\begin{figure}
\centering
\includegraphics[width=0.5\textwidth]{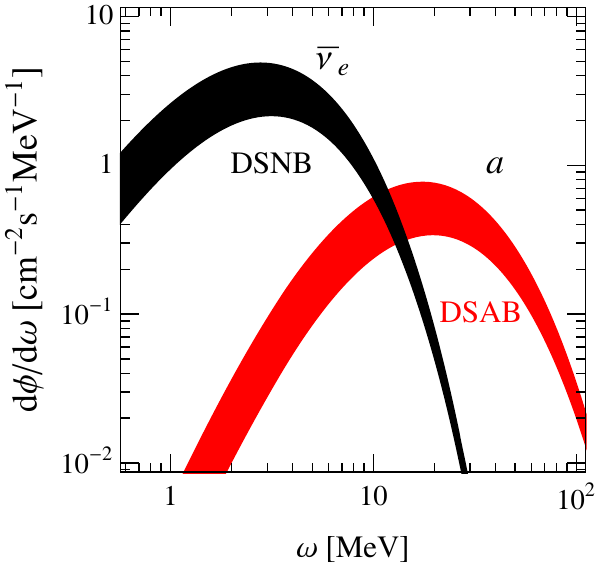}
\caption{Diffuse SN axion background (DSAB) compared with the
DSNB~\cite{Raffelt:2011ft}. It was assumed that either neutrinos or
axions carry away the full SN energy of $3\times10^{53}~{\rm erg}$. The
width of the bands reflects only the uncertainty in the core
collapse rate $R_{\rm cc}$. For $\bar\nu_e$ a thermal spectrum
with $T=4$~MeV is assumed, carrying away 1/6 of the total energy, whereas
for axions a bremsstrahlung-inspired spectrum with $T_{\rm core}=30$~MeV
was assumed.\label{fig:dsab}}
\end{figure}

Of particular interest are the SN~1987A axion bounds that squeeze
the allowed $m_a$ range to very small values below 10~meV
(fig.~\ref{fig:axionbounds}). These bounds leave open the
possibility that axions with a nonvanishing electron interaction
could account for an additional white-dwarf cooling channel that may
be suggested by observations as discussed earlier (see~fig.
\ref{fig:wd}). If the white-dwarf axion cooling interpretation were
correct, axions would provide a significant energy-loss channel for
SNe, although the axion burst from the next nearby SN would not be
observable due to the extremely weak axion interactions. Still, the
universe would be filled with a diffuse SN axion background (DSAB)
with an energy density comparable to the DSNB~\cite{Raffelt:2011ft}.
Axions would be emitted from the inner SN core and thus have much
larger energies than the emitted neutrinos, reflecting in a harder
DSAB spectrum (fig.~\ref{fig:dsab}). So the universe could be filled
with a significant amount of axion radiation that, however, appears
to be nearly impossible to measure.

Conventional SN simulations are based on standard particle-physics
assumptions that are not necessarily tested in the laboratory. In
particular, lepton-number conservation is crucial in the collapse
process because it ensures that the liberated gravitational energy
is at first stored primarily in the degeneracy energy of electrons
and electron neutrinos, i.e.\ the SN core after collapse is
relatively cold. On the other hand, it is now commonly assumed that
lepton number is not conserved in that neutrino masses are widely
assumed to be of Majorana type. While neutrino Majorana masses would
not suffice for significant lepton-number violating effects in a SN
core, other sources of lepton-number violation may well be strong
enough, e.g.\ R-parity violating supersymmetric models that in turn
can induce Majorana masses. Therefore, it would be intriguing to
study core collapse with ``internal'' deleptonization, leading to a
hot SN core immediately after collapse.

In a SN core, the matter potentials are so large that flavor
conversion by oscillation is strongly suppressed even though some of
the mixing angles are large. Therefore, the initial $\nu_e$ Fermi
sea is conserved---in a SN core, flavor lepton number is effectively
conserved. On the other hand, certain non-standard interactions
(NSI) \cite{Biggio:2009nt} that are not diagonal in flavor space
would allow for flavor lepton number violation in collisions and
therefore lead to a quick equipartition among flavors of the trapped
lepton number. The required interaction strength is much smaller
than what is typically envisioned for NSI effects on long-baseline
neutrino oscillation experiments. In other words, a SN core is
potentially the most sensitive laboratory for NSI effects. While it
has been speculated that such effects would strongly modify the
physics of core collapse~\cite{Amanik:2006ad,Lychkovskiy:2010ue}, a
numerical simulation including the quick equipartition of flavors
has never been performed.

\subsection{Flavor oscillations of SN neutrinos}

Flavor conversion by neutrino oscillations is a large effect, for
example for solar neutrinos, and will also be important for SN
neutrinos, but not in the inner SN core. In this nuclear-density
environment, the Wolfenstein matter effect is huge and propagation
eigenstates are almost identical with weak interaction eigenstates,
in spite of the large mixing angles. The weak potential difference
of eq.~(\ref{eq:weakpotential}) between $\nu_e$ and other flavors,
that is around 0.2~peV in normal matter, is 14 orders of magnitude
larger and thus a few tens of eV. As a consequence, the trapped
electron lepton number is conserved on all time scales relevant for
SN dynamics. Unless nonstandard flavor lepton number violating
effects operate in a SN core, lepton number can disappear only on
the neutrino diffusion time scale of seconds.

\begin{figure}
\centering
\includegraphics[width=1.0\textwidth]{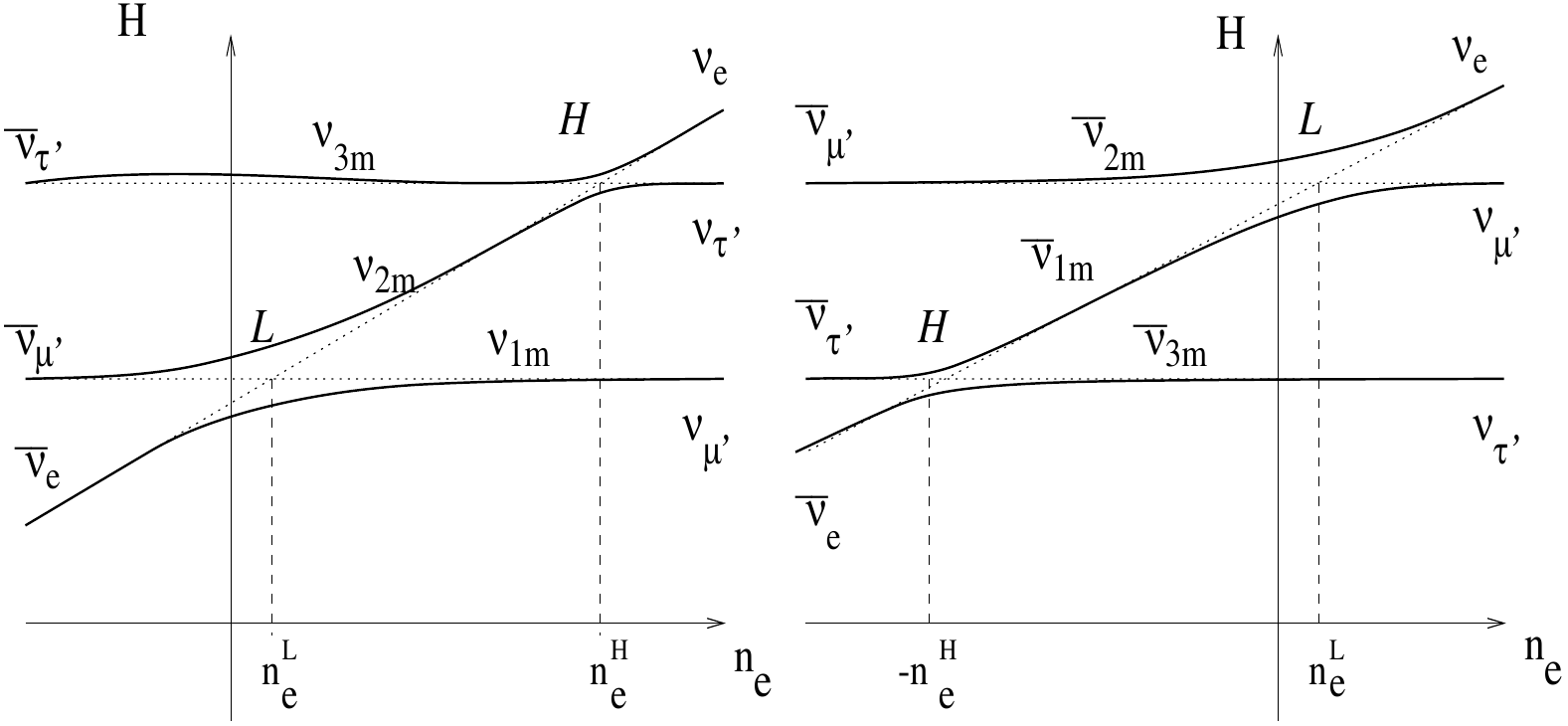}
\caption{Three-flavor level diagram for neutrino propagation eigenmodes, in
analogy to fig.~\ref{fig:eigenvalue},
relevant for neutrinos streaming from a SN core~\cite{Dighe:1999bi}
for normal hierarchy (left) and inverted hierarchy (right).
\label{fig:levelcrossing}}
\end{figure}

Of course, as neutrinos stream from the SN core through the stellar
envelope, they will eventually encounter MSW resonances
corresponding to the atmospheric mass difference (H~resonance) and
the solar mass difference (L~resonance). The corresponding level
diagram for the two mass hierarchies (fig.~\ref{fig:levelcrossing})
allows one to determine in which mass eigenstate a neutrino will
emerge that was produced in a given interaction eigenstate. Of
particular interest is the MSW effect at the H-resonance driven by
the 13-mixing angle. This resonance occurs in the neutrino sector
for the normal mass hierarchy, and among anti-neutrinos for the
inverted hierarchy. It is adiabatic for $\sin^2\theta_{13}\agt
10^{-3}$ and non-adiabatic for $\sin^2\theta_{13}\alt 10^{-5}$.
Therefore, the neutrino burst is, in principle, sensitive to the
mass hierarchy and the 13-mixing angle~\cite{Dighe:1999bi,
Dighe:2004xy}.

What arrives at Earth after propagation are mass eigenstates that
need to be projected on interaction eigenstates to determine the
detector response. The arriving flux relevant for detection can then
be expressed in terms of the energy-dependent $\nu_e$ survival
probability $p(E)$ in the form
\begin{equation}
F_{\nu_e}=p(E)\,F_{\nu_e}^0(E)+[1-p(E)]\,F_{\nu_x}^0(E)\,,
\end{equation}
where the subscript 0 denotes the primary fluxes at emission. An
analogous expression pertains to $\bar\nu_e$ with the survival
probability $\bar p(E)$. Table~\ref{tab:survival} summarizes the
survival probabilities for different mixing scenarios, assuming that
collective flavor conversions are not important (see below). The
recent hints for a ``large'' value for $\theta_{12}$ discussed
earlier suggest that the H resonance is adiabatic and we are in
scenario A or B. The $\nu_e$ and $\bar\nu_e$ survival probabilities
then distinguish between the normal and inverted mass hierarchy. How
can this effect be measured?

\begin{table}
\caption{\label{tab:survival}Survival probabilities for neutrinos,
  $p$, and antineutrinos, $\bar{p}$, in various mixing scenarios,
  assuming collective flavor conversion plays no
  role~\cite{Dighe:1999bi, Dighe:2004xy}.}
  \centering
\begin{tabular}{llllll}
  \hline
  Scenario&Hierarchy& $\sin^2\theta_{13}$ & $p$ & $\bar{p}$&Earth effects\\
  \hline
  A & Normal &${\agt}\,10^{-3}$  & 0  & $\cos^2\theta_{12}$&$\bar\nu_e$\\
  B & Inverted &  $\agt\, 10^{-3}$ &  $\sin^2\theta_{12}$ &  0&$\nu_e$\\
  C & Any & $\alt\, 10^{-5}$  & $\sin^2\theta_{12}$ &
  $\cos^2\theta_{12}$ &$\nu_e$ and $\bar\nu_e$\\
  \hline
\end{tabular}
\end{table}

The most pronounced flavor-dependent feature in the SN neutrino
signal is the prompt $\nu_e$ burst, which in addition is rather
model independent~\cite{Kachelriess:2004ds}. In the normal
hierarchy, it would completely oscillate into the $\nu_x$ flavor so
that it could not be seen in the charged-current (CC) channel of a
liquid argon detector, whereas the electron-scattering signal would
be reduced by about a factor of~7. On the other hand, in the
inverted hierarchy, we would have $p=\sin^2\theta_{12}\sim0.30$ and
thus a significant CC signal. Existing detectors, however, do not
have a sufficient $\nu_e$ sensitivity for a clear detection.

\begin{figure}
\centering
\includegraphics[width=1.0\textwidth]{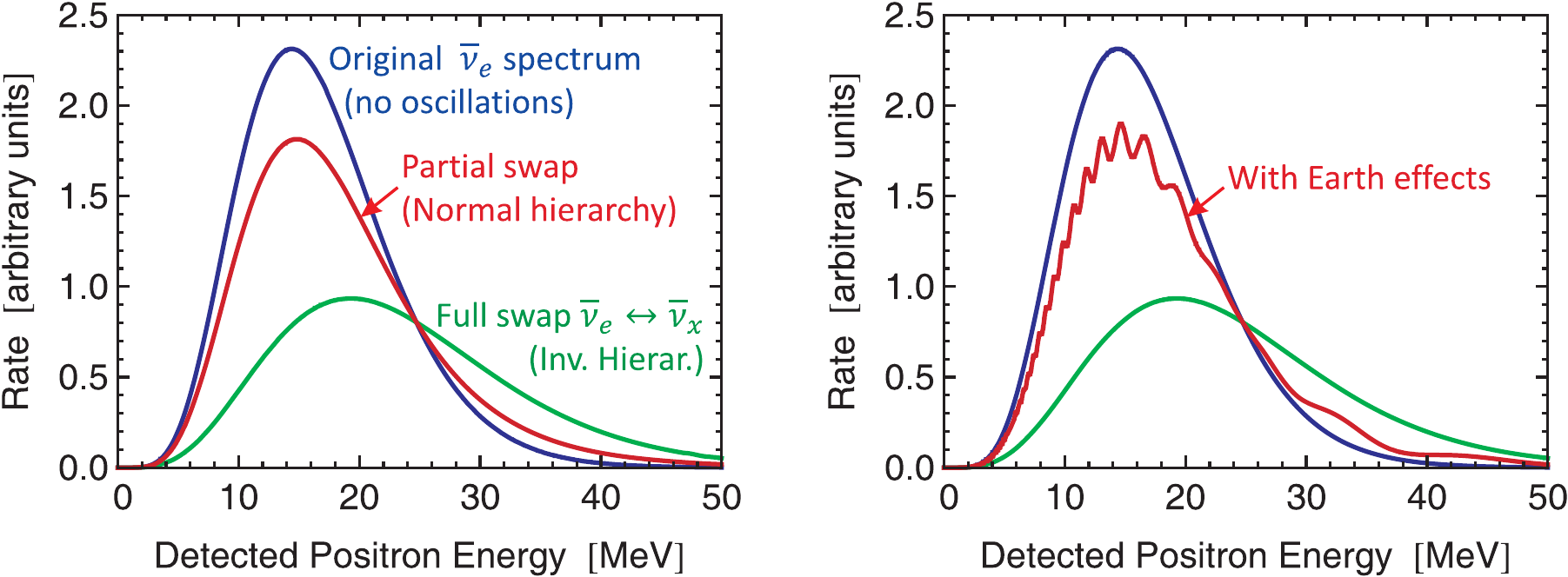}
\caption{Accretion-phase $\bar\nu_e$ signal in water Cherenkov
or scintillator detectors for different oscillation scenarios.
For the regeneration effect (right panel) an 8000~km path length in
the Earth is assumed.\label{fig:eartheffects}}
\end{figure}

Another option is to look for a signature in the $\bar\nu_e$
channel. During the accretion phase, the expected $\bar\nu_e$ and
$\bar\nu_x$ fluxes are very different (fig.~\ref{fig:basel}) so that
the expected detection signal depends strongly on the oscillation
scenario (fig.~\ref{fig:eartheffects}). However, in the absence of a
quantitatively reliable prediction of the flavor-dependent fluxes
and spectra it is difficult to distinguish between these cases. One
model-independent signature would be the matter regeneration effect
if the SN signal is received through the Earth in a ``shadowed''
position~\cite{Mirizzi:2006xx}. The Earth effect would imprint
energy-dependent modulations on the received $\bar\nu_e$ signal
(right panel of fig.~\ref{fig:eartheffects}) with a frequency that
depends on the distance traveled through the Earth. In principle,
these ``wiggles'' can be resolved, but not with present-day
detectors~\cite{Dighe:2003jg}. With the water Cherenkov technique
one would need a megaton class detector, whereas with a scintillator
detector a few thousand events would be enough due to the superior
energy resolution. One may also compare the signals between a
shadowed and an unshadowed detector~\cite{Dighe:2003be}. The signal
rise time is generically different between $\bar\nu_e$ and
$\bar\nu_x$ (fig.~\ref{fig:promptburst}) so that the rise time of
the oscillated $\bar\nu_e$ signal depends on the mixing scenario.
Conceivably, this signature can be used to determine the hierarchy,
although the effect is subtle~\cite{Chakraborty:2011ir}.

During the SN cooling phase, the shock wave propagates through the
envelope, eventually disturbs the resonance region, and may imprint
detectable features on the time-dependent neutrino
flux~\cite{Schirato:2002tg, Tomas:2004gr, Fogli:2004ff,
Dasgupta:2005wn, Choubey:2006aq}. Of course, the expectation of
strong signatures was originally driven by the perception of a
strong flavor dependence of the cooling fluxes that is not borne out
by modern simulations with the full range of neutrino interaction
channels. Therefore, any such signatures are likely somewhat subtle.
Moreover, the matter behind the shock wave will exhibit stochastic
density fluctuations from turbulent matter flows that can lead to
flavor equilibration~\cite{Friedland:2006ta, Fogli:2006xy}.

A major new issue was recognized only a few years ago, the impact of
collective or self-induced flavor conversions. The neutrinos
streaming from the SN core are so dense that they provide a large
matter effect for each other. The nonlinear nature of this
neutrino-neutrino effect renders its consequences very different
from the ordinary matter effect in that it results in collective
oscillation phenomena~\cite{Pantaleone:1992eq, Samuel:1993uw,
Samuel:1996ri, Pastor:2001iu, Sawyer:2005jk} that can be of
practical interest in the early universe for the oscillation of
neutrinos with hypothetical primordial
asymmetries~\cite{Dolgov:2002ab, Wong:2002fa, Abazajian:2002qx,
Pastor:2008ti, Mangano:2010ei}. These effects are also important in
SNe in the region up to a few 100~km above the neutrino
sphere~\cite{Duan:2005cp, Duan:2006an}, an insight that has
triggered a torrent of recent activities~\cite{Duan:2010bg}.

Collective effects are important in regions where the effective
neutrino-neutrino interaction energy $\mu$ exceeds a typical vacuum
oscillation frequency $\Delta m^2/2E$. In an isotropic ensemble we
have $\mu\sim\sqrt2 G_{\rm F}n_\nu$ with $n_\nu$ the neutrino
density. The current-current nature of low-energy weak interactions
implies that a factor $1-\cos\theta$ appears in the interaction
potential where $\theta$ is the angle between neutrino trajectories.
If the background is isotropic (approximately true for ordinary
matter), this term averages to~1. On the other hand, neutrinos
streaming from a SN core become more and more collinear with
distance, so the average interaction potential is reduced by a
suitable average $\langle1-\cos\theta\rangle$. One finds that $\mu$
effectively decreases with distance as $r^{-4}$ where two powers
derive from the geometric flux dilution, another two powers from the
increasing collinearity. Therefore, collective effects are important
only fairly close to the neutrino sphere.

\begin{figure}
\centering
\includegraphics[width=0.7\textwidth]{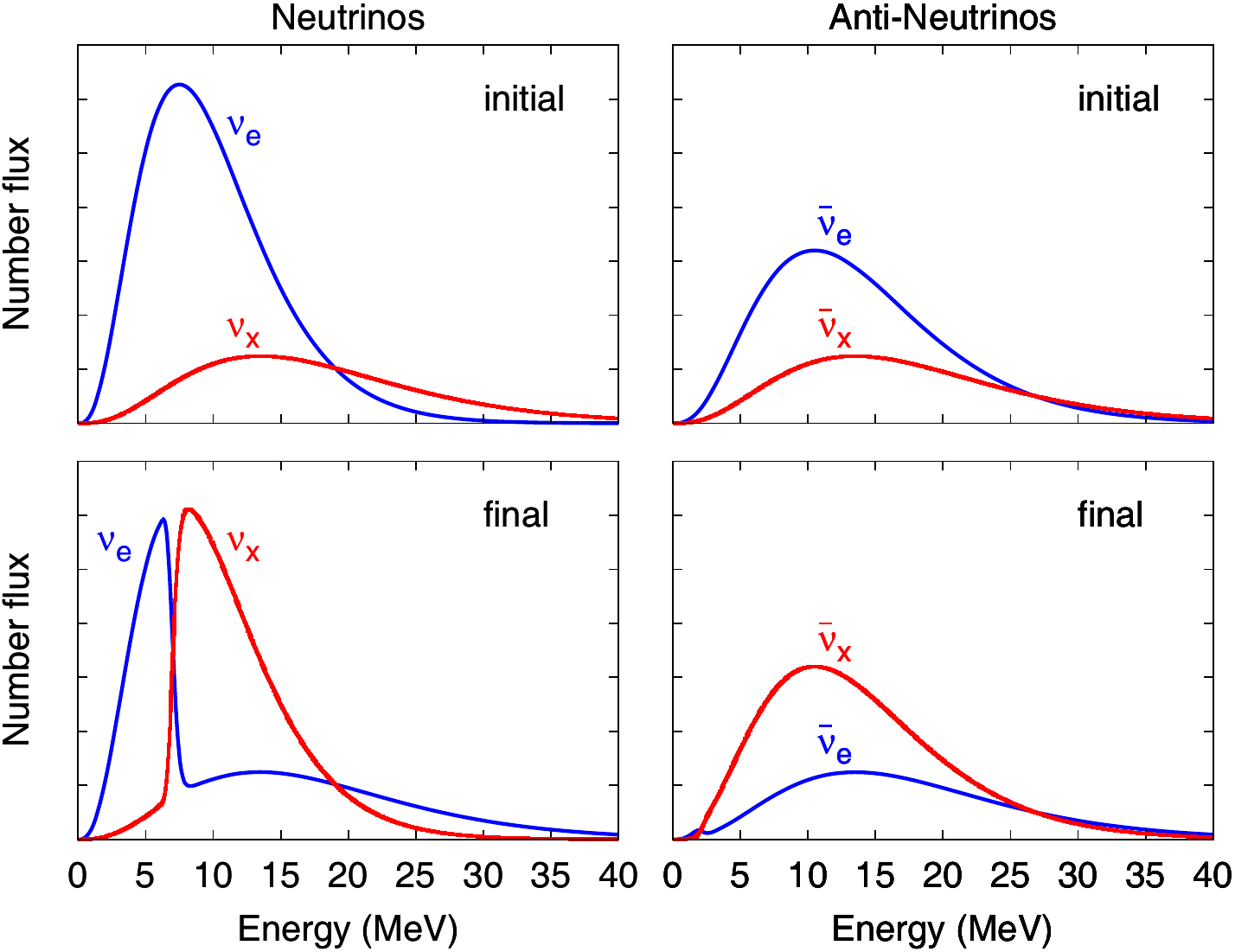}
\caption{Example for SN neutrino spectra before and after collective oscillation,
assuming inverted hierarchy and that
ordinary matter does not suppress self-induced conversions.
\label{fig:split}}
\end{figure}

Let us assume for now that collective effects are not affected by
matter. Let us further assume that we have a pronounced hierarchy of
number fluxes $F_{\nu_x}\ll F_{\bar\nu_e}<F_{\nu_e}$ that certainly
applies after bounce and during the accretion phase, but probably
does not apply during the cooling phase. In this scenario the impact
of collective oscillations is straightforward. Nothing new happens
for normal hierarchy (NH), whereas for the inverted hierarchy (IH)
the $\bar\nu_e$ flux is swapped with the $\bar\nu_x$ flux. In
addition, the $\nu_e$ flux is swapped with the $\nu_x$ flux, but
only for $E>E_{\rm split}$ where the energy $E_{\rm split}$ marks a
sharp ``spectral split,'' separating the swapped part of the
spectrum from the unswapped part (fig.~\ref{fig:split}). $E_{\rm
split}$ is fixed by the condition that the net $\nu_e$ flux
$F_{\nu_e}-F_{\bar\nu_e}$ is conserved~\cite{Raffelt:2007cb}. In
other words, there is no net flavor conversion: essentially one has
self-induced collective pair conversions
$\nu_e\bar\nu_e\to\nu_x\bar\nu_x$.

Collective oscillations at first seemed unaffected by matter because
its influence does not depend on neutrino
energies~\cite{Duan:2006an}. However, depending on emission angle,
neutrinos accrue different matter-induced flavor-dependent phases
until they reach a given radius. This ``multi-angle matter effect''
can suppress self-induced flavor
conversion~\cite{EstebanPretel:2008ni}. Based on schematic flux
spectra, this was numerically confirmed for accretion-phase SN
models where the density near the core is
large~\cite{Chakraborty:2011gd}. Self-induced conversion requires
that part of the spectrum is prepared in one flavor, the rest in
another. The collective mode consists of pendulum-like flavor
exchange between these parts without changing the overall flavor
content~\cite{Samuel:1993uw, Hannestad:2006nj, Duan:2007mv}. The
inevitable starting point is a flavor instability of the neutrino
distribution caused by neutrino-neutrino refraction. An
exponentially growing mode can be detected with a linearized
analysis of the evolution equations~\cite{Sawyer:2008zs,
Banerjee:2011fj}. This method was applied to realistic multi-angle
multi-energy neutrino fluxes and also confirm the suppression of
self-induced conversion for the investigated accretion-phase models
\cite{Sarikas:2011jc, Sarikas:2011am}. If these results turn out to
be generic, then for the accretion phase the survival probabilities
of table~\ref{tab:survival} remain applicable.

Likewise, the prompt $\nu_e$ burst should not be affected by
collective oscillations with the possible exception of very low-mass
progenitor stars. In this case the matter density is so low even at
shock break out that the MSW region is very close to the possible
collective oscillation region. In this case, interesting combined
effects between MSW and collective conversion have been
identified~\cite{Duan:2007sh, Dasgupta:2008cd, Cherry:2011fm}.

\begin{figure}
\centering
\includegraphics[width=0.6\textwidth]{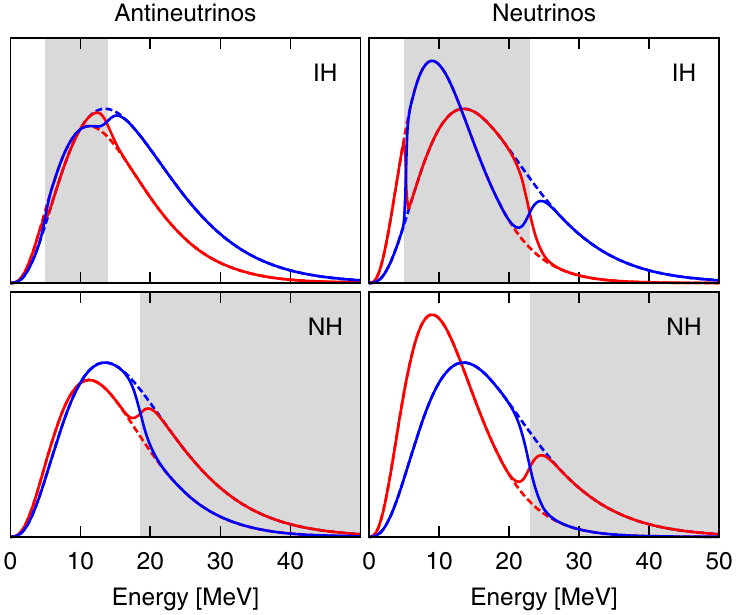}
\caption{Example for possible cooling phase SN neutrino spectra before
(dashed lines) and after
(solid lines) collective oscillations, but before possible MSW
conversions~\cite{Dasgupta:2009mg}. The panels are for $\nu$ and $\bar\nu$, each time for
inverted hierarchy (IH) and normal hierarchy (NH). Red lines
$e$--flavor, blue $x$--flavor. Shaded regions mark swap intervals.
\label{fig:multisplits}}
\end{figure}

During the cooling phase, the matter profile has become so low that
self-induced flavor conversions can operate unimpeded. The flavor
hierarchy of fluxes and spectral energies is not large, allowing for
more complicated conversion patterns---see
fig.~\ref{fig:multisplits} for an example. Multiple spectral swaps
and splits are possible~\cite{Dasgupta:2009mg}, where however
multi-angle effects play a crucial role~\cite{Banerjee:2011fj,
Raffelt:2008hr, Duan:2010bf}. At the present time it is not obvious
if one can arrive at generic predictions for what happens during the
cooling phase. The interacting neutrino gas, however, remains a
fascinating system for collective motions of what is effectively an
interacting spin system~\cite{Balantekin:2006tg, Raffelt:2010za,
Raffelt:2011yb}, with analogies in the area of
superconductivity~\cite{Pehlivan:2011hp, Yuzbashyan:2008}.

\section{Conclusion}                            \label{sec:conclusion}

The physics of stars is inseparably intertwined with that of
neutrinos and we have discussed some of the many fascinating topics
at the interface of these fields. Solar neutrinos play a special
role in that the measured $\nu_e$ flux provided first evidence for
flavor oscillations: a deep particle-physics issue was directly
connected to low-energy neutrino astronomy. Learning about the exact
chemical composition of the solar interior is the next frontier of
solar neutrino spectroscopy. Beyond the Sun, neutrinos play a
crucial role as an energy-loss channel that is a necessary
ingredient for understanding stellar evolution. In the same spirit,
we can use observed properties of stars, notably in globular
clusters, to learn about neutrinos, such as their electromagnetic
properties, or about other low-mass particles such as axions. These
hypothetical particles would also emerge from the Sun. The search
for solar axions with the CAST experiment has provided important
constraints and a next-generation axion helioscope is being
discussed. It would probe deeply into realistic parameter space.

Of course, the royal discipline of neutrino astrophysics is their
role in stellar core collapse and the dynamics of supernova
explosions. SN~1987A remains the only observed astrophysical
neutrino source other than the Sun. It has confirmed our basic
understanding of supernova physics and has provided several
particle-physics limits that remain of topical interest to date.
Detecting the diffuse neutrino background from all past supernovae
in the universe is the next milestone for low-energy neutrino
astronomy. A high-statistics neutrino observation of the next nearby
supernova with one of the operating or future large-scale
experiments will provide a bonanza of astrophysical, neutrino and
particle-physics information. Many questions remain open about
supernova dynamics, nucleosynthesis in the neutrino-driven wind, and
flavor oscillations in an environment of dense matter and neutrinos.
The next generation of large-scale detectors remains to be developed
and built. So while we wait for the next supernova neutrino
observation, a lot of numerical, theoretical, and experimental work
remains to be done.

\acknowledgments Partial support by the Deutsche
Forschungsgemeinschaft under Grants No.\ TR~27 and EXC~153 is
acknowledged.


\end{document}